\documentclass[preprint,prd,singlecolumn,superscriptaddress,nofootinbib]{revtex4-2}

\usepackage{geometry}
\usepackage{amsmath,amssymb}
\usepackage{amssymb}
\usepackage{rotating}
\usepackage{mathtools}
\usepackage{color}
\usepackage{comment}
\usepackage{amsmath,amssymb}
\usepackage{dcolumn}
\usepackage{bm}
\usepackage{longtable}
\usepackage{multirow}

\newcommand\op[1]{\hat{#1}}						  %operator
\newcommand\vecl[1]{{\bf{#1}}}         		  %vector letter
\newcommand\vecs[1]{\boldsymbol{#1}}     		  %vector symbol
\newcommand\sqb[1]{\left[#1\right]}     		  %square brackets
\newcommand\nba[1]{\left(#1\right)}     		  %normal brackets adaptable
\newcommand{\ket}[1]{\left| #1 \right>} 		  %for Dirac bras
\newcommand{\bra}[1]{\left< #1 \right|}  		  %for Dirac kets
\newcommand{\signs}[3]{(#1\ #2\ #3)\ }			  %3 signs
\newcommand{\foursigns}[4]{(#1\ #2\ #3 \ #4)\ }  %4 signs

\newcommand{\func}[1]{\operatorname{#1}}
\begin{document}

%\title{Long-Range Neutrinoless Double Beta Decay Mechanisms} 

\title{Long-range neutrinoless double beta decay mechanisms}

\author{J. Kotila}
\email{jenni.kotila@jyu.fi}
\affiliation{Finnish Institute for Educational Research, University of Jyv\"askyl\"a, P.O. Box 35, Jyv\"askyl\"a FI-40014, Finland}
\affiliation{Center for Theoretical Physics, Sloane Physics Laboratory, Yale University, New Haven, Connecticut 06520-8120, USA}

\author{J. Ferretti}
\email{jacopo.j.ferretti@jyu.fi}
\affiliation{Department of Physics, University of Jyv\"askyl\"a, P.O. Box 35, Jyv\"askyl\"a FI-40014, Finland}
\author{F. Iachello}
\email{francesco.iachello@yale.edu}
\affiliation{Center for Theoretical Physics, Sloane Physics Laboratory, Yale University, New Haven, Connecticut 06520-8120, USA}

%\author{XXX}
%\email{YYY}
%\affiliation{ZZZ}

\begin{abstract}
Understanding the origin of lepton number violation is one of the
fundamental questions in particle physics today. Neutrinoless double beta
decay provides a way in which this violation can be tested. In this article,
we derive the form of hadronic and leptonic matrix elements for all possible
long-range mechanisms of neutrinoless double beta decay. With these, we
calculate the numerical values of the nuclear matrix elements (NME) and
phase space factors (PSF) by making use of the interacting boson model of
the nucleus (IBM-2) for NMEs and of exact Dirac wave functions for the PSFs.
We show that:\ (I) lepton number violation can occur even with zero (or very
small) neutrino mass and (II) the angular correlations of the emitted
electrons can distinguish between different models of non-standard (NS)
mechanisms. We set limits on the coupling constants of some NS models, in
particular Left-Right models and SUSY models.\end{abstract}

\maketitle

\section{Introduction}
\begin{figure}[htbp]
\centering
\includegraphics[width=7cm]{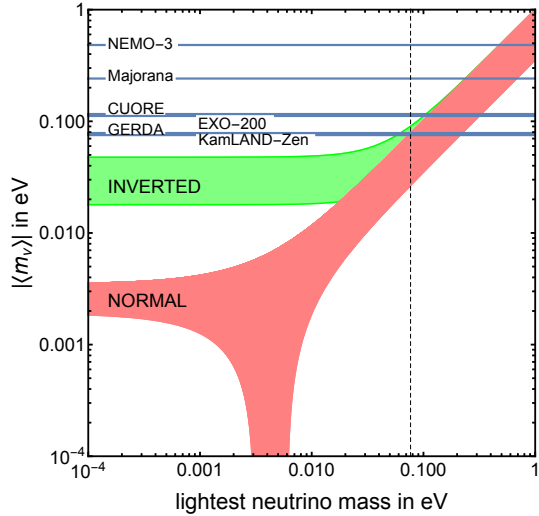} 
\caption{Current limits to $\langle m_\nu \rangle$ from CUORE \cite{Alfonso:2015wka}, GERDA \cite{Agostini:2013mzu}, EXO \cite{Auger:2012ar}, KAMLAND-Zen \cite{Gando:2012zm}, NEMO-3 \cite{Arnold:2015wpy}, and Majorana \cite{Aalseth:2017btx}, and using IBM-2, Argonne SRC, isospin restored NMEs, and $g_{\rm A} = 1.269$. The limit from the Planck Collaboration \cite{Ade:2015xua} is shown by a vertical line.}
\label{fig:EXP-LIMITS}
\end{figure}

In spite of several attempts by many groups \cite{Alfonso:2015wka,Agostini:2013mzu,Auger:2012ar,Gando:2012zm,Arnold:2015wpy,Aalseth:2017btx,Azzolini:2019}, neutrinoless double beta ($0\nu\beta\beta$) decay has not yet been observed.
This observation is crucial for understanding lepton number violation.
After the discovery of neutrino oscillations \cite{Fukuda:1998mi,Ahmad:2002jz,Eguchi:2002dm}, attention has been mostly concentrated on the mass mechanism of $0\nu\beta\beta$, wherein the three species of neutrinos have masses $m_i$ and couplings to the electron neutrino $U_{ei}$.
In this case, the inverse half-life is given by
\begin{equation}
	\left[\tau_{1/2}^{0\nu}(0^+ \rightarrow 0^+)\right]^{-1} = G_{0\nu} \left| M_{0\nu} \right|^2 \left| \frac{\langle m_\nu \rangle}{m_e} \right|^2  \mbox{ },
\end{equation}
where $G_{0\nu}$ is the so-called phase space factor, PSF, $M_{0\nu}$ the nuclear matrix element, NME, and $\langle m_\nu \rangle = \displaystyle \sum_i U_{ei}^2 m_i$, where $U_{ei}$ are the charged-current leptonic mixing matrix elements.
The allowed values of the masses consistent with oscillation experiments are given in Fig. \ref{fig:EXP-LIMITS}.
A limit to the sum of the masses, $\displaystyle \sum_i m_i$, is also provided by cosmology \cite{Ade:2015xua}, with current results $\displaystyle \sum_i m_i < 0.23$ eV. This limit is also shown in Fig. \ref{fig:EXP-LIMITS}.
%%%%%%%%%%%%%%%%%%%%%
%%%%%%%%%%%%%%%%%%%%%

In view of the difficulties to observe the mass mechanism, other mechanisms need to be investigated. They can be divided into short-range and long-range. 
In  previous papers \cite{Graf:2018ozy, Graf:2020ozy}, we have investigated all possible short-range non-standard mechanisms. In this paper, we investigate all possible long-range non-standard mechanisms.
The aim of the paper is: I) to provide explicit formulas for the nuclear matrix elements (NMEs) and phase-space factors (PSFs) from which the decay rate can be calculated; II) to provide numerical values of the NMEs and PSFs obtained by making use of the interacting boson model for the NMEs \cite{Barea:2009zza,Barea:2013bz,Barea:2015kwa} and of exact Dirac wave functions for the PSFs \cite{Kotila:2012zza}.
Previous approaches to non-standard mechanisms of $0\nu\beta\beta$ are by Doi {\it et al. } \cite{Doi:1981mi} and Tomoda \cite{Tomoda:1990rs}, who investigated L-R models \cite{Pati:1974yy,Mohapatra:1974hk,Senjanovic:1975rk,Doi:1985dx,Hirsch:1996qw}, by Ali {\it et al.}  \cite{Ali:2007ec}, and Cirigliano {\it et al.} \cite{cir2017} who provided a general framework for the investigation of non-standard models.

In the last part of this paper, we provide explicit formulas for two models of non-standard mechanisms, a L-R model \cite{Pati:1974yy,Mohapatra:1974hk,Senjanovic:1975rk,Doi:1985dx,Hirsch:1996qw} and a SUSY model \cite{Mohapatra:1986su,Vergados:1986td,Hirsch:1995zi,Babu:1995vh,Hirsch:1995cg,Faessler:1996ph}. We will show explicitly that even if the neutrino masses were to be very small these models would give rise to neutrinoless double beta decay, and thus to lepton number violation.

\section{General effective Lagrangian}
The most general effective Lagrangian for long-range mechanisms is the Lorenz-invariant combination of leptonic, $j_\alpha$, and hadronic, $J_\alpha$, currents with definite tensor structure and chirality \cite{Ali:2007ec,Pas:1999fc},
\begin{equation}
	\label{eqn:effective-L}
	{\mathcal L} = \frac{G_{\rm F} \cos \theta_{\rm c}}{\sqrt 2} \left[ U_{ei} ~ j_{V-A}^{\mu,i} J_{V-A,\mu}^\dag
	+ \displaystyle \sum_{\alpha,\beta} \epsilon_{\alpha,i}^\beta ~ j_\beta^i J_\alpha^\dag + {\rm h.c.} \right] \mbox{ },
\end{equation}
where the hadronic and leptonic currents are defined as $J_\alpha^\dag = \bar u \mathcal O_\alpha d$ and $j_\beta^i = \bar e \mathcal O_\beta \nu_i$, where the index $i$ runs over the neutrino mass eigenstates.
The indices $\alpha,\beta$ are $V\mp A$, $S\mp P$, $T\mp T_5$.
For the tensor operators $T\mp T_5$, we use ${\mathcal O}_{{\rm T}_\rho} = 2 \sigma^{\mu\nu} P_\rho$, $\sigma^{\mu\nu} = \frac{i}{2} \left[\gamma^\mu,\gamma^\nu\right]$, with $P_\rho = \frac{1 \mp \gamma_5}{2}$ the L-R projector.
In Eq. (\ref{eqn:effective-L}) we have isolated the standard model contribution proportional to $U_{ei}$, where $U_{ei}$ is the PMNS mixing matrix element \cite{Pontecorvo:1957qd,Maki:1962mu}, from non-standard contributions.

For purposes of derivation of the appropriate formulas for the calculation of the decay rate, it is convenient to re-write the Lagrangian ${\mathcal L}$ in the form
\begin{equation}
	\label{eqn:rewritten-L}
	\begin{array}{rcl}	
	{\mathcal L} & = & \frac{G_{\rm F} \cos \theta_{\rm c}}{\sqrt 2} \left[ j_{V-A}^{\mu, i} \tilde J_{V-A,\mu}^\dag 
	+ j_{V+A}^{\mu, i} \tilde J_{V+A,\mu}^\dag + j_{S-P}^{i} \tilde J_{S-P}^\dag  \right. \\ 
	& + & \left. j_{S+P}^{i} \tilde J_{S+P}^\dag + j_{T - T_5}^{\mu \nu, i} \tilde J_{T - T_5,\mu \nu}^\dag + 
	j_{T + T_5}^{\mu \nu, i} \tilde J_{T + T_5,\mu \nu}^\dag 
	+ {\rm h.c.} \right] \mbox{ },
	\end{array}
\end{equation}
where 
\begin{subequations}
\label{eqn:hadronic01}
\begin{equation}
%	\label{eqn:hadronic01}
	\tilde J_{V-A,\mu}^\dag = \left( U_{ei} + \epsilon_{V-A,i}^{V-A} \right) J_{V-A,\mu}^\dag + \epsilon_{V+A,i}^{V-A} J_{V+A,\mu}^\dag,
\end{equation}
\begin{equation}
	\tilde J_{V+A,\mu}^\dag = \epsilon_{V+A,i}^{V+A} J_{V+A,\mu}^\dag + \epsilon_{V-A,i}^{V+A} J_{V-A,\mu}^\dag,
\end{equation}
\end{subequations}

\begin{subequations}
\begin{equation}
	\tilde J_{S-P}^\dag = \epsilon_{S-P,i}^{S-P} J_{S-P}^\dag + \epsilon_{S-P,i}^{S+P} J_{S+P}^\dag,
\end{equation}
\begin{equation}
	\tilde J_{S+P}^\dag = \epsilon_{S+P,i}^{S+P} J_{S+P}^\dag + \epsilon_{S+P,i}^{S-P} J_{S-P}^\dag,
\end{equation}
\end{subequations}
\begin{subequations}
\label{eqn:hadronic03}
\begin{equation}
	\tilde J_{T-T_5,\mu\nu}^\dag = \epsilon_{T-T_5,i}^{T-T_5} J_{T-T_5,\mu\nu}^\dag + \epsilon_{T-T_5,i}^{T+T_5} J_{T-T_5,\mu\nu}^\dag,
\end{equation}
\begin{equation}
%	\label{eqn:hadronic03}
	\tilde J_{T+T_5,\mu\nu}^\dag = \epsilon_{T+T_5,i}^{T+T_5} J_{T+T_5,\mu\nu}^\dag + \epsilon_{T+T_5,i}^{T-T_5} J_{T+T_5,\mu\nu}^\dag,
\end{equation}
\end{subequations}
In the following section, the hadronic currents will be explicitly given by going from quarks to nucleons and by taking their non-relativistic limit.

Each non-standard model of long-range $\beta\beta$ decay is defined by the set of 12 coefficients, $\epsilon_{V\mp A}^{V\mp A}$, $\epsilon_{S\mp P}^{S\mp P}$, and $\epsilon_{T\mp T_5}^{T\mp T_5}$, which are those of Eqs. (\ref{eqn:hadronic01})-(\ref{eqn:hadronic03}) summed over $i$ with appropriate weighting factors. 
Non-zero coefficients for L-R models \cite{Pati:1974yy,Mohapatra:1974hk,Senjanovic:1975rk,Hirsch:1996qw,Doi:1985dx} and $R$-parity-violating SUSY models \cite{Mohapatra:1986su,Vergados:1986td,Hirsch:1995zi,Babu:1995vh,Hirsch:1995cg,Faessler:1996ph} are listed in Table \ref{tab:LR-SUSY-coeffs}.
%%%%%%%%%%%%%%%%%%%%%%%%%%%%%%%%%%%%%%%
\begin{table}[htbp]
\centering
\begin{tabular}{c|c}
\hline
\hline
Model & Non-zero $\epsilon$ \\
\hline
L-R with $W_{\rm R}$ & $\epsilon_{V+A}^{V-A}$, $\epsilon_{V\mp A}^{V+A}$ \\
RPV SUSY & $\epsilon_{S+P}^{S\mp P}$, $\epsilon_{V-A}^{V-A}$, $\epsilon_{T+T_5}^{T+T_5}$ \\
\hline
\hline
\end{tabular}
\caption{Non-zero coefficients $\epsilon_\alpha^\beta$ for the two models discussed in this paper.}
\label{tab:LR-SUSY-coeffs}
\end{table}
%%%%%%%%%%%%%%%%%%%%%%%%%%%%%%%%%%%%%%%
The coefficients $\epsilon_{\alpha, i}^\beta$ can also be expressed as
\begin{equation}
	\epsilon_{\alpha, i}^\beta = \hat \epsilon_\alpha^\beta U_{ei}^{(\alpha,\beta)}  \mbox{ },
\end{equation}
where $U_{ei}^{(\alpha,\beta)}$ are mixing parameters for non-standard model neutrinos.
Since the Lagrangian $\mathcal L$ describes also ordinary $\beta$-decay, the parameters $\hat \epsilon_\alpha^\beta$ are constrained by precision measurements in allowed nuclear beta decays \cite{Severijns:2006dr}.

%%%%%%%%%%%%%%%%%%%%%%%%%%%%%%%%%%%%%%%%
\begin{figure*}[htbp]
\centering
\begin{minipage}{9pc}
\includegraphics[width=9pc]{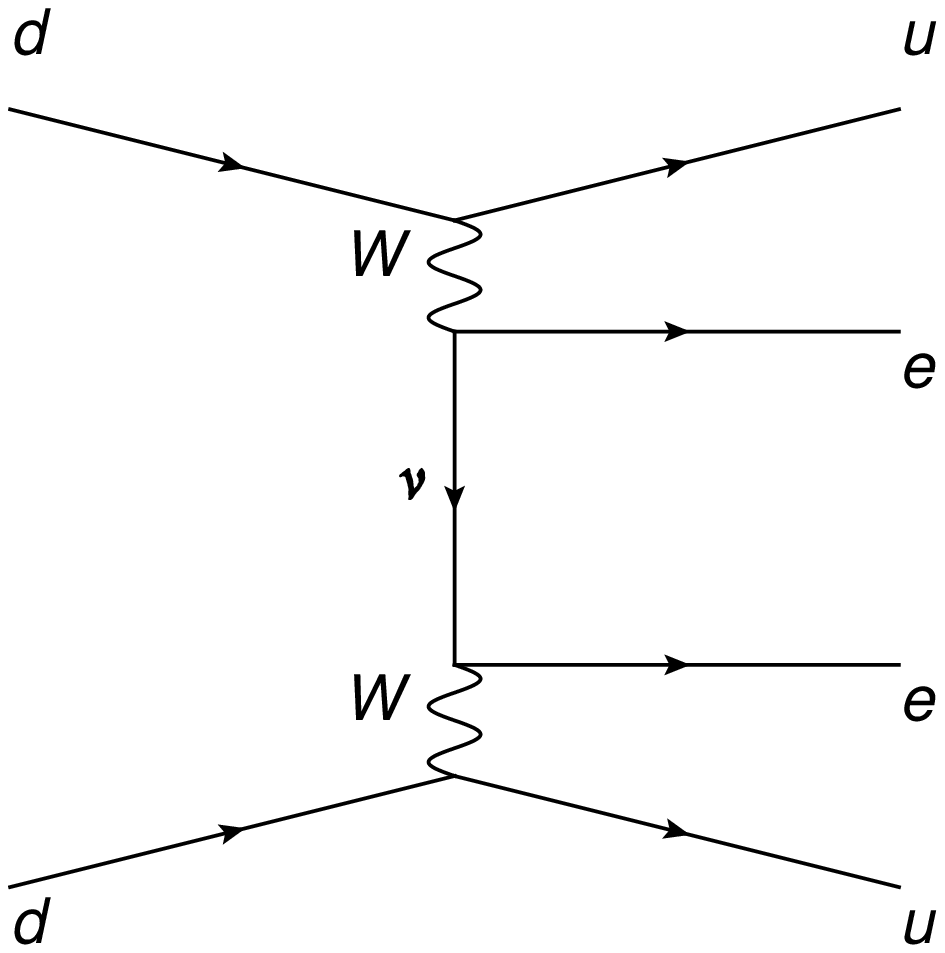}
\end{minipage}
\hspace{2pc}
\begin{minipage}{9pc}
\includegraphics[width=9pc]{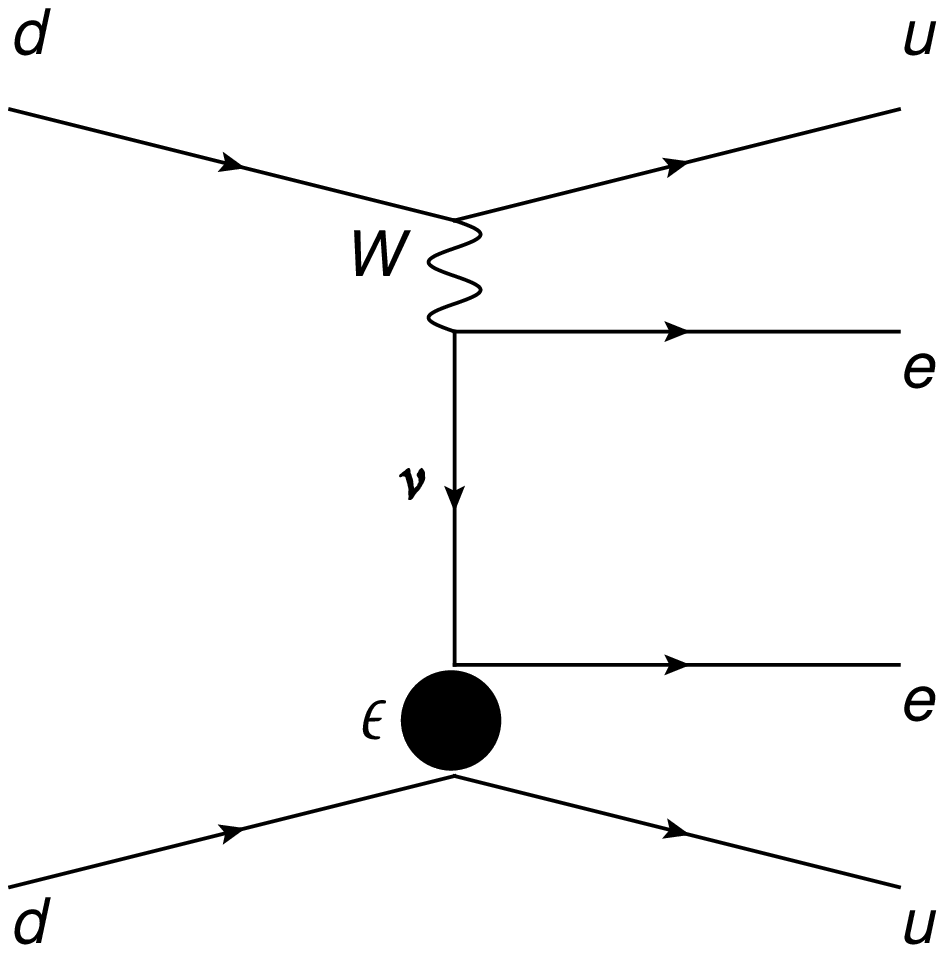}
\end{minipage}
\hspace{2pc}
\begin{minipage}{9pc}
\includegraphics[width=9pc]{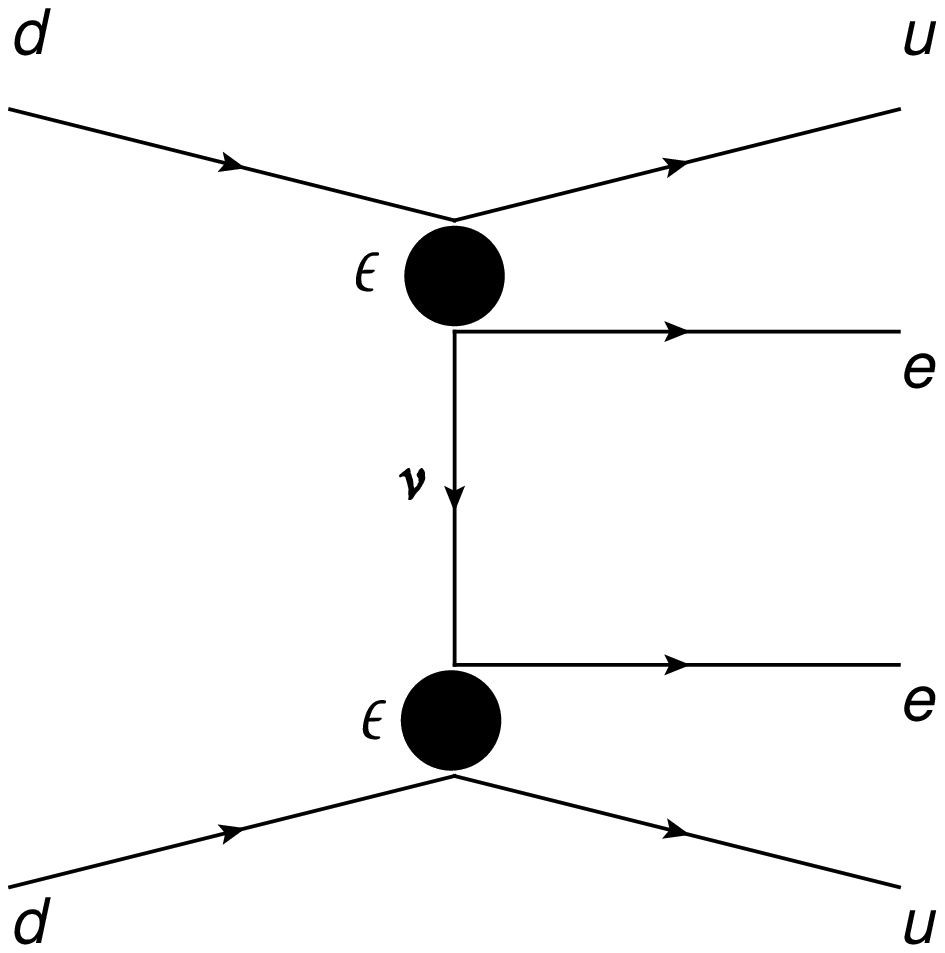}
\end{minipage}
\caption{Contributions to long-range $0\nu\beta\beta$ decay: standard $V$-$A$-like neutrino exchange (left); product of an ordinary ($V$-$A$-like) interaction and a non-standard one (centre); product of two non-standard interactions (right).}
\label{fig:Feynman}
\end{figure*}
%%%%%%%%%%%%%%%%%%%%%%%%%%%%%%%%%%%%%%%%

For double beta decay, one has to consider the product of two effective Lagrangians, Eq. (\ref{eqn:effective-L}), which can be written as
\begin{equation}
	\label{eqn:time-ordered}
	\begin{array}{rcl}
	({\mathcal L}_1 {\mathcal L}_2) & = & \frac{G_{\rm F}^2 \cos^2 \theta_{\rm c}}{2} \left[ U_{ei}^2 j_{V-A}^{\mu,i} J_{V-A,\mu}^\dag 
	j_{V-A,\nu}^i J_{V-A}^{\dag,\nu}  \right. \\
	& + & \left. U_{ei} j_{V-A}^{\mu,i} J_{V-A,\mu}^\dag \displaystyle \sum_{\alpha,\beta}
	\epsilon_{\alpha,i}^\beta j_\beta^i J^{\dag,\alpha} + \displaystyle \sum_{\alpha,\beta} \epsilon_{\alpha,i}^\beta j_\beta^i J^{\dag,\alpha}
	\displaystyle \sum_{\gamma,\delta} \epsilon_{\gamma,i}^\delta j_\delta^i J^{\dag,\gamma }\right] + {\rm h.c.} \mbox{ },
	\end{array}
\end{equation}
where the first term on the right-hand side of Eq. (\ref{eqn:time-ordered}) is the standard $V-A$ neutrino exchange contribution, the second term is the product of an ordinary standard interaction with a non standard one of order $\epsilon$, and the third term is the product of two non-standard interactions of order $\epsilon^2$, see Fig. \ref{fig:Feynman}.

\section{Differential decay rate}
Starting with the Lagrangian of Eq. (\ref{eqn:effective-L}), one can calculate the differential rate for both $2\nu\beta\beta$ and $0\nu\beta\beta$ decays. Here, we discuss $0\nu\beta\beta$ decays.
The differential rate for this process was derived by Doi {\it et al. } \cite{Doi:1981mi} and Tomoda \cite{Tomoda:1990rs}. It can be written as \cite{Tomoda:1990rs}
\begin{equation}
	\label{eqn:diff-rate}
	{\rm d}W_{0\nu\beta\beta} = 2 \pi \displaystyle \sum_{\rm spin} \left| R_{0\nu\beta\beta} \right|^2 
	\delta(E_1 + E_2 + E_{\rm F} - E_{\rm I}) \frac{{\rm d} {\bf p}_1}{(2 \pi)^3} \frac{{\rm d} {\bf p}_2}{(2 \pi)^3}  \mbox{ },
\end{equation}
where $E_{\rm I, F}$ are the initial/final nuclear state energies, $E_{1,2}$ and ${\bf p}_{1,2}$ the electron energies and momenta, respectively. 
In Eq. (\ref{eqn:diff-rate}), $R_{0\nu\beta\beta}$ is the full matrix element of the $0\nu\beta\beta$ decay process and the sum over spins is over the projections $s_1, s_2$ of the electrons and the final nuclear states. The full matrix element $R_{0\nu\beta\beta}$ can be formally expressed as
\begin{equation}
	R_{0\nu\beta\beta} = \left\langle {\rm F} e_{{\bf p}_1, s_1} e_{{\bf p}_2, s_2} \right| {\mathcal L}_1 
	{\mathcal L}_2 \left| {\rm I} \right\rangle \mbox{ },
\end{equation}
where F and I denote the final and initial nuclear states and $e_{{\bf p}_1, s_1}, e_{{\bf p}_2, s_2}$ the two emitted electrons.	
Since the Lagrangian of Eq. (\ref{eqn:time-ordered}) contains the product of two currents, symbolically denoted by $J_\beta^\dag J_\alpha^\dag$, one should in principle calculate all the matrix elements by going through all the intermediate states in the odd-odd nucleus. 
This is a daunting task since all the states up to an energy $E \approx 100$ MeV contribute.
It is therefore customary to treat the summation over intermediate states in the closure approximation, i.e. summing over a complete set of states
\begin{equation}
	\displaystyle \sum_{\mathcal N} \left\langle {\rm F} \right| J_\beta^\dag \left| {\mathcal N} \right\rangle 
	\left\langle {\mathcal N} \right| J_\alpha^\dag \left| {\rm I} \right\rangle
	\simeq \left\langle {\rm F} \right| J_\beta^\dag J_\alpha^\dag \left| {\rm I} \right\rangle  \mbox{ }.
\end{equation}
The approximation is well justified for $0\nu\beta\beta$ decays because the energy of the virtual neutrinos exchanged between nucleons ($\omega \approx 100$ MeV) is much larger than the typical excitation energy of the intermediate nuclear states ($E_{\rm N} - E_{\rm I} \approx 10$ MeV).
In the closure approximation, the matrix elements $R_{0\nu\beta\beta}$ can be written as \cite{Doi:1981mi,Tomoda:1990rs}
\begin{equation}
	\label{eqn:rate-Tomoda}
	\begin{array}{l}
	R_{0\nu\beta\beta} = \frac{4}{\sqrt 2} \frac{G_{\rm F}^2 \cos^2 \theta_{\rm c}}{2} \displaystyle \sum_{i,\alpha,\beta} 
	\int {\rm d}{\bf x} {\rm d}{\bf y} \int \frac{{\rm d}{\bf k}}{(2\pi)^3} \left\langle {\rm F} \right| 
	\tilde J^{\dag}_{\beta i} ({\bf y}) \tilde J^{\dag}_{\alpha i} ({\bf x}) \left| {\rm I} \right\rangle 
	{\rm e}^{i {\bf k}\cdot({\bf y}-{\bf x})} \\ 
	\hspace{0.5cm} \times \mbox{ } \bar e_{{\bf p}_2 s_2'} ({\bf y}) {\mathcal O}_\beta \frac{1}{2 \omega} 
	\left[ \frac{\omega \gamma^0 - {\bf k} \cdot {\bm \gamma} + m_i}{\omega + A_1} 
	- \frac{\omega \gamma^0 + {\bf k} \cdot {\bm \gamma} - m_i}{\omega + A_2}\right] 
	{\mathcal O}_\alpha e_{{\bf p}_1 s_1'}^{\rm c} ({\bf x}) \mbox{ }.
	\end{array}
\end{equation}
Here, $\tilde J^{\mu\dag}_{\alpha i}$ denote the hadronic currents of Eqs. (\ref{eqn:hadronic01})-(\ref{eqn:hadronic03}), and ${\mathcal O}_{\alpha}$ the corresponding leptonic currents coupled to them as in Eq. (\ref{eqn:rewritten-L}).
For the case of the standard $V-A$ current, ${\mathcal O}_\alpha = \gamma_\mu \frac{1}{2} (1 - \gamma_5)$, for the non-standard $V+A$, ${\mathcal O}_\alpha = \gamma_\mu \frac{1}{2} (1 + \gamma_5)$, for $S\mp P$, ${\mathcal O}_\alpha =  \frac{1}{2} (1 \mp \gamma_5)$, and for $T \mp T_5$, ${\mathcal O}_\alpha = \sigma_{\mu\nu} \frac{1}{2} (1 \mp \gamma_5)$.
${\bf k}$ is the neutrino momentum, $m_i$ its mass, and $e_{{\bf p}s}$ the electron wave functions \cite{Doi:1981mi,Tomoda:1990rs}.
The quantities $A_j =  E_j + \left\langle E_{\rm N} \right\rangle - E_{\rm I}$ ($j = 1,2$) are called closure energies. 
Note also that the $Q$ value of the process is given by 
\begin{equation}
	Q_{\beta\beta} = E_{\rm I} - E_{\rm F} - 2 m_e \mbox{ }.
\end{equation}	

The evaluation of the matrix element $R_{0\nu\beta\beta}$ is still rather complicated, since, in general, the leptonic part is nested with the hadronic part. However, as shown in \cite{Doi:1981mi,Tomoda:1990rs}, a good approximation is to evaluate the electron wave functions at the nuclear radius, in which case the evaluation factorizes into two parts, the hadronic and leptonic parts. The evaluation of the hadronic part will be discussed in Sec. \ref{Hadronic matrix elements}, that of the leptonic part in Sec. \ref{Leptonic matrix elements}, and the combination of the two previous parts in Sec. \ref{Results}.

The combination of the hadronic nuclear part (NME) and the leptonic part (PSF) will produce the final results for the differential rate for the $0^+ \rightarrow 0^+$ $0\nu\beta\beta$ decay \cite{Doi:1981mi,Tomoda:1990rs},
\begin{equation}
	\label{eqn:dW0vdE1dcostheta12}
	\frac{{\rm d}^2 W_{0\nu}}{{\rm d}E_1 {\rm d}\cos \theta_{12}} = \left(a^{(0)} + a^{(1)} \cos \theta_{12} \right) w_{0\nu}  \mbox{ },
\end{equation}
where $E_1$ is the energy of the first electron, $\theta_{12}$ the angle between the two emitted electrons, $p_i=|\vec{p_i}|$
\begin{equation}
	\label{eqn:w0nu}
	w_{0\nu} = \frac{(G_{\rm F} \cos \theta_{\rm c})^4 m_e^2}{16\pi^5} \mbox{ } p_1 p_2 E_1 E_2  \mbox{ },
\end{equation}	
and $E_1 + E_2 + E_{\rm F} = E_{\rm I}$. The quantities $a^{(0)}$ and $a^{(1)}$ will be given in Sec. \ref{Results} for specific models.
Integrating Eq. (\ref{eqn:dW0vdE1dcostheta12}) over $E_1$, we obtain
\begin{equation}
	\label{eqn:dW0vdcostheta12}
	\frac{{\rm d} W_{0\nu}}{{\rm d}\cos \theta_{12}} = \frac{{\rm ln}~2}{2} \left[A^{(0)} + A^{(1)} \cos \theta_{12}\right] 
	= \frac{{\rm ln}~2}{2} A^{(0)} \left[ 1 + {\mathcal K} \cos \theta_{12} \right]  \mbox{ },
\end{equation}
where
\begin{equation}
	A^{(i)} = \displaystyle \int a^{(i)} w_{0\nu} {\rm d}E_1 \mbox{ , } \mbox{  } i = 1,2 \mbox{ }.
\end{equation}	
and ${\mathcal K}=A^{(1)}/A^{(0)}$.
Further integration over $\cos \theta_{12}$ yields the half-life for $0\nu\beta\beta$ decays
\begin{equation}
	\label{eqn:half-life-tau}
	\left[ \tau_{1/2}^{0\nu} (0^+ \rightarrow 0^+) \right]^{-1} = \frac{W_{0\nu}}{{\rm ln} ~ 2} = A^{(0)}  \mbox{ }.
\end{equation}	

\section{Hadronic matrix elements}
\label{Hadronic matrix elements}
We are interested here in the matrix elements of the products of the currents $\tilde J_\alpha$ of Eqs. (\ref{eqn:hadronic01})-(\ref{eqn:hadronic03}), which are the sum of currents with chirality $V\mp A$, $S \mp P$, $T \mp T_5$,
\begin{equation}
	\left\langle {\rm F} \right| \tilde J_\beta \tilde J_\alpha \left| {\rm I} \right\rangle  \mbox{ }.
\end{equation}
In order to calculate these matrix elements we need to write down the explicit non-relativistic form of the product of currents.
In order to do so, we need first to write down the nuclear matrix elements of the quark currents, take their non-relativistic limit and then their product.

\subsection{Nucleon matrix elements of quark currents}
\label{Nucleon matrix elements of quark currents}
Introducing the notation $N = \left(\begin{array}{c} p \\ n\end{array}\right)$ for nucleon iso-doublets, the nucleon matrix elements of color-singlet quark currents have the following structure \cite{Graf:2018ozy,Tomoda:1990rs,Ali:2007ec,Adler:1975he}
\begin{equation}
	\label{eqn:nucleon-ME01}
	\bra{p}\bar{u}(1\pm\gamma_5)d\ket{n} = \bar{N} \tau^+ \sqb{F_S(q^2) \pm F_{P}(q^2)\gamma_5} N' \mbox{ }, 
\end{equation}
\begin{equation}
	\label{eqn:nucleon-ME02}
	\begin{array}{rcl}		
	\bra{p}\bar{u}\gamma^\mu(1\pm\gamma_5)d\ket{n} & = & \bar{N} \tau^+ 
	\sqb{F_V(q^2)\gamma^\mu-i\frac{F_W(q^2)}{2m_p}\sigma^{\mu\nu}q_\nu}N' \\
	& \pm & \bar{N} \tau^+ \sqb{F_A(q^2)\gamma^\mu\gamma_5 
	- \frac{F_{P'}(q^2)}{2m_p}\gamma_5q^\mu}N'  
	\end{array}
\end{equation}	
and
\begin{equation}
	\label{eqn:nucleon-ME03}
	\bra{p}\bar{u}\sigma^{\mu\nu}(1\pm\gamma_5)d\ket{n} = \bar{N} \tau^+ \sqb{J^{\mu\nu} \pm \frac{i}{2} 
	\epsilon^{\mu\nu\rho\sigma} J_{\rho\sigma}} N' \mbox{ },
\end{equation}
where $m_p$ is the nucleon mass, $\tau^+$ the isospin-raising operator, and
\begin{equation}
	\label{eqn:Jmunu}
	J^{\mu\nu} = F_{T_1}(q^2)\sigma^{\mu\nu} + i\frac{F_{T_2}(q^2)}{m_p}(\gamma^\mu q^\nu 
	- \gamma^\nu q^\mu) + \frac{F_{T_3}(q^2)}{m_p^2}(\sigma^{\mu\rho} q_\rho q^\nu 
	- \sigma^{\nu\rho} q_\rho q^\mu) \mbox{ }.
\end{equation}
In the previous expressions, Eqs. (\ref{eqn:nucleon-ME01})-(\ref{eqn:Jmunu}), the $q^2$-dependent form factors, $F_X(q^2)$, are introduced, with $X = S, P, V, W, A, P', T_1, T_2$ or $T_3$. 
Following Ref. \cite{Graf:2018ozy}, the previous form-factors, with the exception of $F_{P}(q^2)$ and $F_{P'}(q^2)$, are parametrized by means of a dipole form,
\begin{align}
\label{eq:formfactor_dipole}
	F_X(q^2) = \frac{g_X}{\left(1 + q^2 / m_X^2\right)^2}  \mbox{ },
\end{align}
where the coupling constants $g_X$ give the value of the form factor at zero momentum transfer, $g_X = F_X(0)$. 
The values of $g_X$ and $m_X$ used in this article are given in Table \ref{tab:form-factors-parameters}.
%%%%%%%%%%%%%%%%%%%%%%%%%%%%%%%%%%%%%%%
\begin{table}[htbp]
\centering
\begin{tabular}{ccc}
\hline
\hline
Form factor & $m_X$ [GeV] & $g_X$ \\
\hline
$F_S(q^2)$ & 0.84 & 1 \\
$F_V(q^2)$ & 0.84 & 1 \\
$F_W(q^2)$ & 0.84 & 3.70 \\
$F_A(q^2)$ & 1.09 & 1.269 \\
$F_{T_1}(q^2)$ & 0.84 & 1 \\
$F_{T_2}(q^2)$ & 0.84 & $-3.30$ \\
$F_{T_3}(q^2)$ & 0.84 & 1.34 \\
\hline
\hline
\end{tabular}
\caption{Values of $g_X$ and $m_X$ for the form factors $F_X(q^2)$, with $X = S, V, W, A, T_1, T_2$ and $T_3$.}
\label{tab:form-factors-parameters}
\end{table}
%%%%%%%%%%%%%%%%%%%%%%%%%%%%%%%%%%%%%%%

The value of $m_V = 0.84$ GeV is taken from experiments on the electromagnetic form factor of the nucleon \cite{Iachello:1972nu,Bijker:2004yu} and $g_V = 1$ from the conserved vector current (CVC) hypothesis.
The form factor $F_W(q^2)$ parameters, $m_W$ and $g_W$, can also be extracted from the experimental data, due to the relation of $F_W(q^2)$ with the Pauli form factor $F_2(q^2)$ \cite{Iachello:1972nu,Bijker:2004yu} and the isovector anomalous magnetic moment of the nucleon, $g_W = \mu_p - \mu_n = 3.70$. The value of $g_A$ is determined from neutron decays \cite{Tanabashi:2018oca} and $m_A$ from neutrino scattering data \cite{Schindler:2006jq}. No experimental information is available for the tensor form factors, $F_{T_i}(q^2)$ ($i = 1,2,3$). Ref.~\cite{Gonzalez-Alonso:2018omy} quotes a value of $0.987\pm 0.055$ for $F_{T_1}(0) \equiv g_{T_1}$. An old calculation \cite{Adler:1975he} estimates, from the MIT bag model, $F_{T_1}(0) \equiv g_{T_1} = 1.38$, $F_{T_2}(0) \equiv g_{T_2}=-3.30$ and $F_{T_3}(0) \equiv g_{T_3} = 1.34$. Here, we take $g_{T_1} = 1$, $g_{T_2} = -3.30$, and $g_{T_3} = 1.34$, and $m_{T_1} = m_{T_2} = m_{T_3} = m_V$.
In recent years, particular attention has been devoted to the scalar, $S$, and the pseudoscalar form factors, $P$ and $P'$.
The value $g_S = 1.02\pm0.11$ has been quoted \cite{sch2007}. In Table \ref{tab:form-factors-parameters}, we have taken $g_S = 1$ and $m_S = m_V = 0.84$ GeV.
The pseudoscalar form factors $P$ and $P'$ cannot be parametrized in the simple dipole form, Eq. (\ref{eq:formfactor_dipole}). 
For the intrinsic form factor $P$, we take the form suggested in \cite{Severijns:2006dr}
\begin{equation}
	\label{eq:formfactor_PS}
	F_{P}(q^2) = \frac{g_{P}}{\left(1 + q^2/m_V^2\right)^2} \frac{1}{1 + q^2/m^2_\pi}  \mbox{ }.
\end{equation}
This form factor diverges at $q^2 = 0$ in the chiral limit, $m_\pi = 0$. A value of $g_{P} = 349\pm9$ is suggested in \cite{Severijns:2006dr}.
%We take $g_{P'} = 349$ and $m_P = m_A = 1.09$ GeV.

For the induced pseudoscalar form factor $P'$, we make use of the parametrization suggested in \cite{Simkovic:1999re}, based on the partially conserved axial-vector current (PCAC) hypothesis,
\begin{equation}
	\label{eq:formfactor_P}
	F_{P'}(q^2) = \frac{g_A}{\left(1 + q^2 / m_A^2\right)^2} \frac{1}{1 + q^2 / m_\pi^2}\frac{4m_p^2}{m_\pi^2}
	\left(1-\frac{m_\pi^2}{m_A^2}\right) \equiv \frac{g_{\rm P'}}{\left(1 + q^2 / m_A^2\right)^2} 
	\frac{1}{1 + q^2 / m_\pi^2} \mbox{ },
\end{equation}
where $m_\pi = 0.138$~GeV is the pion mass, and $g_{\rm P'} = 182$ $g_A$.
With this parametrization, the two form factors $F_P(q^2)$ and $F_{P'}(q^2)$, are consistent with each other.

\subsection{Non-relativistic expansion}
The nuclear currents are obtained first by performing a non-relativistic expansion of the nucleon currents, Eqs. (\ref{eqn:nucleon-ME01})-(\ref{eqn:nucleon-ME03}).
The expansion, in powers of $\frac{q}{m_p}$, is obtained by means of a Foldy-Wouthuysen transformation of the nuclear currents \cite{Tomoda:1990rs,Foldy:1949wa,Rose,Bhalla}. See Appendix \ref{Nuclear currents}.

In particular, the nonrelativistic expansion of the scalar/pseudoscalar nucleon current, $J_{S\pm P} = (1\pm\gamma_5)$, is given by 
\begin{align} 
	\label{eq:sbilinear}
	J_{S\pm P} = 
	F_S(q^2) I \pm \frac{F_{P}(q^2)}{2m_p} \mbox{ } \boldsymbol{\sigma} \cdot \mathbf{q} + \dots \mbox{ }.
\end{align}
The nonrelativistic expansion of the vector current, $J^\mu_{V\pm A} = \gamma^\mu(1\pm\gamma_5)$, can be written as
\begin{equation}
	\begin{array}{rcl} 
	\label{eq:vbilinear}
	J^\mu_{V\pm A} &= &
	g^{\mu i}\left[\mp F_A(q^2){\bm \sigma}_i - \frac{F_V(q^2)}{2m_p} {\bf Q}_i I 
	+ \frac{F_V(q^2)+F_W(q^2)}{2m_p} \mbox{ } i (\boldsymbol{\sigma} \times \mathbf{q})_i \right. \\
	& \pm & \left. \frac{F_{P'}(q^2)}{4m_p^2} \mbox{ } {\bf q}_i \boldsymbol{\sigma} \cdot \mathbf{q}\right] 
	+ g^{\mu 0}\left[F_V(q^2) I \pm \frac{F_A(q^2)}{2m_p} \boldsymbol{\sigma} \cdot \mathbf{Q} \right. \\
	& \mp & \left. \frac{F_{P'}(q^2)}{4m_p^2} q_0 \boldsymbol{\sigma} \cdot \mathbf{q} 
	- \frac{1}{4m_p^2} F_W(q^2) {\bf q} \cdot ({\bf q} - i \boldsymbol{\sigma} \times \mathbf{Q}) \right] 
	\end{array}  \mbox{ }.
\end{equation}
The nonrelativistic reduction of the tensor bilinears, $J^{\mu\nu}_{T\pm T_5} = \sigma^{\mu\nu}(1\pm\gamma_5)$, is
\begin{equation} 
	\label{eq:tbilinear}
	\begin{array}{rcl}	
	J^{\mu\nu}_{T\pm T_{5}} & = & F_{T_1}(q^2) g^{\mu j} g^{\nu k}\varepsilon_{ijk}\sigma^i + (g^{\mu i} g^{\nu 0} - 
	g^{\mu 0} g^{\nu i})T_i \\ 
	& \pm & \frac{i}{2} \varepsilon^{\mu\nu\rho\sigma} \left[ (g_{\rho i}g_{\sigma 0} 
	- g_{\rho 0} g_{\sigma i}) T^i + F_{T_1}(q^2) g_{\rho m} g_{\sigma n} 
	\varepsilon^{mni}\sigma_i \right] + \dots \mbox{ },
	\end{array}
\end{equation}
where
\begin{align} \label{eq:tbilinearterm}
	T^i &= \frac{i}{2m_p}\left[\left(F_{T_1}(q^2) - 2 F_{T_2}(q^2) \right)q^i I
	+ F_{T_1}(q^2) (\boldsymbol{\sigma}_a \times \mathbf{Q})^i \right].
\end{align}
In the above expressions, $I$ is the $2\times 2$ identity matrix, $\boldsymbol{\sigma}$ a Pauli matrix, the convention for $g^{\mu \nu}$ is that of Eq. (\ref{eqn:gmunu}), $q_\mu = p_\mu - p_\mu'$ the 4-momentum transfer and $Q_\mu = p_\mu + p_\mu'$ the momentum sum.
The terms depending explicitly on $\vecl{Q}=\vec{p} + \vec{p'}$ are called recoil terms \cite{Tomoda:1990rs}.

%\color{red}
The nuclear currents can be obtained from Eqs. (\ref{eq:sbilinear})-(\ref{eq:tbilinearterm}) by summing over all neutrons, located at positions $\vecl{r}_a$, as
\begin{equation}
	\mathcal{I}_K^\Xi(\vecl{x}) = \sum_a\tau^a_+\delta(\vecl{x}-\vecl{r}_a) J_{K,a}^\Xi  \mbox{ },
\end{equation}
where $J_{K,a}^{\Xi }$ denotes any of the nucleon currents. For positron
emission, $\tau _{+}$ is replaced by $\tau _{-}$ and the sum is over all
protons. This current is needed for the calculation of single $\beta $ decay
and of double $\beta $ decays with the emission of two neutrinos, $2\nu
\beta \beta $, when the closure approximation is not used.

The transformation properties of the currents under Lorentz transformations, 
$SO(3,1)$, and rotations, $SO(3)$, are given in Table \ref{table:III}.

%%%%%%%%%%%%%%%%%%%%%%%%%%%%%%%%%%%%%%%
\begin{table}[htbp]
\centering
\begin{tabular}{ccccc}
\hline
\hline
\text{Current} & $SO(3,1)$ &&\multicolumn{2}{c}{$SO(3)$} \\ 
\hline
$J$ 	&rank-0 &&\multicolumn{2}{c}{ \text{rank-0}} \\ 
\hline
\multirow{ 2}{*}{$J^{\mu }$}&\multirow{ 2}{*}{\text{rank-1}} & &$\mu =0$ &rank-0 \\ 
&  && $\mu =k$&rank-1 \\ 
\hline
\multirow{ 4}{*}{$J^{\mu \nu }$}&\multirow{ 4}{*}{\text{rank-2}}  && $\mu =\nu =0$&rank-0 \\ 
&&&$\mu =k,\nu =0 $ &\multirow{ 2}{*}{rank=1} \\ 
&&&$\mu =0,\nu =k$\\
&  & & $\mu =k,\nu =j$&rank-2 \\
\hline
\hline
\end{tabular}
\caption{Transformation properties of the currents under Lorentz
transformations and rotations.}
\label{table:III}\end{table}
%%%%%%%%%%%%%%%%%%%%%%%%%%%%%%%%%%%%%%%

For $0\nu \beta \beta $ decays in the closure approximation we need to
consider products of these currents denoted schematically by%
\begin{equation}
\Pi _{1}=JJ,\Pi _{2}=J^{\mu \nu }J_{\rho \sigma },\Pi _{3}=J^{\mu }J_{\nu
},\Pi _{4}=J^{\mu }J_{\rho \sigma },\Pi _{5}=J^{\mu }J,
\end{equation}%
the rank of which under Lorentz transformations is given in Table \ref{table:IV}.

%%%%%%%%%%%%%%%%%%%%%%%%%%%%%%%%%%%%%%%
\begin{table}[htbp]
\centering
\begin{tabular}{ccccc}
\hline
\hline
\text{Product} & \text{Rank }$SO(3,1)$ \\ 
\hline
$\Pi _{1}=JJ $& 0 \\ 
$\Pi _{2}=J^{\mu \nu }J_{\rho \sigma }$ & 0,1,2,3,4 \\ 
$\Pi _{3}=J^{\mu }J_{\nu }$ & 0,1,2 \\ 
$\Pi _{4}=J^{\mu }J_{\rho \nu } $& 1,2,3 \\ 
$\Pi _{5}=J^{\mu }J $& 1\\
\hline
\hline
\end{tabular}
\caption{Properties of products of currents under Lorentz transformations.}
\label{table:IV}\end{table}
%%%%%%%%%%%%%%%%%%%%%%%%%%%%%%%%%%%%%%%

(If we restrict ourselves to terms of order $\varepsilon $, see Fig. \ref{fig:Feynman}, only
the products $\Pi _{3},\Pi _{4},\Pi _{5}$ are needed). For each rank-$r$
under the Lorentz group, we also have a rank-$s$ under the rotation group,
given by $s=r,r-1,...,0$. Understanding which products contribute to the
calculation of the rates is more complicated here than in the case of
short-range non-standard mechanisms \cite{Graf:2018ozy} because in the latter there is no
propagator between the two vertices in Fig. \ref{fig:Feynman}, while for long-range the
expression%
\begin{equation}
	\label{eqn:nu-propagator}
\frac{1}{\omega }\left[ \frac{\omega \gamma ^{0}-\mathbf{k\cdot \gamma }%
+m_{i}}{\omega +A_{1}}-\frac{\omega \gamma ^{0}+\mathbf{k\cdot \gamma }-m_{i}%
}{\omega +A_{2}}\right]
\end{equation}%
appears \cite{Doi:1981mi,Tomoda:1990rs}. In this expression, we have three terms, which we call
respectively $m_{i}$ terms, $\omega $ terms and $\mathbf{k}$ terms. While
the $m_{i}$ and $\omega $ terms are rank-0 under rotation, the $\mathbf{k}$
term is rank-1. The transition operators for the $m_{i}$ and $\omega $ terms
are then the same as for short-range which, for $0_{I}^{+}\rightarrow
0_{F}^{+}$ transitions, are the Lorentz rank-0 tensors $\Pi _{1}=JJ,\Pi
_{2}=J^{\mu \nu }J_{\mu \nu },\Pi _{3}=J^{\mu }J_{\mu }$, and the zeroth
component of the rank-1 tensors $\Pi _{4}=J^{\mu }J_{\mu \nu }$ and $\Pi
_{5}=J^{\mu }J$ summed over the three directions of $\mathbf{k}$, plus
additional terms obtained from the rank-1 components of $\Pi _{2},\Pi
_{3},\Pi _{4},$ and $\Pi _{5}$ dotted with $\mathbf{k}$. Some of these
additional terms were considered in \cite{Doi:1981mi,Tomoda:1990rs} for the special case of L-R
models, and will be also considered by us separately.

Apart from these additional terms, the transition operators for $0\nu \beta
\beta $ decay in the closure approximation are then given, for $m_{i}$ and $%
\omega $ terms, by%
\begin{equation}
\mathcal{H}_{K}(\mathbf{x,y)=}\sum_{a\neq b}\tau _{+}^{a}\tau _{+}^{b}\delta
\left( \mathbf{x-r}_{a}\right) \delta \left( \mathbf{y-r}_{b}\right) \Pi
_{K,ab}^{\Xi },
\end{equation}%
with the products $\Pi $ symmetrized with respect to the nucleon indices $%
a,b $, and given in Appendix \ref{Nucleon current products}. For $\mathbf{k}$ terms, they are instead
given by%
\begin{equation}
\mathcal{H}_{K}\left( \mathbf{x,y}\right) =\sum_{a\neq b}\tau _{+}^{a}\tau
_{+}^{b}\delta \left( \mathbf{x-r}_{a}\right) \delta \left( \mathbf{y-r}%
_{b}\right) \Pi _{K,ab}^{\prime \Xi },
\end{equation}%
where the prime indicates a sum over the direction of $\mathbf{k\equiv q}$ and a change in sign due to the minus sign for the $\mathbf{k}$ term relative to the $m_i$ and $\omega$ terms in Eq. (\ref{eqn:nu-propagator}). Splitting into the parallel and perpendicular direction of $\mathbf{q}$, we then
have%
\begin{equation}
\label{modification}
	\begin{array}{rcl}
\overline{\left( {\bm\sigma _{a}}\mathbf{\cdot} {\bm\sigma_{b}}\right) }
&=&-\left[\left( {\bm\sigma_{a} }\mathbf{\cdot \hat{q}}\right) \cdot \left( 
{\bm\sigma_{b} }\mathbf{\cdot \hat{q}}\right) +\left( {\bm\sigma%
_{a} }\mathbf{\times \hat{q}}\right) \cdot \left( {\bm\sigma_{b} }\mathbf{%
\times \hat{q}}\right)\right]   \\
&=&\frac{1}{3}\left({ \bm\sigma_{a} }\mathbf{\cdot}{\bm \sigma_{b} }\right) -%
\frac{2}{3}S_{ab} \\
\overline{S_{ab}} &=&-\frac{4}{3}\left( {\bm\sigma_{a} }\mathbf{\cdot}
{\bm\sigma_{b} }\right) -\frac{1}{3}S_{ab},  
\end{array}%
\end{equation}
where the relations (\ref{eqn:recoupling01}) and (\ref{eqn:recoupling02}) have been used. The operators $\Pi
^{\prime }$ are then the same as $\Pi $ but with $\left( {\bm\sigma_{a} }%
\mathbf{\cdot}{\bm \sigma_{b} }\right) $ and $S_{ab}$ replaced by $\overline{%
\left( {\bm\sigma_{a} }\mathbf{\cdot}{\bm \sigma_{b} }\right) }$ and $%
\overline{S_{ab}}$.

The additional terms to be considered here are the rank-1 tensors under
Lorentz transformations and rotations. For the $\Pi _{1}$ product there are
no such terms since $\Pi _{1}=JJ$ is rank-0 under Lorentz transformations
and rotations. For the product $\Pi _{2}=J^{\mu \nu }J_{\rho \sigma }$ the
construction of the rank-1 tensors is rather complicated and since we do not
use it here, where only terms of order $\varepsilon $ are kept, we omit it.
We consider instead the explicit construction for the product $J^{\mu
}J_{\nu }$. The rank-0 part of this product is $J^{\mu }J_{\mu }$ and we
have considered it previously. The rank-1 part is $J^{\mu }J^{\nu }-J^{\nu
}J^{\mu }$ as discussed in \cite{Doi:1981mi}. For $\mu =\nu =0$ this is obviously zero,
while for $\mu =0,\nu =k$ it is $J^{0}J^{k}-J^{k}J^{0}$ and for $\mu =k,\nu
=j$ it is $J^{k}J^{j}-J^{j}J^{k}$, which is the vector product of the two
currents $\mathbf{J}_{a}\times \mathbf{J}_{b}$. The additional rank-1 terms
of the product $J^{\mu }J^{\nu }$ are given in Appendix \ref{Nucleon current products}. We consider next
the product $\Pi _{4}=J^{\mu }J_{\rho \nu }$. The rank-1 Lorentz tensor is $%
J^{\mu }J_{\mu \nu }$. This tensor has under rotations a rank-0 component
and a rank-1 component. Both of them have been given in Appendix \ref{Nucleon current products}, Eq. (\ref{eq:rank1}), 
$\nu =0$ rank-0 and $\nu =j$ rank-1. Additional terms are here the $\nu =j$
components of $J^{\mu }J_{\mu \nu }$. The construction of the rank-1 terms
coming from the Lorentz rank-2 and -3 tensors is rather complicated and it
will not be given here. Finally, we consider the product $\Pi _{5}=J^{\mu }J$%
. This product is rank-1 Lorentz tensor and the additional terms are here
its $\mu =j$ components given in Appendix \ref{Nucleon current products}. The transition operators for
the additional terms will be denoted by%
\begin{equation}
	\label{eq:transitionop}
\overline{\mathcal{H}}_{K}\left( \mathbf{x,y}\right) =\sum_{a\neq b}\tau
_{+}^{a}\tau _{+}^{b}\delta \left( \mathbf{x-r}_{a}\right) \delta \left( 
\mathbf{x-r}_{b}\right) \overline{\Pi }_{K,ab}^{\Xi }
\end{equation}%
where the bar over $\Pi $ indicates scalar product in the direction of $%
\mathbf{k=q}$.

\color{black}%ends the red color

\subsection{Nuclear matrix elements}
\label{Nuclear matrix elements}
%\color{red}
In view of Eq.~(\ref{eqn:nu-propagator}), while for short-range only one hadronic and one leptonic
matrix element need to be calculated, in the case of long-range several
hadronic and leptonic matrix elements need to be calculated. For rank-0
tensors there are three hadronic and three leptonic matrix elements. We
denote these three hadronic matrix elements by%
\begin{equation}
M_{K,j}=\left\langle F\left\vert \mathcal{H}_{K,j}\right\vert I\right\rangle
,
\end{equation}%
where $K=1,...,5$ denotes the product of currents and $j=1,3,4$ three types
of matrix elements that need to be calculated. Here we follow the notation
of Tomoda \cite{Tomoda:1990rs} that we have followed in all our previous calculations of $%
0\nu \beta \beta $ decays \cite{Graf:2018ozy} and that is used in our computer programs. It
should be understood that in the calculation of the matrix elements for $j=4$%
, $\left( {\bm\sigma }_{a}\mathbf{\cdot}{\bm \sigma }_{b}\right) $ and $%
S_{ab} $ should be replaced by $\overline{\left( {\bm\sigma }_{a}\mathbf{%
\cdot}{\bm\sigma }_{b}\right) }$ and $\overline{S}_{ab}$.

The three terms that need to be calculated are identified by the so-called
neutrino potential, $v(q)$. These terms, called $m_{i},\omega $ and $\mathbf{%
k}$-terms, respectively, in the previous subsection, are denoted here,
following Tomoda \cite{Tomoda:1990rs}, by indices $j=1,3$ and $4$. The three neutrino
potentials $v_{i}$ ($i=1,3,4$) are written in momentum space as:

$m_{i}$ terms%
\begin{equation}
	\label{eqn:vpot1}
v_{1}(q)=\frac{2}{\pi }\frac{1}{q\left( q+\tilde{A}\right) }
\end{equation}

$\omega $ terms%
\begin{equation}
	\label{eqn:vpot3}
v_{3}(q)=\frac{2}{\pi }\frac{1}{\left( q+\tilde{A}\right) ^{2}}
\end{equation}

$\mathbf{k}$ terms%
\begin{equation}
	\label{eqn:vpot4}
v_{4}(q)=\frac{2}{\pi }\frac{q+2\tilde{A}}{q\left( q+\tilde{A}\right) ^{2}}
\end{equation}%
Here $\tilde{A}$ is the average closure energy, $\tilde{A}=\frac{1}{2}\left(
A_{1}+A_{2}\right) $. Details of the neutrino potentials are given in \cite[Appendix 2]{Tomoda:1990rs}, where the approximation $\omega =\left\vert \mathbf{k}%
\right\vert $ is used. For comparison, in the case of the short-range
mechanisms, the neutrino potential is given by%
\begin{equation}
	\label{eqn:vpotSR}
v_{short}(q)=\frac{2}{\pi }\frac{1}{m_{e}m_{p}}.
\end{equation}

Although only matrix elements for $K=3,4,5$ are needed here (see Fig. \ref{fig:Feynman}), we
give for future reference the explicit expressions of the matrix elements $%
\mathcal{M}_{K}$ for all five products of currents $\Pi _{K},K=1,2,3,4,5$.
They are given by 
\begin{equation}
	\label{eq:nme1}
	{\mathcal M}_1 = g_S^2 {\mathcal M}_F \mbox{ } \mbox{ } \signs{+}{+}{-} \frac{g_{P}^2}{12}
	\left({\mathcal M}_{GT}^{'PP}+{\mathcal M}_T^{'PP}\right)
	\mbox{ },
\end{equation}
\begin{equation}
	\label{eq:nme2}
	{\mathcal M}_2 = -2g_{T_1}^2{\mathcal M}_{GT}^{T_1 T_1}
	\mbox{ },
\end{equation}
\begin{equation}
	\label{eq:nme3}
	\begin{array}{rcl}
	{\mathcal M}_3 & = & g_V^2 {\mathcal M}_{F} \mbox{ } \mbox{ }  \signs{-}{-}{+} g_A^2 {\mathcal M}_{GT}^{AA}
	\mbox{ } \mbox{ } \signs{+}{+}{-} \frac{g_A g_{P'}}{6} \nba{{\mathcal M}^{\prime AP'}_{GT} 
	+ {\mathcal M}^{\prime AP'}_{T}}  \\
	& + & \frac{\nba{g_V+g_{W}}^2}{12}\nba{-2{\mathcal M}^{\prime WW}_{GT} + {\mathcal M}^{\prime WW}_{T}}  
	 \mbox{ } \mbox{ } \signs{-}{-}+{} \frac{g_{P'}^2}{48} \nba{{\mathcal M}^{\prime\prime P'P'}_{GT} 
	 + {\mathcal M}^{\prime\prime P'P'}_{T}}
	\end{array}
	\mbox{ },
\end{equation}
\begin{equation}
	\label{eq:nme4}
	{\mathcal M}_4 = \foursigns{-}{-}{+}{+} i g_A g_{T_1} {\mathcal M}_{GT}^{A T_1} \mbox{ } \mbox{ } 
	\foursigns{+}{+}{-}{-} i \frac{g_{P'}g_{T_1}}{12}\nba{{\mathcal M}^{\prime P'T_1}_{GT} 
	+ {\mathcal M}^{\prime P'T_1}_{T}}
	\mbox{ },
\end{equation}
\begin{equation}
	\label{eq:nme5}
	\begin{array}{rcl}
	{\mathcal M}_5 & = & g_S g_V{\mathcal M}_F \mbox{ } \mbox{ } 
	\foursigns{+}{+}{-}{-} \frac{g_A g_{P}}{12} \nba{\tilde{{\mathcal M}}^{AP}_{GT} 
	+ \tilde{{\mathcal M}}^{AP}_{T}} \\
	& & \foursigns{-}{-}{+}{+} \frac{g_{P}g_{P'}}{24}\nba{{\mathcal M}^{\prime q_0 PP'}_{GT} 
	+ {\mathcal M}^{\prime q_0 PP'}_{T}}
	\end{array} \mbox{ },
\end{equation}
where for ${\mathcal M}_4$ and ${\mathcal M}_5$ only the zero component of the products $\Pi_4$ and $\Pi_5$ is included.
In each case we keep track of signs corresponding to different combinations of chiralities.
For the first three operators, three sign possibilities are presented and they correspond to the following combinations of chiralities (in this order): $RR$, $LL$ and $(1/2)\nba{RL + LR}$. For the fourth and fifth operators, a row of four signs is shown, as in those cases the two hadronic currents have different Lorentz structures, thus all four possible combinations of chiralities have to be considered (in this order): $RR$, $LL$, $RL$ and $LR$.
To keep the expressions simple, when three/four signs are the same, we show only a single sign.
In Eqs. (\ref{eq:nme1})-(\ref{eq:nme5}), we have separated out from the form factors $F_X(q^2)$, the so-called charges, i.e., the values at $q^2= 0$, $F_X(0) = g_X$ given in Sec. \ref{Nucleon matrix elements of quark currents}. The $q$-dependence is then included in the matrix elements.

Defining the form factors
\begin{equation}
	\label{eqn:hVVq2}
	\tilde{h}_{VV}(q^2) = \frac{1}{\left(1+q^2/m_V^2\right)^4}  \mbox{ },
\end{equation}
\begin{equation}
	\tilde{h}_{AA}(q^2) = \frac{1}{\left(1+q^2/m_A^2\right)^4}  \mbox{ },
\end{equation}
\begin{equation}
	\tilde{h}_{AX}(q^2) =	\frac{1}{\left(1 + q^2/m_A^2\right)^2}\frac{1}{\left(1 + q^2/m_V^2\right)^2}  \mbox{ },
\end{equation}
\begin{equation}
	\tilde{h}_{AP}(q^2) = \tilde{h}_{AP'}(q^2) =	\frac{1}{\left(1 + q^2/m_A^2\right)^4}\frac{1}{\left(1 + q^2/m_\pi^2\right)}  
	\mbox{ },
\end{equation}
\begin{equation}
	\tilde{h}_{XP}(q^2) = \tilde{h}_{XP'}(q^2) =	\frac{1}{\left(1 + q^2/m_V^2\right)^2} \frac{1}{\left(1 + q^2/m_A^2\right)^2}
	\frac{1}{\left(1 + q^2/m_\pi^2\right)}  
	\mbox{ },
\end{equation}
\begin{equation}
	\label{eqn:hPPq2}
	\tilde{h}_{PP}(q^2) = \tilde{h}_{P'P'}(q^2) =  \tilde{h}_{PP'}(q^2) 
	= \frac{1}{\left(1 + q^2/m_A^2\right)^4}\frac{1}{\left(1 + q^2/m_\pi^2\right)^2}  \mbox{ },
\end{equation}
with $m_V = 0.84$ GeV, $m_A = 1.09$ GeV, $m_\pi = 0.138$ GeV, $X=S,V,W,T_1,T_2,T_3$, and making use of the neutrino potentials, Eqs. (\ref{eqn:vpot1})-(\ref{eqn:vpotSR}), we can write all the matrix elements appearing in Eqs. (\ref{eq:nme1})-(\ref{eq:nme5}) in the form
\begin{equation}
\label{eq:nme_gen}
	{\mathcal M}_j = \left\langle \left(\frac{q}{m_p}\right)^k v_j \tilde h(q^2) \Sigma_{12} \right\rangle  \mbox{ },
\end{equation}	
where $k = 0,1,2,3, ...$ and
\begin{equation}
	\begin{array}{c}
	\Sigma_{12}^F = 1 \mbox{ },  \\
	\Sigma_{12}^{GT} = {\bm \sigma}_1 \cdot {\bm \sigma}_2 \mbox{ },  \\
	\Sigma_{12}^T = S_{12} \mbox{ }
	\end{array}
\end{equation}
for Fermi ($F$), Gamow-Teller ($GT$) and tensor ($T$), respectively.

Introducing the notation
\begin{equation}
	h(q) = v(q) \tilde h(q)  \mbox{ },
\end{equation}
the matrix elements of interest are \cite{Graf:2018ozy}: 
\\for $k=0$,
\begin{equation}
	\label{eqn:M53}
	\begin{array}{c}
	{\mathcal M}_F = \left\langle h_{VV}(q^2) \right\rangle \\
	{\mathcal M}_{GT}^{AA} = \left\langle h_{AA}(q^2) ({\bm \sigma}_1 \cdot {\bm \sigma}_2) \right\rangle
	\end{array} \mbox{ },
\end{equation}
\begin{equation}
	\begin{array}{c}
	{\mathcal M}_{GT}^{T_1T_1} = \left\langle h_{VV}(q^2) ({\bm \sigma}_1 \cdot {\bm \sigma}_2) \right\rangle \\
	{\mathcal M}_{GT}^{AT_1} = \left\langle h_{AX}(q^2) ({\bm \sigma}_1 \cdot {\bm \sigma}_2) \right\rangle
	\end{array} \mbox{ },
\end{equation}		
for $k=2$,
\begin{equation}
	\begin{array}{c}
	{\mathcal M}'^{P'P'}_{GT} = {\mathcal M}'^{P'P}_{GT} = {\mathcal M}'^{PP}_{GT} = 
	\left\langle \frac{q^2}{m_p^2} h_{PP}(q^2) 	({\bm \sigma}_1 \cdot {\bm \sigma}_2) \right\rangle \\
	{\mathcal M}'^{P'P'}_{T} = {\mathcal M}'^{P'P}_{T} = {\mathcal M}'^{PP}_{T} = 
	\left\langle \frac{q^2}{m_p^2} h_{PP}(q^2) S_{12} \right\rangle
	\end{array} \mbox{ },
\end{equation}
and
\begin{equation}
	\begin{array}{c}
	{\mathcal M}'^{AP'}_{GT} = {\mathcal M}'^{AP}_{GT} = \left\langle \frac{q^2}{m_p^2} h_{AP}(q^2) 
	({\bm \sigma}_1 \cdot {\bm \sigma}_2) \right\rangle \\
	{\mathcal M}'^{AP'}_{T} = {\mathcal M}'^{AP}_{T} = \left\langle \frac{q^2}{m_p^2} h_{AP}(q^2) S_{12} \right\rangle
	\end{array} \mbox{ },
\end{equation}
\begin{equation}
	\begin{array}{c}
	{\mathcal M}'^{P'T_1}_{GT} = {\mathcal M}'^{PT_1}_{GT} = \left\langle \frac{q^2}{m_p^2} h_{XP}(q^2) 
	({\bm \sigma}_1 \cdot {\bm \sigma}_2) \right\rangle \\
	{\mathcal M}'^{P'T_1}_{T} = {\mathcal M}'^{PT_1}_{T} = \left\langle \frac{q^2}{m_p^2} h_{XP}(q^2) S_{12} \right\rangle
	\end{array} \mbox{ },
\end{equation}
and for $k=4$,
\begin{equation}
	\begin{array}{c}
	{\mathcal M}''^{P'P'}_{GT} = \left\langle \frac{q^4}{m_p^4} h_{PP}(q^2) 
	({\bm \sigma}_1 \cdot {\bm \sigma}_2) \right\rangle \\
	{\mathcal M}''^{P'P'}_{T} = \left\langle \frac{q^4}{m_p^4} h_{PP}(q^2) S_{12} \right\rangle
	\end{array} \mbox{ }.
\end{equation}
In addition, there are two so-called recoil terms of order $k=2$,
\begin{equation}
	\begin{array}{c}
	\tilde {\mathcal M}^{AP}_{GT} = \left\langle \frac{{\bf Q} \cdot {\bf q}}{m_p^2} h_{AP}(q^2) 
	({\bm \sigma}_1 \cdot {\bm \sigma}_2) \right\rangle \\
	\tilde {\mathcal M}^{AP}_{T} = \left\langle \frac{{\bf Q} \cdot {\bf q}}{m_p^2} h_{AP}(q^2) S_{12} \right\rangle
	\end{array} \mbox{ },
\end{equation}
and two terms of order $k=3$,
\begin{equation}
	\begin{array}{c}
	{\mathcal M}'^{q_0PP'}_{GT} = \left\langle \frac{q_0q^2}{m_p^3} h_{PP}(q^2) 
	({\bm \sigma}_1 \cdot {\bm \sigma}_2) \right\rangle \\
	{\mathcal M}'^{q_0PP'}_{T} = \left\langle \frac{q_0q^2}{m_p^3} h_{PP}(q^2) S_{12} \right\rangle
	\end{array} \mbox{ }.
\end{equation}
No term of order $k=1$ appears for $0^+ \rightarrow 0^+$ transitions.

The recoils terms and the terms of order 3 are difficult to evaluate.
A good approximation is to take in the recoil terms ${\bf Q} \simeq {\bf q}$, as discussed in \cite{Doi:1981mi,Tomoda:1990rs}, and in the terms of order $k=3$, $q_0/m_p \simeq 0.01$ \cite[page 67]{Tomoda:1990rs}. 
With these approximations,
\begin{equation}
	\tilde {\mathcal M}_{GT}^{AP} = {\mathcal M}_{GT}'^{AP}  \mbox{ , } \mbox{ }
	\tilde {\mathcal M}_{T}^{AP} = {\mathcal M}_{T}'^{AP}  \mbox{ }
\end{equation}
and
\begin{equation}
	{\mathcal M}_{GT}'^{q_0PP'} = 0.01 {\mathcal M}_{GT}'^{PP'}  \mbox{ , } \mbox{ }
	{\mathcal M}_{T}'^{q_0PP'} = 0.01 {\mathcal M}_{T}'^{PP'}  \mbox{ }.
\end{equation}
In addition to these, two more terms play an important role \cite{Tomoda:1990rs}. They are of order $k = 2$ and are given by
\begin{equation}
	\label{eqn:M63}
	\begin{array}{c}
	{\mathcal M}'^{WW}_{GT} = \left\langle \frac{q^2}{m_p^2} h_{WW}(q^2) 
	({\bm \sigma}_1 \cdot {\bm \sigma}_2) \right\rangle \\
	{\mathcal M}'^{WW}_{T} = \left\langle \frac{q^2}{m_p^2} h_{WW}(q^2) S_{12} \right\rangle
	\end{array} \mbox{ },
\end{equation}
where $h_{WW}(q^2) \equiv h_V(q^2)$. These terms are called $MM$ in \cite{Barea:2013bz}.

In all formulas, Eqs. (\ref{eqn:M53})-(\ref{eqn:M63}), we have incorporated the dependence on $m_p$ in the matrix elements to make them dimensionless. We have also used the notation ${\mathcal M}$, ${\mathcal M}'$ and ${\mathcal M}''$ for matrix elements of order $k=0,2$ and 4, respectively. Note that the matrix elements ${\mathcal M}'$ are suppressed by a factor $\frac{\langle q^2 \rangle}{m_p^2} \simeq 0.01$, and those of ${\mathcal M}''$ by a factor $\frac{\langle q^4 \rangle}{m_p^4} \simeq 10^{-4}$ relative to the matrix elements ${\mathcal M}$.
We include these terms here, because some of them are enhanced by large values of the coupling constants $g_P$ and $g_{P'}$.

%\color{red}
The bracket denotes the matrix element%
\begin{equation}
\left\langle \mathcal{H}_{ab}\right\rangle =\left\langle F\left\vert
\sum_{a\neq b}\tau _{+}^{a}\tau _{+}^{b}\mathcal{H}_{ab}\right\vert
I\right\rangle .
\end{equation}%
It should be understood that in the calculation of the matrix elements $%
\mathcal{M}_{3,4}^{\prime }$ for $j=4$, $\left( {\bm\sigma_{a}}\mathbf{%
\cdot} {\bm\sigma_{b}}\right) $ and $S_{ab}$ should be replaced by $\overline{%
\left( {\bm\sigma_{a}}\mathbf{\cdot}{\bm\sigma_{b}}\right) }$ and $%
\overline{S}_{ab}$.

For rank-1 tensors under rotation several new terms appear. We denote the
matrix elements of these new terms by%
\begin{equation}
\label{eq:xx}
\overline{\mathcal{M}}_{K,j}=\left\langle F\left\vert\overline{ \mathcal{H}}%
_{K,j}\right\vert I\right\rangle ,
\end{equation}%
where $K=2,3,4,5$ denotes the product of currents and $j=5,6$ in Tomoda's
notation. In this article, we consider only rank-1 matrix elements for $K=3$%
, which are of importance for L-R models as explicitly discussed in \cite{Doi:1981mi} and
\cite{Tomoda:1990rs}. The matrix elements that we include are, in the notation of
Eqs.(\ref{eq:nme1})-(\ref{eq:nme5}),%
\begin{equation}
	\begin{array}{rcl}
\overline{\mathcal{M}}_{3,5} &=&\left( +--\right) g_{A}g_{V}\mathcal{M}%
_{\Sigma ^{\prime }}^{AV}\left( +--\right) g_{A}g_{V}\mathcal{M}_{\Sigma
^{\prime \prime }}^{AV}   \\
\overline{\mathcal{M}}_{3,6} &=&\left( +--\right) g_{A}\left(
g_{V}+g_{W}\right) \mathcal{M}_{R}^{AV}.
\end{array}
\end{equation}%
The matrix element $\mathcal{M}_{R}^{AV}$, called recoil term in \cite{Doi:1981mi,Tomoda:1990rs},
can be written in a form similar to Eq. (\ref{eq:nme_gen}) and further split into GT and T
components%
\begin{equation}
\mathcal{M}_{R}^{AV}=-\frac{1}{3}\mathcal{M}_{R,GT}^{AV}+\frac{1}{6}\mathcal{%
M}_{R,T}^{AV}
\end{equation}%
with 
\begin{equation}
\label{eq:72}
\mathcal{M}_{R,GT}^{AV} =m_{p}R\left\langle \frac{q^{2}}{m_{p}^{2}}%
h_{AV}(q^{2})\left( {\bm\sigma_{a}}\mathbf{\cdot}{\bm \sigma_{b}}\right)
\right\rangle 
\end{equation}
\begin{equation}
\label{eq:73}
\mathcal{M}_{R,T}^{AV} =m_{p}R\left\langle \frac{q^{2}}{m_{p}^{2}}%
h_{AV}(q^{2})S_{ab}\right\rangle ,
\end{equation}%
with $h_{AV}(q^{2})=v(q)\tilde{h}_{AV}(q^{2})$ where we have normalized
these matrix elements as in \cite{x}. These matrix elements are to be calculated
with a neutrino potential 
\begin{equation}
v(q)=\frac{2}{\pi }\frac{1}{q(q+\tilde{A})}.
\end{equation}%
The other two matrix elements can be written as%
\begin{equation}
\label{eq:75}
\mathcal{M}_{\Sigma ^{\prime }}^{AV} =m_{p}R\left\langle \frac{q}{m_{p}}%
h_{AV}(q^{2})\left[ \left( {\bm\sigma_{a}}-{\bm\sigma_{b}}\right)
\cdot \mathbf{\hat{q}}\right] \right\rangle 
\end{equation}
\begin{equation}
\label{eq:76}
\mathcal{M}_{\Sigma ^{\prime \prime }}^{AV} =\frac{m_{p}R}{2}\left\langle 
\frac{Qq}{m_{p}^{2}}h_{AV}(q^{2})\left[ \left( {\bm\sigma_{a}}%
-{\bm\sigma_{b}}\right) \cdot \left( \mathbf{\hat{Q}\times \hat{q}}\right) %
\right] \right\rangle
\end{equation}%
with $h_{AV}(q^{2})=v(q)\tilde{h}_{AV}(q^{2})$ and $v(q)=\frac{2}{\pi }\frac{%
1}{q(q+\tilde{A})}$ as before. Tomoda \cite{Tomoda:1990rs} and Doi \textit{et al. } \cite{Doi:1981mi}
considered only a term of the type $\mathcal{M}_{\Sigma ^{\prime \prime
}}^{AV}$ and called it P-term. The matrix elements in Eqs. (\ref{eq:75})-(\ref{eq:76}) are
difficult to calculate. Following Tomoda \cite{Tomoda:1990rs} we approximate 
\begin{equation}
\left\langle \left( {\bm\sigma_{a}}-{\bm\sigma_{b}}\right) \cdot
\left( \mathbf{\hat{Q}\times \hat{q}}\right) \right\rangle \cong -\frac{1}{3}
\end{equation}%
and using similar arguments as in \cite{Tomoda:1990rs} we also approximate%
\begin{equation}
\left\langle \left( {\bm\sigma_{a}}-{\bm\sigma_{b}}\right) \cdot 
\mathbf{\hat{q}}\right\rangle \cong -\frac{1}{3}.
\end{equation}%
Further approximating $Q\simeq q$ in $\mathcal{M}_{\Sigma ^{\prime \prime
}}^{AV}$ we then consider%
\begin{eqnarray}
\mathcal{M}_{\Sigma ^{\prime }}^{AV} &=&\frac{m_{p}R}{3}\left\langle \frac{q%
}{m_{p}}h_{AV}(q^{2})\right\rangle \\
\mathcal{M}_{\Sigma ^{\prime \prime }}^{AV} &=&\frac{m_{p}R}{6}\left\langle 
\frac{q^{2}}{m_{p}^{2}}h_{AV}(q^{2})\right\rangle ,
\end{eqnarray}%
with total contribution%
\begin{equation}
\mathcal{M}_{\Sigma }^{AV}=\mathcal{M}_{\Sigma ^{\prime }}^{AV}+\mathcal{M}%
_{\Sigma ^{\prime \prime }}^{AV}.
\end{equation}

The matrix elements appearing in Eqs. (\ref{eq:nme1})-(\ref{eq:nme5}) and in Eqs.(\ref{eq:72})-(\ref{eq:73}),(\ref{eq:75})-(\ref{eq:76}) can
be calculated in any nuclear model \cite{Barea:2013bz,Simkovic:2007vu,Caurier:2007wq}. Here we use the interacting
boson model, IBM2 \cite{Barea:2009zza,Barea:2013bz} with isospin restoration \cite{Barea:2015kwa}. In this model, the
calculation of all matrix elements, when formulated in momentum space, is
straightforward. The calculation includes short-range correlations. This
inclusion is done by multiplying the neutrino potential $v(r)$ in coordinate
space by a correlation function $f(r)$ squared. The most commonly used
correlation function is the Jastrow function,%
\begin{equation}
f_{J}(r)=1-ce^{-ar^{2}}(1-br^{2}),
\end{equation}%
with $a=1.1$ fm$^{-2}$, $b=0.68$ fm$^{-2}$ and $c=1$ for the phenomenological
Miller-Spencer parametrization \cite{Miller:1975hu}, and $a=1.59$ fm$^{-2}$, $b=1.45$ fm$%
^{-2}$ and $c=0.92$ for the Argonne parametrization \cite{Simkovic:2009pp}. Since the
formulation described above is in momentum space, we take SRCs into account
by using the Fourier-Bessel transform of $f_{J}(r)$. Short-range
correlations are of crucial importance for short-range non-standard
mechanisms but not as much for long-range mechanisms. Nonetheless, they are
included here.

%%%%%%%%%%%%%%%%%%%%%
\begin{table}[htbp]
%\footnotesize
\centering
\begin{tabular}{lcccccccc}
\hline
\hline 
A	&$M_{F}$ 	&$M^{AA}_{GT} $	&$M^{T_1T_1}_{GT} $	&$M^{AT_1}_{GT} $	&$M'^{PP}_{GT}$	&$M'^{PP}_{T}$	&$M'^{AP'}_{GT}$	\\
\hline 
76	&-0.780	&6.062	&5.880	&5.970	&0.017	& $-3.8\times10^{-3}$		&0.036 \\
82	&-0.667	&4.928	&4.778	&4.852	&0.014	&$-3.7\times10^{-3}$ 		&0.030 \\
100	&-0.511	&5.553	&5.356	&5.454	&0.017	& $4.1\times10^{-3}$			&0.038 \\
130	&-0.651	&3.894	&3.769	&3.831	&0.011	& $-2.1\times10^{-3}$		&0.024 \\
136	&-0.522	&3.203	&3.104	&3.153	&0.009	& $-1.6\times10^{-3}$		&0.019 \\
\hline 
A	& $M'^{AP'}_{T}$ &$M'^{P'T_1}_{GT}$	&$M'^{P'T_1}_{T}$	&$M'^{WW}_{GT}$	&$M'^{WW}_{T}$ &$M''^{P'P'}_{GT}$  		&$M''^{P'P'}_{T}$\\
\hline 
76	& $-1.0\times10^{-2}$		&0.035	& $-9.9\times10^{-3}$		&0.089	& $-3.5\times10^{-2}$	&$4.1\times10^{-4}$	&$-1.4\times10^{-4}$ \\
82	& $-9.9\times10^{-3}$		&0.029	& $-9.5\times10^{-3}$		&0.073	& $-3.4\times10^{-2}$	& $3.4\times10^{-4}$	& $-1.3\times10^{-4}$ \\
100	& $1.2\times10^{-2}$		&0.037	& $1.1\times10^{-2}$		&0.096	& $4.1\times10^{-2}$	& $4.7\times10^{-4}$	& $1.6\times10^{-4}$ \\
130	& $-5.9\times10^{-3}$		&0.023	& $-5.4\times10^{-3}$		&0.061	& $-2.1\times10^{-2}$	& $2.8\times10^{-4}$	& $-8.3\times10^{-5}$ \\
136	& $-4.5\times10^{-3}$		&0.019	& $-4.2\times10^{-3}$		 &0.048	& $-1.6\times10^{-2}$	& $2.2\times10^{-4}$	& $-6.3\times10^{-5}$ \\
\hline
\hline								
\end{tabular}
\caption{Nuclear matrix elements of neutrinoless $\beta\beta$-decay with neutrino potential $v_1(q)=\frac{2}{\pi}\frac{1}{q(q+\tilde{A})}$ of some nuclei of interest.}
\label{tab:NME1}
\end{table}

%%%%%
\begin{table}[htbp]
\centering
\begin{tabular}{lccccccc}
\hline
\hline 
A	&$M_{F}$ 	&$M^{AA}_{GT} $	&$M^{T_1T_1}_{GT} $	&$M^{AT_1}_{GT} $	&$M'^{PP}_{GT}$	&$M'^{PP}_{T}$	&$M'^{AP'}_{GT}$\\ 
\hline
76	&-0.730	&5.032	&4.861	&4.946	&0.016	& $-3.6\times10^{-3}$		&0.034	 \\
82	&-0.621	&4.049	&3.908	&3.978	&0.013	& $-3.4\times10^{-3}$		&0.027	 \\
100	&-0.479	&4.561	&4.377	&4.468	&0.015	& $3.7\times10^{-3}$			&0.035	 \\
130	&-0.589	&3.067	&2.951	&3.008	&0.010	& $-1.9\times10^{-3}$		&0.022	 \\
136	&-0.472	&2.516	&2.424	&2.470	&0.008	& $-1.4\times10^{-3}$		&0.018	 \\
\hline 
A	& $M'^{AP'}_{T}$	&$M'^{P'T_1}_{GT}$	&$M'^{P'T_1}_{T}$		&$M'^{WW}_{GT}$	&$M'^{WW}_{T}$ &$M''^{P'P'}_{GT}$  		&$M''^{P'P'}_{T}$ \\
\hline
76	& $-9.9\times10^{-3}$	&0.033	& $-9.5\times10^{-3}$			&0.084	& $-3.4\times10^{-2}$	& $3.9\times10^{-4}$	& $-1.4\times10^{-4}$ \\
82	& $-9.5\times10^{-3}$	&0.026	& $-9.0\times10^{-3}$			&0.069	& $-3.2\times10^{-2}$		& $3.2\times10^{-4}$	& $-1.3\times10^{-4}$ \\
100	& $1.1\times10^{-2}$	&0.034	& $1.0\times10^{-2}$			&0.090	& $4.0\times10^{-2}$		& $4.4\times10^{-4}$	& $1.6\times10^{-4}$ \\
130	& $-5.4\times10^{-3}$	&0.021	& $-5.0\times10^{-3}$			&0.056	& $-2.0\times10^{-2}$		& $2.6\times10^{-4}$	& $-7.7\times10^{-5}$ \\
136	& $-4.2\times10^{-3}$	&0.017	& $-5.4\times10^{-3}$			&0.044	& $-1.5\times10^{-2}$		& $2.1\times10^{-4}$	&$-5.9\times10^{-5}$ \\
\hline								
\hline
\end{tabular}
\caption{Nuclear matrix elements of neutrinoless $\beta\beta$-decay with neutrino potential $v_3(q)=\frac{2}{\pi}\frac{1}{(q+\tilde{A})^2}$ of some nuclei of interest.}
\label{tab:NME2}
\end{table}%%%%%%%%%%%%%%%%%%%%%

\section{Numerical values of the nuclear matrix elements}
\label{Numerical values of the nuclear matrix elements}
%\color{red}
The results of our calculation of the nuclear matrix elements $\mathcal M$ of Eqs. (\ref{eqn:M53})-(\ref{eqn:M63}) for the nuclei $^{76}$Ge, $^{82}$Se, $^{100}$Mo, $^{130}$Te and $^{136}$Xe within the framework of IBM-2 \cite{Barea:2009zza,Barea:2013bz,Barea:2015kwa} are shown in Tables \ref{tab:NME1}-\ref{tab:NME3} for rank-0 and in Table \ref{tab:NME1b} for rank-1. For completeness, we include in Table \ref{tab:NME4} also the values of the matrix elements $\mathcal{M}$ for
short-range non-standard mechanisms \cite{Graf:2018ozy}.

%%%%%%%%%%%%%%%%%%%%%
\begin{table}[htbp]
\centering
\begin{tabular}{lccccccc}
\hline 
\hline
A	&$M_{F}$ 	&$M^{AA}_{GT} $	&$M^{T_1T_1}_{GT} $	&$M^{AT_1}_{GT} $	&$M'^{PP}_{GT}$	&$M'^{PP}_{T}$	&$M'^{AP'}_{GT}$	\\
\hline 
76	& -0.830	&7.092	&6.900	&6.995	&0.019	& $-4.1\times10^{-3}$		&0.039	 \\
82	&-0.713	&5.807	&5.648	&5.727	&0.015	& $-4.0\times10^{-3}$		&0.032	 \\
100	&-0.543	&6.545	&6.336	&6.439	&0.018	& $4.5\times10^{-3}$			&0.041	 \\
130	&-0.712	&4.722	&4.588	&4.654	&0.013	& $-2.3\times10^{-3}$		&0.027	 \\
136	&-0.571	&3.890	&3.783	&3.836	&0.010	& $-1.8\times10^{-3}$		&0.021	 \\
\hline
A	& $M'^{AP'}_{T}$	&$M'^{P'T_1}_{GT}$	&$M'^{P'T_1}_{T}$		&$M'^{WW}_{GT}$	&$M'^{WW}_{T}$		&$M''^{P'P'}_{GT}$  		&$M''^{P'P'}_{T}$\\
\hline
76	& $-1.1\times10^{-2}$	&0.038	& $-1.0\times10^{-2}$			&0.094	& $-3.7\times10^{-2}$	& $4.3\times10^{-4}$	& $-1.5\times10^{-4}$ \\
82	& $-1.0\times10^{-2}$	&0.031	& $-9.9\times10^{-3}$			&0.078	& $-3.5\times10^{-2}$	& $3.6\times10^{-4}$	& $-1.4\times10^{-4}$ \\
100	& $1.2\times10^{-2}$	&0.040	& $1.2\times10^{-2}$			&0.102	& $4.3\times10^{-2}$	& $4.9\times10^{-4}$	& $1.7\times10^{-4}$ \\
130	& $-6.3\times10^{-3}$	&0.026	& $-6.3\times10^{-3}$			&0.066	& $-2.2\times10^{-2}$ & $3.0\times10^{-4}$	& $-8.7\times10^{-5}$ \\
136	& $-5.0\times10^{-3}$	&0.021	& $-5.0\times10^{-3}$			&0.052	& $-1.7\times10^{-2}$	& $2.4\times10^{-4}$	& $-6.7\times10^{-5}$ \\
\hline								
\hline
\end{tabular}
\caption{Nuclear matrix elements of neutrinoless $\beta\beta$-decay with neutrino potential $v_4(q)=\frac{2}{\pi}\frac{q+2\tilde{A}}{q(q+\tilde{A})^2}$ of some nuclei of interest.}
\label{tab:NME3}
\end{table}

%%%%%%NEW TABLE rank 1 NMEs
\begin{table}[htbp]
\centering
\begin{tabular}{lcccc}
\hline
\hline 
A	&$M_{R,GT}^{AV}$ 	&$M^{AV}_{R,T} $	&$M^{AV}_{\Sigma'} $	&$M^{AV}_{\Sigma''} $\\ 
\hline
76	&10.140	&-4.384		& $-1.119$		&-0.102	 \\
82	&8.583	&-4.283		& $-0.973$		&-0.087	 \\
100	&11.900	&-5.713		& $-1.043$		&-0.122	 \\
130	&8.350	&-3.132		& $-1.053$		&-0.092	 \\
136	&6.674	&-2.439		& $-0.845$		&-0.073	 \\
\hline 
\hline
\end{tabular}
\caption{Nuclear matrix elements of rank-1 operators with neutrino
potential $\frac{2}{\pi }\frac{1}{q(q+\tilde{A})}$.}
\label{tab:NME1b}
\end{table}

\begin{table}[htbp]
\centering
\begin{tabular}{lccccccc}
\hline 
\hline
A	&$M_{F}$ 	&$M^{AA}_{GT} $	&$M^{T_1T_1}_{GT} $	&$M^{AT_1}_{GT} $	&$M'^{PP}_{GT}$	&$M'^{PP}_{T}$	&$M'^{AP'}_{GT}$	\\
\hline 
76	&$-48.89$	&170		&173.5	&174.3	&0.798	&$-0.271$		&2.111	 \\
82	&$-41.22$	&140.7	&143.6	&144.3	&0.660	&$-0.259$		&1.758	 \\
100	&$-51.96$	&181.9	&188.6	&188.1	&0.910	&0.317		&2.273	 \\
130	&$-38.05$	&119.7	&121.9	&122.6	&0.561	&$-0.160$		&1.514	 \\
136	&$-29.83$	&94.18	&96.09	&96.56	&0.442	&$-0.123$		&1.177	 \\
\hline
A & $M'^{AP'}_{T}$	&$M'^{P'T_1}_{GT}$	&$M'^{P'T_1}_{T}$		&$M'^{WW}_{GT}$	&$M'^{WW}_{T}$		&$M''^{P'P'}_{GT}$  		&$M''^{P'P'}_{T}$\\
\hline
76	&$-1.310$	&2.255	&$-1.183$			&$-2.945$	&$-6.541$	&0.028	& $-2.2\times10^{-2}$ \\
82	&$-1.249$	&1.878	&$-1.128$			&$-2.456$	&$-6.206$	&0.024	& $-2.1\times10^{-2}$ \\
100	&1.590	&2.464	&1.433			&$-4.590$	&8.055	&0.029	& $2.7\times10^{-2}$  \\
130	&$-0.807$	&1.613	&$-0.726$			&$-1.951$	&$-4.105$	&0.021	& $-1.4\times10^{-2}$  \\
136	&$-0.620$	&1.257	&$-0.558$			&$-1.625$	&$-3.158$	&0.016	& $-1.1\times10^{-2}$  \\
\hline								
\hline
\end{tabular}
\caption{Nuclear matrix elements of neutrinoless $\beta\beta$-decay with neutrino potential $v(q)=\frac{2}{\pi}\frac{1}{m_em_p}$ of some nuclei of interest.}
\label{tab:NME4}
\end{table}
%%%%%%%%%%%%%%%%%%%%%
\color{black}
\section{Leptonic matrix elements}
\label{Leptonic matrix elements}
The leptonic matrix elements associated with the three neutrino potentials $v_1$, $v_3$ and $v_4$ are
\\ $m_i$ terms
\begin{equation}
	\label{eqn:LME1}
	\bar e_{{\bf p}_1, s_1} {\mathcal O}_\beta {\mathcal O}_\alpha e_{{\bf p}_2, s_2}^{\rm c}  \mbox{ } 
	\mbox{ } \mbox{ } (\alpha = \beta)
\end{equation}
\\ $\omega$ terms
\begin{equation}
	\label{eqn:LME2}
	\bar e_{{\bf p}_1, s_1} {\mathcal O}_\beta \gamma^0 {\mathcal O}_\alpha e_{{\bf p}_2, s_2}^{\rm c}  \mbox{ } 
	\mbox{ } \mbox{ } (\alpha \neq \beta)
\end{equation}	
\\ ${\bf k}$ terms
\begin{equation}
	\label{eqn:LME3}
	\bar e_{{\bf p}_1, s_1} {\mathcal O}_\beta \gamma^j {\mathcal O}_\alpha e_{{\bf p}_2, s_2}^{\rm c}  \mbox{ } 
	\mbox{ } \mbox{ } (\alpha \neq \beta)
\end{equation}
The operators ${\mathcal O}_{\alpha,\beta}$ are ${\mathcal O}_\alpha = \gamma_\mu \frac{1}{2} (1 \mp \gamma_5)$ for $V\mp A$, ${\mathcal O}_\alpha = \frac{1}{2} (1 \mp \gamma_5)$ for $S \mp P$ and ${\mathcal O}_\alpha = \sigma_{\mu\nu} \frac{1}{2} (1 \mp \gamma_5)$ for $T \mp T_5$.
The Lorentz indices of these operators are contracted with the corresponding Lorentz indices of the currents $\tilde J^\mu_{\alpha i}$ in Eq. (\ref{eqn:rate-Tomoda}).
Because the operators ${\mathcal O}_\alpha$ contain the projectors ${\mathcal P}_\alpha = \frac{1}{2} (1 \mp \gamma_5)$, $\alpha = $ L, R, the matrix elements in (\ref{eqn:LME1}) vanish unless $\alpha = \beta$, while those in (\ref{eqn:LME2}) and (\ref{eqn:LME3}) require $\alpha \neq \beta$. A list of matrix elements associated with the products of currents $\Pi_i$ ($i = 1, ..., 5$) is given in Appendix \ref{Leptonic matrix elements and phase space factors}.

In the evaluation of the matrix elements of the leptonic currents it is convenient to expand the electron wave functions $e_{{\bf p},s}$ in spherical waves
\begin{equation}
	\label{eqn:electron-WF}
	e_{\vecl{p},s}(\vecl{r}) = e_{\vecl{p},s}^{S_{1/2}}(\vecl{r}) + e_{\vecl{p},s}^{P_{1/2}}(\vecl{r}) + \dots  \mbox{ },
\end{equation}
where $\vecl{p}$ is the asymptotic momentum of the electron and $s$ its spin projection.
We include in our calculations $S_{1/2}$- and $P_{1/2}$-waves, the explicit form of which is \cite{Tomoda:1990rs}
\begin{equation}
	\label{eq:radwaves-S}
	e^{S_{1/2}}_{\vecl{p},s}(\vecl{r}) = \begin{pmatrix} g_{-1}^{(-)}(E, r) \chi_s \\ f_1^{(-)}(E, r)(\vecs{\sigma}\cdot\vecl{\op p}) \chi_s 
	\end{pmatrix}
\end{equation}
and
\begin{equation}	
	\label{eq:radwaves-P}
	e^{P_{1/2}}_{\vecl{p},s}(\vecl{r}) = 
	i\begin{pmatrix}
	        g_1^{(-)}(E, r)(\vecs{\sigma}\cdot\vecl{\op r}) (\vecs{\sigma}\cdot\vecl{\op p})\chi_s \\
	        - f_{-1}^{(-)}(E, r)(\vecs{\sigma}\cdot\vecl{\op r})\chi_s 
	\end{pmatrix} \mbox{ },
\end{equation}
where $g_\kappa^{(-)}(E, r)$ and $f_\kappa^{(-)}(E, r)$ are the radial wavefunctions of the ``large'' and ``small'' components. The electron energy at asymptotically large distances is $E = \sqrt{\vecl{p}^2 + m_e^2}$ and its spin state is described by the two-dimensional spinor $\chi_s$. The Pauli matrices $\vecs{\sigma}$ here operate in the electron spin space.
The radial wavefunctions satisfy the asymptotic boundary condition
\begin{equation}
	\begin{pmatrix}
		g_\kappa^{(-)}(E, r) \\ 
		f_\kappa^{(-)}(E, r)
	\end{pmatrix}
	\xrightarrow{r\to\infty}
	\frac{e^{-i\Delta_\kappa^c}}{pr}
	\begin{pmatrix}
		\sqrt{\frac{E+m_e}{2E}}\sin\nba{pr+y\ln(2pr)
			- \tfrac{1}{2}\pi l_\kappa + \Delta_\kappa^c} \\ 
		\sqrt{\frac{E-m_e}{2E}}\cos\nba{pr+y\ln(2pr)-\tfrac{1}{2}\pi l_{\kappa}+\Delta_{\kappa}^c}
	\end{pmatrix},
\end{equation}
where $\kappa=\pm(j + \frac{1}{2})$, $l_{\kappa} = j\pm\frac{1}{2}$, $y = \alpha Z_F E/p$, $\Delta^c_{\kappa}$ is a phase shift, $\alpha$ the fine structure constant, and $j$ is the total angular momentum of the electron.

In most calculations, the radial wave functions are expanded as a function of $r$. We instead calculate the radial wave functions by numerical solution of the Dirac equation with potential \cite{Kotila:2012zza}
\begin{align}
	V(r) = 
	\begin{cases}
		- \alpha Z_F\frac{3-(r/R_A)^2}{2R_A} \times \varphi(r) \mbox{ }, &r < R \mbox{ }, \\
		- \frac{\alpha Z_F}{r} \times \varphi(r) \mbox{ },               &r \geq R \mbox{ },
	\end{cases}
\end{align}
which includes finite size corrections to the Coulomb potential of the final nucleus with charge $Z_{\rm F}$ and electron screening due to the electronic cloud described in the Thomas-Fermi approximation by the function $\varphi(r)$.

From the leptonic matrix elements one can calculate the phase space factors (PSF), denoted by $f_{ij}$, by squaring and summing over spin. 
In doing this calculation, the electron radial wave functions are evaluated at the nuclear radius $R_A$, given by $R_A = R_0 A^{1/3}$, with $R_0 = 1.2$ fm. 
Details of this calculation, as well as a list of all $f_{ij}$, $f_{ij}'$ and $f_{ij}''$, are given in Appendix \ref{Leptonic matrix elements and phase space factors}.
From the $f_{ij}$'s we calculate the integrated PSF by \cite{Kotila:2012zza}
\begin{equation}
	\label{eqn:integrated-PSF}
	F_{jk}^{(i)} = \frac{2}{{\rm ln}~2} \displaystyle \int f_{jk}^{(i)} w_{0\nu} {\rm d}E_1  \mbox{ , } \mbox{ } i = 0,1 \mbox{ },
\end{equation}	
where $w_{0\nu}$ is given by Eq. (\ref{eqn:w0nu}) and $E_2 = E_{\rm I} - E_{\rm F} - E_1$, and their normalized values
\begin{equation}
	\label{eqn:integrated-normalized-PSF}
	G_{jk}^{(i)} = \frac{F_{jk}^{(i)}}{4 R_A^2}  \mbox{ , } \mbox{ } i = 0,1 \mbox{ },
\end{equation}	
where $R_A$ is the nuclear radius.

\section{Numerical values of the PSF}
\label{Numerical values of the PSF}
There are three types of PSF: 1) those associated with the products of currents $\Pi_1$, $\Pi_2$ and $\Pi_3$; 2) those associated with the products of currents $\Pi_4$ and $\Pi_5$; 3) interference terms $\Pi_{1,2,3} - \Pi_{4,5}$. We list them in Tables \ref{tab:Gij-a}-\ref{tab:Gij-b}.
%%%%%%%%%%%%%%%%%%%%%%%%%%%%%%%%%%%%%%%
%\newpage
%\LTcapwidth=\textwidth
%\begin{center}
%\begin{longtable}{lccccccc}
\begin{table}[htbp]
\centering
\begin{tabular}{lccccccc}
\hline 
\hline
&$G_{11}^{(0)}$	&$G_{33}^{(0)}$	&$G_{44}^{(0)}$	&$G_{13}^{(0)}$	&$G_{14}^{(0)}$	&$G_{34}^{(0)}$ & \\
\hline
$^{76}$Ge	&2.36	&3.87	&1.22	& $-0.65$	&0.92	& $-1.76$	&	\\
$^{82}$Se		&10.19	&35.29	&7.89	& $-3.45$	&4.37	& $-14.43$	&	 \\
$^{100}$Mo	&15.91	&57.62	&12.91	& $-5.46$	&6.91	& $-23.72$	&	 \\
$^{130}$Te	&14.20	&37.34	&9.82	& $-4.45$	&6.00	& $-16.23$	&	 \\
$^{136}$Xe	&14.55	&36.48	&9.85	& $-4.48$	&6.13	& $-16.01$	&	 \\
\hline
& $G_{11}^{(1)}$	&$G_{33}^{(1)}$	&$G_{44}^{(1)}$	&$G_{13}^{(1)}$	&$G_{14}^{(1)}$&$G_{34}^{(1)}$ & \\
\hline
$^{76}$Ge	& $-1.95$	&3.26	& $-0.06$		& 0 & $-0.66$	& $-1.11$ & \\
$^{82}$Se		& $-9.08$	&31.08	&1.61	& 0	& $-3.05$	& $-10.66$ & \\
$^{100}$Mo	& $-14.25$	&50.91	&2.91 & 0		& $-4.81$	& $-17.70$ & \\
$^{130}$Te	& $-12.45$	&31.61	&1.22 & 0		& $-4.22$	& $-11.55$ & \\
$^{136}$Xe	& $-12.72$	&31.85	&1.09	& 0	& $-4.32$	& $-11.32$ & \\
\hline%% 8.12.2020 $G'^{(0)}_{34}$, $G'^{(1)}_{34}$ corrected
&$G'^{(0)}_{11}$	&$G'^{(0)}_{33}$	&$G'^{(0)}_{44}$	&$G'^{(0)}_{13}$	&$G'^{(0)}_{14}$	&$G'^{(0)}_{34}$	&	\\
\hline
$^{76}$Ge		&2.36	&3.87	&308.11	& -0.65	& $-2.83\times10^{-3}$ 	& $-11.17$	&	 \\
$^{82}$Se			&10.19	&35.29	&1597.97	& -3.45	& $-3.35\times10^{-2}$ 	& $-63.39$ 	&	 \\
$^{100}$Mo		&15.91	&57.62	&3197.55	& -5.46	& $-7.11\times10^{-2}$	& $-113.32$	&	 \\
$^{130}$Te		&14.20	&37.34	&3393.87	& -4.45	& $-5.60\times10^{-2}$ 	& $-101.86$	&	 \\
$^{136}$Xe		&14.55	&36.48	&3602.01	& -4.48	& $-4.43\times10^{-2}$	 & $-104.55$	&	 \\
\hline
&$G'^{(1)}_{11}$	&$G'^{(1)}_{33}$	&$G'^{(1)}_{44}$	&$G'^{(1)}_{13}$	&$G'^{(1)}_{14}$	&$G'^{(1)}_{34}$	&	\\
\hline
$^{76}$Ge	& 1.95	& $-3.26$	        &$289.45$	& 0	& 0 	& $0$ & \\
$^{82}$Se		& 9.08	& $-31.08$	&$1531.07$	& 0	& 0	& $0$ & \\
$^{100}$Mo	& 14.25	& $-50.91$	&$3066.50$	& 0 & 0 	&$0$ & \\
$^{130}$Te	& 12.45	& $-32.61$	&$3234.28$	& 0 & 0	& $0$ & \\
$^{136}$Xe	& 12.72	& $-31.85$	&$3430.54$	& 0 & 0 	& $0$ & \\
%\hline%% 8.12.2020 $G''^{(0)}_{44}$, $G''^{(1)}_{44}$ corrected
%&$G''^{(0)}_{11}$	&$G''^{(0)}_{33}$	&$G''^{(0)}_{44}$	&$G''^{(0)}_{13}$	&$G''^{(0)}_{14}$	&$G''^{(0)}_{34}$	&$G''^{(0)}_{43}$ \\
%\hline
%$^{76}$Ge	&0.87	& $-2.59\times10^{-3}$	&15.81	& $-6.85\times10^{-4}$	&$-17.90$	& $-11.11$	& $1.76$ \\
%$^{82}$Se		&2.92	& $-2.68\times10^{-2}$	&80.22	& $-4.80\times10^{-3}$ 	& $-87.01$	& $-63.33$ & $14.40$  \\
%$^{100}$Mo	&4.45	& $-1.20\times10^{-1}$	& 143.40	& $-2.16\times10^{-2}$ 	&$-154.13$	& $-113.33$	& $23.72$	  \\
%$^{130}$Te	&4.36	& $-5.97\times10^{-2}$	&136.86	& $-1.34\times10^{-2}$ &$-148.87$	& $-101.40$	& $16.28$  \\
%$^{136}$Xe	&4.52	& $-4.38\times10^{-2}$	&142.30	& $-1.02\times10^{-2}$ & $-155.10$	& $-104.06$ & $16.06$ \\
%\hline
%& $G''^{(1)}_{11}$ & $G''^{(1)}_{33}$ & $G''^{(1)}_{44}$		&$G''^{(1)}_{13}$	&$G''^{(1)}_{14}$	&$G''^{(1)}_{34}$&$G''^{(1)}_{43}$		\\
%\hline
%$^{76}$Ge	& 0 & 0 & $-11.29$	& $4.18\times10^{-4}$	& $-16.81$	& $-6.28\times10^{-4}$	& $1.11$ \\
%$^{82}$Se		& 0 & 0 & $-56.04$	& $2.86\times10^{-3}$	& $-83.37$	& $-6.75\times10^{-3}$	& $10.62$ \\
%$^{100}$Mo	& 0 & 0 & $-99.74$	& $1.48\times10^{-2}$	& $-147.82$	& $-1.43\times10^{-2}$	& $17.70$ \\
%$^{130}$Te	& 0 & 0 & $-96.22$	& $1.16\times10^{-2}$	& $-141.87$	& $-1.22\times10^{-2}$	& $11.56$ \\
%$^{136}$Xe	& 0 & 0 & $-100.30$	& $9.23\times10^{-3}$ 	& $-147.72$	& $-1.25\times10^{-2}$	& $11.34$ \\
\hline
\hline
\end{tabular}
\caption{Integrated rank-0 phase-space factors $G_{ij}^{(a)}$ and $G'^{(a)}_{ij}$ in units of $10^{-15}y^{-1}$.}
\label{tab:Gij-a}
%\end{longtable}
\end{table}

\begin{table}[htbp]
\centering
\begin{tabular}{lccccccc}
\hline 
\hline
&$G''^{(0)}_{11}$	&$G''^{(0)}_{33}$	&$G''^{(0)}_{44}$	&$G''^{(0)}_{13}$	&$G''^{(0)}_{14}$	&$G''^{(0)}_{34}$	&$G''^{(0)}_{43}$ \\
\hline
$^{76}$Ge	&0.87	& $-2.59\times10^{-3}$	&15.81	& $6.85\times10^{-4}$	&$17.90$	& $-11.11$	& $1.76$ \\
$^{82}$Se		&2.92	& $-2.68\times10^{-2}$	&80.22	& $1.11\times10^{-2}$ 	& $87.01$	& $-63.33$ & $14.40$  \\
$^{100}$Mo	&4.45	& $-1.20\times10^{-1}$	&143.40	& $2.32\times10^{-2}$ 	&$154.13$	& $-113.33$	& $23.72$	  \\
$^{130}$Te	&4.36	& $-5.97\times10^{-2}$	&136.86	& $1.34\times10^{-2}$ &$148.87$	& $-101.40$	& $16.28$  \\
$^{136}$Xe	&4.52	& $-4.38\times10^{-2}$	&142.30	& $1.02\times10^{-2}$ & $155.10$	& $-104.06$ & $16.06$ \\
\hline
& $G''^{(1)}_{11}$ & $G''^{(1)}_{33}$ & $G''^{(1)}_{44}$		&$G''^{(1)}_{13}$	&$G''^{(1)}_{14}$	&$G''^{(1)}_{34}$&$G''^{(1)}_{43}$		\\
\hline
$^{76}$Ge	& 0 & 0 & $-11.29$	& $3.85\times10^{-4}$	& $16.81$	& $-6.28\times10^{-4}$		& $1.11$ \\
$^{82}$Se		& 0 & 0 & $-56.04$	& $5.85\times10^{-3}$	& $83.37$	& $-6.75\times10^{-3}$		& $10.62$ \\
$^{100}$Mo	& 0 & 0 & $-99.74$	& $1.48\times10^{-2}$	& $147.82$	& $-1.43\times10^{-2}$	& $17.70$ \\
$^{130}$Te	& 0 & 0 & $-96.22$	& $1.19\times10^{-2}$	& $141.87$	& $-1.22\times10^{-2}$	& $11.56$ \\
$^{136}$Xe	& 0 & 0 & $-100.30$	& $8.93\times10^{-3}$ 	& $147.72$	& $-1.25\times10^{-2}$	& $11.34$ \\
\hline
\hline
\end{tabular}
\caption{Integrated rank-0 phase-space factors $G''^{(a)}_{ij}$ in units of $10^{-15}y^{-1}$.}
\label{tab:Gij-c}
%\end{longtable}
\end{table}

%%%%%%%%%%%%%%%%%%%%%%%%%%%%%%%%%%%%%%%
\begin{table}[htbp]
\centering
\begin{tabular}{lccccccc}
\hline 
\hline
%%%% These are new 8.12.2020
&$G_{55}^{(0)}$	&$G_{66}^{(0)}$	&$G_{56}^{(0)}$	&$G_{15}^{(0)}$	&$G_{16}^{(0)}$	& & \\
\hline
$^{76}$Ge	&$1488$	&$122030$	&$-13470$	& $28.21$		&$-264.64$	& 	\\
$^{82}$Se		&$7054$	&$477162$	&$-57986$	& $99.95$		&$-865.21$	&$$ 	&$$	 \\
$^{100}$Mo	&$14118$	&$652315$	&$-95929$	& $173.45$	&$-1236$	&$$ 	&$$	 \\
$^{130}$Te	&$15435$	&$495720$	&$-87444$	& $188.93$	&$-1111$	& $$	&$$	 \\
$^{136}$Xe	&$16448$	&$493857$	&$-90101$	& $199.70$	&$-1134$	& $$	&$$	 \\
\hline
& $G_{55}^{(1)}$	&$G_{66}^{(1)}$	&$G_{56}^{(1)}$	&$G_{15}^{(1)}$	&$G_{16}^{(1)}$& & \\
\hline
$^{76}$Ge	& $1156$		&$90227$		& $-10222$		& $1.31$		& $0$ & \\
$^{82}$Se		& $6094$		&$397122$	&$-49227$		& $6.08$		& $0$ & \\
$^{100}$Mo	& $12253$	&$547968$	&$-81984$ 		& $9.62$		& $0$ & \\
$^{130}$Te	& $12930$	&$401926$	&$-72120$ 		& $8.44$			& $0$ & \\
$^{136}$Xe	& $13715$	&$398660$			&$-73975$		& $8.64$			& $0$ & \\
%%%%
\hline
\hline
\end{tabular}
\caption{Integrated rank-1 phase-space factors  in units of $10^{-15}y^{-1}$.}
\label{tab:Gij-b}
%\end{longtable}
\end{table}

\section{Results}
\label{Results}
%\color{red}
The formulas of the previous sections, \ref{Hadronic matrix elements} and \ref{Leptonic matrix elements}, and the numerical calculations of Secs. \ref{Numerical values of the nuclear matrix elements} and \ref{Numerical values of the PSF} allow one to study any non-standard long-range mechanism.  However, in view of the complicated form of the general problem
in which all $\epsilon _{\beta }^{\alpha }$ are present, we discuss
explicitly here only two classes of models, one in which the non-zero
coefficients are $\epsilon _{V\mp A}^{V\mp A}$ which we simply call L-R
models, and one in which the non-zero coefficients are $\epsilon _{S\mp
P}^{S\mp P},\epsilon _{T\mp T_{5}}^{T\mp T_{5}}$, in addition to the
standard term $\epsilon _{V-A}^{V-A}$ in Eq. (\ref{eqn:effective-L}), which we call SUSY models.

\subsection{L-R models}

The differential rate for these models is obtained by combining the matrix
elements $\mathcal{M}_{3},\mathcal{M}_{3}^{\prime }$ and $\mathcal{\bar{M}}%
_{3}$ with the phase space factors of Eqs. (\ref{eqn:integrated-PSF}),(\ref{eqn:integrated-normalized-PSF}). Introducing the
quantities%
\begin{equation}
\begin{array}{lcr}
\label{eqn:Xij-LR}
X_{1} &=&\frac{\left\langle m_{\nu }\right\rangle }{m_{e}}\mathcal{M}%
_{3,1}^{LL}+\bar{\epsilon}_{V-A}^{V-A}\mathcal{M}_{3,1}^{LL}+\bar{\epsilon}%
_{V+A}^{V-A}\mathcal{M}_{3,1}^{LR}   \\
X_{3} &=&\bar{\epsilon}_{V+A}^{V+A}\mathcal{M}_{3,3}^{RR}+\bar{\epsilon}%
_{V-A}^{V+A}\mathcal{M}_{3,3}^{RL}   \\
X_{4} &=&\bar{\epsilon}_{V+A}^{V+A}\mathcal{M}_{3,4}^{\prime RR}+\bar{%
\epsilon}_{V-A}^{V+A}\mathcal{M}_{3,4}^{\prime RL} \\
X_{5} &=&\bar{\epsilon}_{V-A}^{V+A}\mathcal{M}_{3,5}^{LR}   \\
X_{6} &=&\bar{\epsilon}_{V-A}^{V+A}\mathcal{M}_{3,6}^{LR}  
\end{array}
\end{equation}%
where $\left\langle m_{\nu }\right\rangle =\sum_{i}m_{i}U_{ei}^{2}$, and $%
\bar{\epsilon}_{V\mp A}^{V\mp A}$ are effective coupling constants summed
over neutrino species, $i$, with appropriate weighting factors. After a
lengthy calculation, the details of which are given in \cite{Doi:1981mi} and \cite{Tomoda:1990rs}, we
obtain%
\begin{equation}
	\label{eqn:a-LR}
a^{(i)}=\sum_{j,k=1,3,4,5,6}f_{jk}^{(i)}\func{Re}\left[ X_{j}X_{k}^{\ast }%
\right] \text{, \ \ \ }i=0,1,
\end{equation}%
where $i=0,1$ denotes Tomoda's index in the expression for the differential
rate (see Eq. (\ref{eqn:dW0vdE1dcostheta12})), not to be confused with the index $i$ of the neutrino
species. The first term in $X_1$ is that arising from the standard mass mechanism, while the others arise from L-R models \cite{Pati:1974yy,Mohapatra:1974hk,Senjanovic:1975rk} and the intrinsic properties of these models. The three indices 1, 3, and 4 in $X_j$ denote the fact that they arise from $m_i$ terms, index 1, $\omega$ terms, index 3, and $\bf k$ terms, index 4, Eq. (\ref{eqn:nu-propagator}). The two indices 5,6 denote
rank-1 matrix elements, Eq. (\ref{eq:xx}).  Note that the PSFs in Eq. (\ref{eqn:a-LR}) have been defined in such a way to take into account the definition of the neutrino potentials, Eqs. (\ref{eqn:vpot1})-(\ref{eqn:vpot4}), including all factors of 2 and the prefactors described in Appendix \ref{Leptonic matrix elements and phase space factors}.

This class of models was investigated by Doi \textit{et al. }\cite{Doi:1981mi}, and Tomoda
\cite{Tomoda:1990rs} who used%
\begin{equation}
	\begin{array}{rcl}
\epsilon _{V-A,i}^{V-A} &=&0\text{, \ \ \ }\epsilon _{V+A,i}^{V-A}=\kappa
U_{ei},   \\
\epsilon _{V+A,i}^{V+A} &=&\lambda V_{ei}\text{, \ \ }\epsilon
_{V-A,i}^{V+A}=\eta V_{ei},
\end{array}%
\end{equation}
and%
\begin{equation}
	\begin{array}{rcl}
	\label{eqn:LR-parameters}
\left\langle m_{\nu }\right\rangle &=&\sum_{i}m_{i}U_{ei}^{2},\text{ \ \ }%
\left\langle \lambda \right\rangle =\lambda \sum_{i}U_{ei}V_{ei}\equiv \bar{%
\epsilon}_{V+A}^{V+A}   \\
\left\langle \eta \right\rangle &=&\eta \sum_{i}U_{ei}V_{ei}\equiv \bar{%
\epsilon}_{V-A}^{V+A},\text{ \ \ }\left\langle \kappa \right\rangle =\kappa
\sum_{i}U_{ei}^{2}\equiv \bar{\epsilon}_{V+A}^{V-A},
\end{array}
\end{equation}%
where $U_{ei}$ and $V_{ei}$ are the mixing matrix elements of the PMNS
matrix \cite{Pontecorvo:1957qd,Maki:1962mu} for the standard and non-standard mechanism, respectively.
Furthermore, Tomoda set $\left\langle \kappa \right\rangle =0$, since it
appears always in combination with the mass term. By introducing the
integrated normalized PSF of Eq. (\ref{eqn:integrated-normalized-PSF}), and explicitly writing down the sum
over the indices $j=1,3,4,5,6$, one obtains%
\begin{equation}
\label{eq:capAi}
A^{(i)}=\sum_{j,k=1,3,4,5,6}G_{jk}^{(i)}\func{Re}[X_{j}X_{k}^{\ast }],\text{
\ \ \ }i=0,1.
\end{equation}%
Rearranging Eq. (\ref{eq:capAi}) with respect to $\frac{\left\langle m_{\nu
}\right\rangle }{m_{e}},\left\langle \lambda \right\rangle $ and $%
\left\langle \eta \right\rangle $ we can write%
\begin{equation}
	\begin{array}{rcl}
A^{(i)} &=&C_{mm}^{(i)}\left( \frac{\left\langle m_{\nu }\right\rangle }{%
m_{e}}\right) ^{2}+C_{\lambda \lambda }^{(i)}\left\langle \lambda
\right\rangle ^{2}+C_{\eta \eta }^{(i)}\left\langle \eta \right\rangle
^{2}+2C_{m\lambda }^{(i)}\frac{\left\langle m_{\nu }\right\rangle }{m_{e}}%
\left\langle \lambda \right\rangle  \\
&&+2C_{m\eta }^{(i)}\frac{\left\langle m_{\nu }\right\rangle }{m_{e}}%
\left\langle \eta \right\rangle +2C_{\lambda \eta }^{(i)}\left\langle
\lambda \right\rangle \left\langle \eta \right\rangle
\end{array}
\end{equation}%
where $i=0,1$ and 
\begin{equation}
	\begin{array}{rcl}
C_{mm}^{(i)} &=&G_{11}^{(i)}\left\vert \mathcal{M}_{1}^{LL}\right\vert ^{2}  \\
C_{\lambda \lambda }^{(i)} &=&G_{33}^{(i)}\left\vert \mathcal{M}%
_{3}^{RR}\right\vert ^{2}+G_{44}^{(i)}\left\vert \mathcal{M}_{4}^{\prime
RR}\right\vert ^{2}+2G_{34}^{(i)}\mathcal{M}_{3}^{RR}\mathcal{M}_{4}^{\prime
RR}   \\
C_{\eta \eta }^{(i)} &=&G_{33}^{(i)}\left\vert \mathcal{M}%
_{3}^{RL}\right\vert ^{2}+G_{44}^{(i)}\left\vert \mathcal{M}_{4}^{\prime
RL}\right\vert^{2} +2G_{34}^{(i)}\mathcal{M}_{3}^{RL}\mathcal{M}_{4}^{\prime RL} \\
&&+G_{55}^{(i)}\left\vert \mathcal{M}_{5}^{LR}\right\vert
^{2}+G_{66}^{(i)}\left\vert \mathcal{M}_{6}^{LR}\right\vert
^{2}+2G_{56}^{(i)}\mathcal{M}_{5}^{LR}\mathcal{M}_{6}^{LR} \\
C_{m\lambda }^{(i)} &=&G_{13}^{(i)}\mathcal{M}_{1}^{LL}\mathcal{M}%
_{3}^{RR}+G_{14}^{(i)}\mathcal{M}_{1}^{LL}\mathcal{M}_{4}^{\prime RR}  \\
C_{m\eta }^{(i)} &=&G_{13}^{(i)}\mathcal{M}_{1}^{LL}\mathcal{M}%
_{3}^{RL}+G_{14}^{(i)}\mathcal{M}_{1}^{LL}\mathcal{M}_{4}^{\prime
RL}+G_{15}^{(i)}\mathcal{M}_{1}^{LL}\mathcal{M}_{5}^{LR}+G_{16}^{(i)}%
\mathcal{M}_{1}^{LL}\mathcal{M}_{6}^{LR}   \\
C_{\lambda \eta }^{(i)} &=&G_{33}^{(i)}\mathcal{M}_{3}^{RR}\mathcal{M}%
_{3}^{RL}+G_{44}^{(i)}\mathcal{M}_{4}^{\prime RR}\mathcal{M}_{4}^{\prime
RL}+G_{34}^{(i)}\left( \mathcal{M}_{3}^{RR}\mathcal{M}_{4}^{\prime RL}+%
\mathcal{M}_{3}^{RL}\mathcal{M}_{4}^{\prime RR}\right)  
\end{array}
\end{equation}%
where, for simplicity of notation, we have dropped the index 3 in $\mathcal{M%
}_{3,j}\equiv \mathcal{M}_{j}$, $j=1,3,4,5,6$ from which we can calculate
the half-life 
\begin{equation}
	\label{eqn:tau-A0}
\left[ \tau _{1/2}^{0\nu }\left( 0^{+}\rightarrow 0^{+}\right) \right]
^{-1}=A^{(0)}
\end{equation}%
and the angular coefficient%
\begin{equation}
\label{eqn:angular-coefficient}
\mathcal{K}=\frac{A^{(1)}}{A^{(0)}}.
\end{equation}

\subsection{SUSY models}

The differential rate for these models is obtained by combining the matrix
elements $\mathcal{M}_{4}$ and $\mathcal{M}_{5}$ of Eqs.(\ref{eq:nme4})-(\ref{eq:nme5}) with the
phase space factors of Eqs. (\ref{eqn:D40})-(\ref{eqn:D45}). Introducing the quantities%

\begin{equation}
	\begin{array}{rcl}
\label{eqn:Xi-Yj}
X_{1} &=&\frac{\left\langle m_{\nu }\right\rangle }{m_{e}}\mathcal{M}%
_{3,1}^{LL}+\bar{\epsilon}_{V-A}^{V-A}\mathcal{M}_{3,1}^{LL}   \\
Y_{1} &=&\bar{\epsilon}_{S-P}^{S-P}\mathcal{M}_{5,1}^{LL}+\bar{\epsilon}%
_{S-P}^{S+P}\mathcal{M}_{5,1}^{RL}+\bar{\epsilon}_{T-T_{5}}^{T-T_{5}}%
\mathcal{M}_{4,1}^{LL}+\bar{\epsilon}_{T-T_{5}}^{T+T_{5}}\mathcal{M}%
_{4,1}^{RL} \\
Y_{3} &=&\bar{\epsilon}_{S+P}^{S+P}\mathcal{M}_{5,3}^{RR}+\bar{\epsilon}%
_{S+P}^{S-P}\mathcal{M}_{5,3}^{LR}+\bar{\epsilon}_{T+T_{5}}^{T+T_{5}}%
\mathcal{M}_{4,3}^{RR}+\bar{\epsilon}_{T+T_{5}}^{T-T_{5}}\mathcal{M}%
_{4,3}^{LR}   \\
Y_{4} &=&\bar{\epsilon}_{S+P}^{S+P}\mathcal{M}_{5,4}^{\prime RR}+\bar{%
\epsilon}_{S+P}^{S-P}\mathcal{M}_{5,4}^{\prime LR}+\bar{\epsilon}%
_{T+T_{5}}^{T+T_{5}}\mathcal{M}_{4,4}^{\prime RR}+\bar{\epsilon}%
_{T+T_{5}}^{T-T_{5}}\mathcal{M}_{4,4}^{\prime LR}  
\end{array}
\end{equation}%
where, again, the first term in $X_{1}$ is the standard mass mechanism, and $%
\bar{\epsilon}_{S\mp P}^{S\mp P}$ and $\bar{\epsilon}_{T\mp T_{5}}^{T\mp
T_{5}}$ are effective coupling constants of SUSY models, one obtains%
\begin{equation}
\label{eqn:a0-a1}
a^{(i)}=f_{11}^{(i)}\left\vert X_{1}\right\vert
^{2}+\sum_{j,k=1,3,4}f_{jk}^{\prime (i)}\func{Re}\left[ Y_{j}Y_{k}^{\ast }%
\right] +2\sum_{k=1,3,4}f_{1k}^{\prime \prime (i)}\func{Re}\left[
X_{1}Y_{k}^{\ast }\right] ,\text{ \ \ }i=0,1.
\end{equation}%
The PSFs $f_{jk}^{\prime }$ and $f_{jk}^{\prime \prime }$ are different from
those of the previous subsection and are given in Appendix \ref{Leptonic matrix elements and phase space factors}. The indices 1,
3 and 4 in $Y_{j}$ denote the fact that they arise from the $m_{i}$ term,
index 1, $\omega $ term, index 3, and $\mathbf{k}$ term, index 4. We do not
consider rank-1 terms for this case.

For $R$-parity violating SUSY models \cite{Mohapatra:1986su,Vergados:1986td,Hirsch:1995zi,Babu:1995vh,Hirsch:1995cg,Faessler:1996ph}, the nonzero coefficients are $%
\bar{\epsilon}_{V-A}^{V-A},$ $\bar{\epsilon}_{S+P}^{S+P},$ $\bar{\epsilon}%
_{S+P}^{S-P},$ and $\bar{\epsilon}_{T+T_{5}}^{T+T_{5}}$; see Table \ref{tab:LR-SUSY-coeffs}. This
class of models was investigated by P\"{a}s \textit{et al.} \cite{Pas:1999fc}. Using the
notation%
\begin{equation}
\epsilon _{V-A,i}^{V-A}=U_{ei}\gamma ,\text{ \ \ }\epsilon
_{S+P,i}^{S+P}=V_{ei}^{\prime }\theta ,\text{ \ \ }\epsilon
_{S+P,i}^{S-P}=V_{ei}^{\prime }\tau ,\text{ \ \ }\epsilon
_{T+T_{5},i}^{T+T_{5}}=V_{ei}^{\prime \prime }\varphi ,
\end{equation}%
we have%
\begin{equation}
	\begin{array}{rcl}
\left\langle m_{\nu }\right\rangle &=&\sum_{i}m_{i}U_{ei}^{2},\text{ \ \ }%
\bar{\epsilon}_{V-A}^{V-A}=\gamma \sum_{i}U_{ei}^{2}\equiv \left\langle
\gamma \right\rangle ,\text{ \ \ }\bar{\epsilon}_{S+P}^{S+P}=\theta
\sum_{i}U_{ei}V_{ei}^{\prime }\equiv \left\langle \theta \right\rangle , 
\\
\bar{\epsilon}_{S+P}^{S-P} &=&\tau \sum_{i}U_{ei}V_{ei}^{\prime }\equiv
\left\langle \tau \right\rangle ,\text{ \ \ }\bar{\epsilon}%
_{T+T_{5}}^{T+T_{5}}=\varphi \sum_{i}U_{ei}V_{ei}^{\prime \prime }\equiv
\left\langle \varphi \right\rangle ,
\end{array}
\end{equation}%
where $U_{ei}$ denote the matrix elements of the standard PMNS matrix
\cite{Pontecorvo:1957qd,Maki:1962mu} and $V_{ei}^{\prime }$ and $V_{ei}^{\prime \prime }$ those of the
non-standard scalar and tensor PMNS matrix. Again, $\left\langle \gamma
\right\rangle $ can be set to zero, since this term always appears in
combination with the mass term.

By introducing the integrated normalized PSFs of Eq. (\ref{eqn:integrated-normalized-PSF}), one then obtains%
\begin{equation}
	\begin{array}{rcl}
A^{(i)} &=&G_{11}^{(i)}\left\vert X_{1}\right\vert ^{2}+G_{11}^{\prime
(i)}\left\vert Y_{1}\right\vert ^{2}+G_{33}^{\prime (i)}\left\vert
Y_{3}\right\vert ^{2}+G_{44}^{\prime (i)}\left\vert Y_{4}\right\vert
^{2}+2G_{13}^{\prime (i)}\func{Re}\left[ Y_{1}Y_{3}^{\ast }\right]  
\\
&&+2G_{14}^{\prime (i)}\func{Re}\left[ Y_{1}Y_{4}^{\ast }\right]
+2G_{34}^{\prime (i)}\func{Re}\left[ Y_{3}Y_{4}^{\ast }\right]
+2G_{11}^{\prime \prime (i)}\func{Re}\left[ X_{1}Y_{1}^{\ast }\right] 
\\
&&+2G_{13}^{\prime \prime (i)}\func{Re}\left[ X_{1}Y_{3}^{\ast }\right]
+2G_{14}^{\prime \prime (i)}\func{Re}\left[ X_{1}Y_{4}^{\ast }\right] ,
\end{array}%
\label{eq:susyAi}
\end{equation}
with $i=0,1$. Rearranging Eq. (\ref{eq:susyAi}) with respect to $\frac{\left\langle m_{\nu
}\right\rangle }{m_{e}}, \left\langle \theta \right\rangle, \left\langle \tau \right\rangle $ and $%
\left\langle \varphi \right\rangle $ one has%
\begin{equation}
	\begin{array}{rcl}
A^{(i)} &=&C_{mm}^{(i)}\left( \frac{\left\langle m_{\nu }\right\rangle }{%
m_{e}}\right) ^{2}+C_{\theta \theta }^{(i)}\left\langle \theta\right\rangle ^{2}
+C_{\tau \tau }^{(i)}\left\langle \tau \right\rangle^{2}
+C_{\varphi \varphi }^{(i)}\left\langle \varphi \right\rangle^{2}\\
&&+2C_{\theta \tau }^{(i)}\left\langle
\theta \right\rangle \left\langle \tau \right\rangle
+2C_{\theta \varphi }^{(i)}\left\langle
\theta \right\rangle \left\langle \varphi \right\rangle
+2C_{\tau \varphi }^{(i)}\left\langle
\tau \right\rangle \left\langle \varphi \right\rangle\\
&&+2C_{m\theta }^{(i)}\frac{\left\langle m_{\nu }\right\rangle }{m_{e}}%
\left\langle \theta \right\rangle 
+2C_{m\tau }^{(i)}\frac{\left\langle m_{\nu }\right\rangle }{m_{e}}%
\left\langle \tau \right\rangle
+2C_{m\varphi }^{(i)}\frac{\left\langle m_{\nu }\right\rangle }{m_{e}}%
\left\langle \varphi \right\rangle 
\end{array}
\end{equation}%
with
\begin{equation}
	\begin{array}{rcl}
C_{mm}^{(i)} &=&G_{11}^{(i)}\left\vert \mathcal{M}_{3,1}^{LL}\right\vert ^{2}  \\
C_{\theta \theta }^{(i)} &=&G_{33}^{\prime(i)}\left\vert \mathcal{M}%
_{5,3}^{RR}\right\vert ^{2}+G_{44}^{\prime(i)}\left\vert \mathcal{M}_{5,4}^{\prime
RR}\right\vert ^{2}+2G_{34}^{\prime(i)}\operatorname{Re}[\mathcal{M}_{5,3}^{RR}\mathcal{M}_{5,4}^{\prime
RR*} ]  \\
C_{\tau \tau }^{(i)} &=&G_{33}^{\prime(i)}\left\vert \mathcal{M}%
_{5,3}^{LR}\right\vert ^{2}+G_{44}^{\prime(i)}\left\vert \mathcal{M}_{5,4}^{\prime
LR}\right\vert^{2} +2G_{34}^{\prime(i)}\operatorname{Re}[\mathcal{M}_{5,3}^{LR}\mathcal{M}_{5,4}^{\prime LR*}] \\
C_{\varphi \varphi }^{(i)} &=&G_{33}^{\prime(i)}\left\vert \mathcal{M}%
_{4,3}^{RR}\right\vert ^{2}+G_{44}^{\prime(i)}\left\vert \mathcal{M}_{4,4}^{\prime
RR}\right\vert^{2} +2G_{34}^{\prime(i)}\operatorname{Re}[\mathcal{M}_{4,3}^{RR}\mathcal{M}_{4,4}^{\prime RR}] \\

C_{\theta \tau }^{(i)} &=&G_{33}^{\prime(i)}\operatorname{Re}[\mathcal{M}_{5,3}^{RR}\mathcal{M}_{5,3}^{LR*}]
+G_{44}^{\prime(i)}\operatorname{Re}[\mathcal{M}_{5,4}^{\prime RR}\mathcal{M}_{5,4}^{\prime LR*}]\\
&&+G_{34}^{\prime(i)}\left( \operatorname{Re}[\mathcal{M}_{5,3}^{RR}\mathcal{M}_{5,4}^{\prime LR*}]+%
\operatorname{Re}[\mathcal{M}_{5,3}^{LR}\mathcal{M}_{5,4}^{\prime RR*}]\right)  \\

C_{\theta \varphi }^{(i)} &=&G_{33}^{\prime(i)}\operatorname{Re}[\mathcal{M}_{5,3}^{RR}\mathcal{M}_{4,3}^{RR*}]
+G_{44}^{\prime(i)}\operatorname{Re}[\mathcal{M}_{5,4}^{\prime RR}\mathcal{M}_{4,4}^{\prime RR*}]\\
&&+G_{34}^{\prime(i)}\left( \operatorname{Re}[\mathcal{M}_{5,3}^{RR}\mathcal{M}_{4,4}^{\prime RR*}]+%
\operatorname{Re}[\mathcal{M}_{4,3}^{RR}\mathcal{M}_{5,4}^{\prime RR*}]\right)  \\

C_{\tau \varphi }^{(i)} &=&G_{33}^{\prime (i)}\operatorname{Re}[\mathcal{M}_{5,3}^{LR}\mathcal{M}_{4,3}^{RR*}]
+G_{44}^{ \prime(i)}\operatorname{Re}[\mathcal{M}_{5,4}^{\prime LR}\mathcal{M}_{4,4}^{\prime RR*}]\\
&&+G_{34}^{\prime(i)}\left( \operatorname{Re}[\mathcal{M}_{5,3}^{LR}\mathcal{M}_{4,4}^{\prime RR*}]+%
\operatorname{Re}[\mathcal{M}_{4,3}^{RR}\mathcal{M}_{5,4}^{\prime LR*}]\right)  \\
C_{m\theta }^{(i)} &=&G_{13}^{\prime\prime(i)}\operatorname{Re}[\mathcal{M}_{3,1}^{LL}\mathcal{M}_{5,3}^{RR*}]
+G_{14}^{\prime\prime(i)}\operatorname{Re}[\mathcal{M}_{3,1}^{ LL}\mathcal{M}_{5,4}^{\prime RR*}]\\
C_{m\tau }^{(i)} &=&G_{13}^{\prime\prime(i)}\operatorname{Re}[\mathcal{M}_{3,1}^{LL}\mathcal{M}_{5,3}^{LR*}]
+G_{14}^{\prime\prime(i)}\operatorname{Re}[\mathcal{M}_{3,1}^{ LL}\mathcal{M}_{5,4}^{\prime LR*}]\\
C_{m\varphi }^{(i)} &=&G_{13}^{\prime\prime(i)}\operatorname{Re}[\mathcal{M}_{3,1}^{LL}\mathcal{M}_{4,3}^{RR*}]
+G_{14}^{\prime\prime(i)}\operatorname{Re}[\mathcal{M}_{3,1}^{ LL}\mathcal{M}_{4,4}^{\prime RR*}]\\

\end{array}
\end{equation}%
 from which one can obtain the inverse half-life 
\begin{equation}
\left[ \tau _{1/2}^{0\nu }\left( 0^{+}\rightarrow 0^{+}\right) \right]
^{-1}=A^{(0)},
\end{equation}%
and the angular correlation coefficient 
\begin{equation}
\mathcal{K}=\frac{A^{(1)}}{A^{(0)}}.
\end{equation}

\color{black}

\section{Limits on coupling constants}
%\color{red}
\subsection{L-R models}

In this subsection, we analyze in detail L-R models giving explicit formulas for the NMEs and put some limits on
their coupling constants. The matrix elements of interest are the rank-0
matrix elements $\mathcal{M}_{3,j}^{LL}=\mathcal{M}_{3,j}^{RR}$ and $\mathcal{%
M}_{3,j}^{LR}=\mathcal{M}_{3,j}^{RL}$, $j=1,3$, the rank-0 matrix elements $%
\mathcal{M}_{3,4}^{\prime LL}=\mathcal{M}_{3,4}^{\prime RR}$ and $\mathcal{M}%
_{3,4}^{\prime LR}=\mathcal{M}_{3,4}^{\prime RL}$, and the rank-1 matrix
elements $\mathcal{M}_{3,j}^{LR}=\mathcal{M}_{3,j}^{RL},$ $j=5,6$. For the
rank-0 matrix elements, it is, for purpose of calculations, convenient to
group together various terms and write the transition operator as%
\begin{equation}
\tilde{h}^{LL}(q)\equiv \tilde{h}_{F}+\tilde{h}_{GT}^{LL}\left( {\bm%
\sigma_{1}}\mathbf{\cdot}{\bm \sigma_{2}}\right) +\tilde{h}_{T}^{LL}S_{12},
\end{equation}%
where%
\begin{equation}
	\begin{array}{rcl}
\label{eqn:LR01}
\tilde{h}_{F} &=&g_{V}^{2}\tilde{h}_{VV}(q^{2}),   \\
\tilde{h}_{GT}^{LL} &=&-g_{A}^{2}\tilde{h}_{AA}(q^{2})+\frac{%
g_{A}g_{P^{\prime }}}{6m_{p}^{2}}q^{2}\tilde{h}_{AP^{\prime }}(q^{2})-\frac{%
\left( g_{V}+g_{W}\right) ^{2}}{6m_{p}^{2}}q^{2}\tilde{h}_{VV}(q^{2}) 
\\
&&-\frac{g_{P^{\prime }}^{2}}{48m_{p}^{4}}q^{4}\tilde{h}_{P^{\prime
}P^{\prime }}(q^{2}), \\
\tilde{h}_{T}^{LL} &=&\frac{g_{A}g_{P^{\prime }}}{6m_{p}^{2}}q^{2}\tilde{h}%
_{AP^{\prime }}(q^{2})+\frac{\left( g_{V}+g_{W}\right) ^{2}}{12m_{p}^{2}}%
q^{2}\tilde{h}_{VV}(q^{2})-\frac{g_{P^{\prime }}^{2}}{48m_{p}^{4}}q^{4}%
\tilde{h}_{P^{\prime }P^{\prime }}(q^{2}),  
\end{array}%
\end{equation}
and%
\begin{equation}
\label{eqn:LR02}
\tilde{h}^{LR}(q)\equiv \tilde{h}_{F}+\tilde{h}_{GT}^{LR}\left( {\bm%
\sigma_{1}}\mathbf{\cdot}{\bm\sigma_{2}}\right) +\tilde{h}_{T}^{LR}S_{12},
\end{equation}%
where%
\begin{equation}
	\begin{array}{rcl}
	\label{eqn:LR03}
\tilde{h}_{F} &=&g_{V}^{2}\tilde{h}_{VV}(q^{2}),   \\
\tilde{h}_{GT}^{LR} &=&g_{A}^{2}\tilde{h}_{AA}(q^{2})-\frac{%
g_{A}g_{P^{\prime }}}{6m_{p}^{2}}q^{2}\tilde{h}_{AP^{\prime }}(q^{2})-\frac{%
\left( g_{V}+g_{W}\right) ^{2}}{6m_{p}^{2}}q^{2}\tilde{h}_{VV}(q^{2}) 
 \\
&&+\frac{g_{P^{\prime }}^{2}}{48m_{p}^{4}}q^{4}\tilde{h}_{P^{\prime
}P^{\prime }}(q^{2}), \\
\tilde{h}_{T}^{LR} &=&-\frac{g_{A}g_{P^{\prime }}}{6m_{p}^{2}}q^{2}\tilde{h}%
_{AP^{\prime }}(q^{2})+\frac{\left( g_{V}+g_{W}\right) }{12m_{p}^{2}}q^{2}%
\tilde{h}_{VV}(q^{2})+\frac{g_{P^{\prime }}^{2}}{48m_{p}^{4}}q^{4}\tilde{h}%
_{P^{\prime }P^{\prime }}(q^{2}).  
\end{array}
\end{equation}%
The form factors $\tilde{h}(q)$ are given in Eqs.  (\ref{eqn:hVVq2})-(\ref{eqn:hPPq2}). Using the
partially conserved axial-vector current hypothesis (PCAC), we can write \cite{Simkovic:1999re}%
\begin{equation}
g_{P^{\prime }}=\left( \frac{2m_{p}}{m_{\pi }}\right) ^{2}\left( 1-\frac{%
m_{\pi }^{2}}{m_{A}^{2}}\right) g_{A}=181.8g_{A}\equiv \tilde{g}_{P^{\prime
}}g_{A},
\end{equation}%
and factorize $g_{A}^{2}$ in Eqs. (\ref{eqn:LR01}) and (\ref{eqn:LR03}),%
\begin{equation}
	\begin{array}{rcl}
\tilde{h}_{F} &=g_{A}^{2}&\left[ \left( \frac{g_{V}^{2}}{g_{A}^{2}}\right) 
\tilde{h}_{VV}(q^{2})\right]   \\
\tilde{h}_{GT} &=g_{A}^{2}&\left[ 
%\begin{array}{c}
\mp \tilde{h}_{AA}(q^{2})\pm \frac{\tilde{g}_{P^{\prime }}}{6m_{p}^{2}}q^{2}%
\tilde{h}_{AP^{\prime }}(q^{2})-\frac{\left( g_{V}+g_{W}\right) ^{2}}{%
g_{A}^{2}6m_{p}^{2}}q^{2}\tilde{h}_{VV}(q^{2})  \right.\\ 
&&\left. \mp \frac{\tilde{g}_{P^{\prime }}}{48m_{p}^{4}}q^{4}\tilde{h}_{P^{\prime
}P^{\prime }}(q^{2})%
%\end{array}%
 \right] \\
\tilde{h}_{T} &=g_{A}^{2}&\left[ \pm \frac{\tilde{g}_{P^{\prime }}}{%
6m_{p}^{2}}q^{2}\tilde{h}_{AP^{\prime }}(q^{2})+\frac{\left(
g_{V}+g_{W}\right) ^{2}}{g_{A}^{2}12m_{p}^{2}}q^{2}\tilde{h}_{VV}(q^{2})\mp 
\frac{\tilde{g}_{P^{\prime }}^{2}}{48m_{p}^{4}}q^{4}\tilde{h}_{P^{\prime
}P^{\prime }}(q^{2})\right]  
\end{array}%
\end{equation}
where, when two signs appear, the upper one is for $LL$ and the lower one
for $LR$. The operator $\tilde{h}^{LL}(q^{2})$ is identical to that given by 
\v{S}imkovic \textit{\ et al. }\cite{Simkovic:1999re}, except for an overall minus sign. The
operators $\tilde{h}^{LL}(q^{2})$ and $\tilde{h}^{LR}(q^{2})$ differ from
each other by flipping of some signs as given in Eq. (\ref{eq:nme3}) for the combination
of chiralities $LL$ and $\frac{1}{2}(LR+RL)$.

In order to calculate the matrix elements of interest, we first consider the
matrix elements $j=1$,$3$. In this case, as discussed in Sec. \ref{Nuclear matrix elements}, one
multiplies $\tilde{h}$ by the neutrino potentials $v_{1},v_{3}$%
\begin{equation}
h_{j}^{LL}(q)=v_{j}(q)\tilde{h}^{LL}(q),\text{ \ \ }h_{j}^{LR}(q)=v_{j}(q)%
\tilde{h}^{LR}(q),\text{ \ \ }j=1,3,
\end{equation}%
and calculates the matrix elements%
\begin{equation}
\mathcal{M}_{3,j}^{LL}=\left\langle h_{j}^{LL}(q)\right\rangle ,\text{ \ \ }%
\mathcal{M}_{3,j}^{LR}=\left\langle h_{j}^{LR}(q)\right\rangle ,\text{ \ \ }%
j=1,3.
\end{equation}%
The values of the matrix elements are given in Tables \ref{tab:NMEs-LL-LR1}-\ref{tab:NMEs-LL-LR2}.%
%%%%%%%%%%%%%%%%%%%%%%%%%%%%%%%%%%%%%%%
\begin{table}[htbp]
\centering
\begin{tabular}{lcccc|cccc}
\hline
\hline 
		&\multicolumn{4}{c|}{LL NME}								&\multicolumn{4}{c}{LR NME}\\				
A		&$M_F$		&$M^{LL}_{GT}$		&$M^{LL}_T$		&${\cal M}^{LL}_{3,1}$		&$M_F$		&$M^{LR}_{GT}$		&$M^{LR}_T$		&${\cal M}^{LR}_{3,1}$\\
\hline
76		&-0.78		&-8.79		&-0.41		&-9.98		&-0.78		&8.13		&0.28		&7.64\\
82		&-0.67		&-7.12		&-0.40		&-8.18		&-0.67		&6.58		&0.27		&6.18\\
100		&-0.51		&-7.96		&0.47		&-8.00		&-0.51		&7.25		&-0.31		&6.43\\
130		&-0.65		&-5.64		&-0.23		&-6.52		&-0.65		&5.19		&0.15		&4.69\\
136		&-0.52		&-4.65		&-0.18		&-5.36		&-0.52		&4.30		&0.12		&3.90\\
\hline
\hline								
\end{tabular}
\caption{Nuclear matrix elements ${\cal M}^{LL}_{3,1}$ and ${\cal M}^{LR}_{3,1}$  \Big[neutrino potential $v_1(q)=\frac{2}{\pi}\frac{1}{q(q+\tilde{A})}$\Big] of some nuclei of interest. Contributing Fermi (F), Gamow-Teller (GT) and tensor (T) nuclear matrix elements are also shown.}
\label{tab:NMEs-LL-LR1}
\end{table}

\begin{table}[htbp]
\centering
\begin{tabular}{lcccc|cccc}
\hline 
\hline
			&\multicolumn{4}{c|}{LL NME}					&\multicolumn{4}{c}{LR NME}\\				
A		&$M_F$		&$M^{LL}_{GT}$		&$M^{LL}_T$		&${\cal M}^{LL}_{3,3}$		&$M_F$		&$M^{LR}_{GT}$		&$M^{LR}_T$		&${\cal M}^{LR}_{3,3}$\\
\hline
76		&-0.73		&-7.18		&-0.40		&-8.31		&-0.73		&6.57		&0.27		&6.11\\
82		&-0.62		&-5.81		&-0.38		&-6.81		&-0.62		&5.30		&0.26		&4.94\\
100		&-0.48		&-6.45		&0.43		&-6.50		&-0.48		&5.79		&-0.28		&5.03\\
130		&-0.59		&-4.36		&-0.21		&-5.17		&-0.59		&3.95		&0.14		&3.50\\
136		&-0.47		&-3.57		&-0.17		&-4.20		&-0.47		&3.24		&0.11		&2.88\\
\hline								
\hline
\end{tabular}
\caption{Nuclear matrix elements ${\cal M}^{LL}_{3,3}$ and ${\cal M}^{LR}_{3,3}$ \Big[neutrino potential $v_3(q)=\frac{2}{\pi}\frac{1}{(q+\tilde{A})^2}$\Big] of some nuclei of interest. Contributing Fermi (F), Gamow-Teller (GT) and tensor (T) nuclear matrix elements are also shown.}
\label{tab:NMEs-LL-LR2}
\end{table}

For $j=4$, as discussed in Sec. \ref{Hadronic matrix elements}, one has to consider 
\begin{equation}
	\begin{array}{rcl}
\tilde{h}^{LL}(q) &=&-\left[\tilde{h}_{F}+\tilde{h}_{GT}^{LL}\left[ -\frac{1}{3}%
\left( {\bm\sigma_{1}}\mathbf{\cdot} {\bm\sigma_{2}}\right) +\frac{2}{3}%
S_{12}\right] +\tilde{h}_{T}^{LL}\left[ \frac{4}{3}\left( {\bm\sigma%
_{1}}\mathbf{\cdot}{\bm\sigma_{2}}\right) +\frac{1}{3}S_{12}\right]\right] ,  
\\
\tilde{h}^{LR}(q) &=&-\left[\tilde{h}_{F}+\tilde{h}_{GT}^{LR}\left[ -\frac{1}{3}%
\left( {\bm\sigma_{1}}\mathbf{\cdot}{\bm\sigma_{2}}\right) +\frac{2}{3}%
S_{12}\right] +\tilde{h}_{T}^{LR}\left[ \frac{4}{3}\left( {\bm\sigma%
_{1}}\mathbf{\cdot}{\bm\sigma_{2}}\right) +\frac{1}{3}S_{12}\right]\right] ,
\end{array}%
\end{equation}
with potential $v_{4}$ and calculate%
\begin{equation}
\mathcal{M}_{3,4}^{\prime LL}=\left\langle h_{4}^{LL}(q)\right\rangle ,\text{
\ \ }\mathcal{M}_{3,4}^{\prime LR}=\left\langle h_{4}^{LR}(q)\right\rangle
\end{equation}%
obtaining the results of Table \ref{tab:NMEs-LL-LR3}.
\begin{table}[htbp]
\centering
\begin{tabular}{lcccccc|cccccc}
\hline 
\hline
			&\multicolumn{6}{c|}{LL NME}								&\multicolumn{6}{c}{LR NME}\\				
A		&$M_F$		&$M^{LL}_{GT}$		&$M^{LL}_T$ &$M'^{LL}_{GT}$		&$M'^{LL}_T$		&${\cal M}^{'LL}_{3,4}$		&$M_F$		&$M^{LR}_{GT}$		&$M^{LR}_T$ &$M'^{LR}_{GT}$		&$M'^{LR}_T$		&${\cal M}^{'LR}_{3,4}$\\
\hline
76	&-0.83		&-10.34		&-0.43	&1.44 &1.60	&-5.57		&-0.83		&9.65		&0.30	&-1.17&-1.25	&6.40\\
82	&-0.71		&-8.47		&-0.41	&1.39 &1.31	&-4.65		&-0.71		&7.90		&0.28	&-1.13&-1.02	&5.37\\
100	&-0.54		&-9.46		&0.48	&-1.72 &1.64	&-3.82		&-0.54		&8.71		&-0.32	&1.40&-1.27	&4.31\\
130	&-0.71		&-6.86		&-0.25	&0.88 &1.10	&-3.55		&-0.71		&6.38		&0.17	&-0.71&-0.86	&4.41\\
136	&-0.57		&-5.70		&-0.20	&0.66 &0.86	&-2.84		&-0.57		&5.31		&0.14	&-0.54&-0.66	&3.54\\
\hline								
\hline
\end{tabular}
\caption{Nuclear matrix elements ${\cal M}^{'LL}_{3,4}$ and ${\cal M}^{'LR}_{3,4}$ \Big[neutrino potential $v_4(q)=\frac{2}{\pi}\frac{q+2\tilde{A}}{q(q+\tilde{A})^2}$\Big] of some nuclei of interest. Contributing Fermi (F), Gamow-Teller (GT) and tensor (T) nuclear matrix elements are also shown.}
\label{tab:NMEs-LL-LR3}
\end{table}

For completeness, we show in \ref{tab:NMEs-LL-LR4} also the matrix elements $\mathcal{M}%
^{short}$ .

\begin{table}[htbp]
\centering
\begin{tabular}{lcccc|cccc}
\hline 
\hline
			&\multicolumn{4}{c|}{LL NME}								&\multicolumn{4}{c}{LR NME}\\				
A		&$M_F$		&$M^{LL}_{GT}$		&$M^{LL}_T$		&${\cal M}^{LL, \rm short}$		&$M_F$		&$M^{LR}_{GT}$		&$M^{LR}_T$		&${\cal M}^{LR, \rm short}$\\
\hline
76		&$-48.89$		&$-190.91$		&$-51.08$		&$-290.88$	&$-48.89$		&212.60		&27.00		&190.70\\
82		&$-41.22$		&$-158.33$		&$-48.69$		&$-248.24$	&$-41.22$		&176.41		&25.84		&161.03\\
100		&$-51.96$		&$-197.21$		&$61.94$		&$-187.23$	&$-51.96$		&231.01		&$-32.28$		&146.77\\
130		&$-38.05$		&$-134.95$		&$-31.46$		&$-204.46$	&$-38.05$		&149.32		&16.35		&127.62\\
136		&$-29.83$		&$-105.96$		&$-24.16$		&$-159.96$	&$-29.83$		&117.93		&12.54		&100.64\\
\hline								
\hline
\end{tabular}
\caption{Nuclear matrix elements ${\cal M}^{LL, \rm short}$ and ${\cal M}^{LR, \rm short}$ \Big[neutrino potential $v(q)=\frac{2}{\pi}\frac{1}{m_em_p}$\Big] of some nuclei of interest. Contributing Fermi (F), Gamow-Teller (GT) and tensor (T) nuclear matrix elements are also shown.}
\label{tab:NMEs-LL-LR4}
\end{table}

The rank-1 matrix elements $\mathcal{M}_{3,j}^{LR},$ $j=5,6$, can be written
as%
\begin{equation}
	\begin{array}{rcl}
\mathcal{M}_{3,5}^{LR} &=&g_{A}^{2}\left( \frac{g_{V}}{g_{A}}\right) \left[ 
\mathcal{M}_{\Sigma ^{\prime }}^{AV}+\mathcal{M}_{\Sigma ^{\prime \prime
}}^{AV}\right]   \\
\mathcal{M}_{3,6}^{LR} &=&-g_{A}^{2}\left( \frac{g_{V}+g_{W}}{g_{A}}\right) %
\left[ -\frac{1}{3}\mathcal{M}_{R,GT}^{AV}+\frac{1}{6}\mathcal{M}_{R,T}^{AV}%
\right]
\end{array}%
\end{equation}
where the various terms are defined in Eqs.(\ref{eq:72})-(\ref{eq:73}), (\ref{eq:75})-(\ref{eq:76}) and we have again
factorized $g_{A}^{2}$. Their numerical values are given in Table \ref{tab:NMEs-LL-LR5}.

\begin{table}[htbp]
\centering
\begin{tabular}{lcccc|cccc}
\hline 
\hline
%			&\multicolumn{4}{c|}{LL NME}								&\multicolumn{4}{c}{LR NME}\\				
& \text{LR NME} &  \\ 
A & $\mathcal{M}_{3,5}^{LR}$ & $\mathcal{M}_{3,6}^{LR}$ \\ 
\hline
76 		&-1.55  	&-6.72  \\ 
82 		&-1.35  	&-5.45  \\ 
100 		&-1.48  	&-7.65  \\ 
130 		&-1.45  	&-5.74  \\ 
136 		&-1.16  	&-4.61 \\
\hline								
\hline
\end{tabular}
\caption{Nuclear matrix elements $\mathcal{M}_{3,5}^{LR}$ and $\mathcal{M}%
_{3,6}^{LR}$}
\label{tab:NMEs-LL-LR5}
\end{table}

These values are obtained by using $g_{V}=1$, $g_{A}=1.269$ and $g_{W}=3.70$. %,and are the matrix elements $M^{(0\nu )}$ in \cite{Barea:2013bz,Barea:2015kwa}. The matrix elements $%
%M_{0\nu }$ of \cite{Barea:2013bz,Barea:2015kwa} can be obtained from $M^{(0\nu )}$ by multiplying by $%
%g_{A}^{2}$, $M_{0\nu }=g_{A}^{2}M^{(0\nu )}$.
 The factorization of $%
g_{A}^{2} $ allows to obtain the values of the matrix elements for any
value of $g_{A}^{eff}=\frac{g_{A}}{1.269}$ by simply multiplying the values
in Tables \ref{tab:NMEs-LL-LR1}-\ref{tab:NMEs-LL-LR5} by $\left( g_{A}^{eff}\right) ^{2}$. This procedure is
necessary in view of the quenching of $g_{A}$ in nuclei \cite{Barea:2013bz,Barea:2015kwa}. The matrix
elements in \ref{tab:NMEs-LL-LR1}-\ref{tab:NMEs-LL-LR5} are calculated within the framework of IBM-2
\cite{Barea:2009zza,Barea:2013bz,Barea:2015kwa}, with single particle levels given in \cite{Kotila:2016pib}.

\color{black}

From these values one can calculate $X_j$, Eq. (\ref{eqn:Xij-LR}), and $a^{(0)}$ and $a^{(1)}$, Eq. (\ref{eqn:a-LR}), and from these the angular correlation coefficient ${\mathcal K} = A^{(1)}/A^{(0)}$, Eq. (\ref{eqn:dW0vdcostheta12}), and half-life $\tau_{1/2}^{0\nu}$, Eq. (\ref{eqn:half-life-tau}).
The latter quantities depend on four parameters, Eq. (\ref{eqn:LR-parameters}), $\frac{\langle m_\nu \rangle}{m_e}$, $\bar \epsilon_{V+A}^{V+A} \equiv \langle \lambda \rangle$, $\bar \epsilon_{V-A}^{V+A} \equiv \langle \eta \rangle$, and $\bar \epsilon_{V+A}^{V-A} \equiv \langle \kappa \rangle$.
However, $\langle \kappa \rangle$ is always in combination with the standard mass mechanism and can be put to zero.
The half-life in terms of the three parameters $\frac{\langle m_\nu \rangle}{m_e}$, $\langle \lambda \rangle$, $\langle \eta \rangle$ is given explicitly in Eq. (\ref{eqn:tau-A0}).

\begin{figure}[htbp]
%\[
%\text{Fig. 3 } 
%\]
\centering
\begin{minipage}{17.5pc}
\flushleft{a)}
\includegraphics[width=17.5pc]{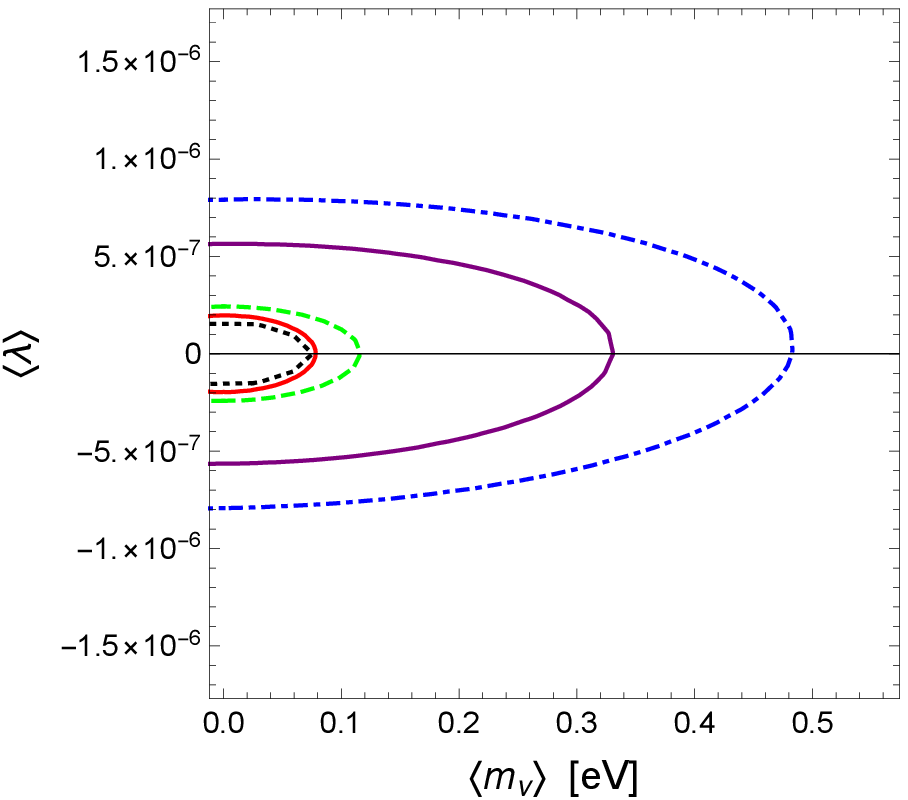} 
\end{minipage}
\hspace{2pc}
\begin{minipage}{17pc}
\flushleft{b)}
\includegraphics[width=17pc]{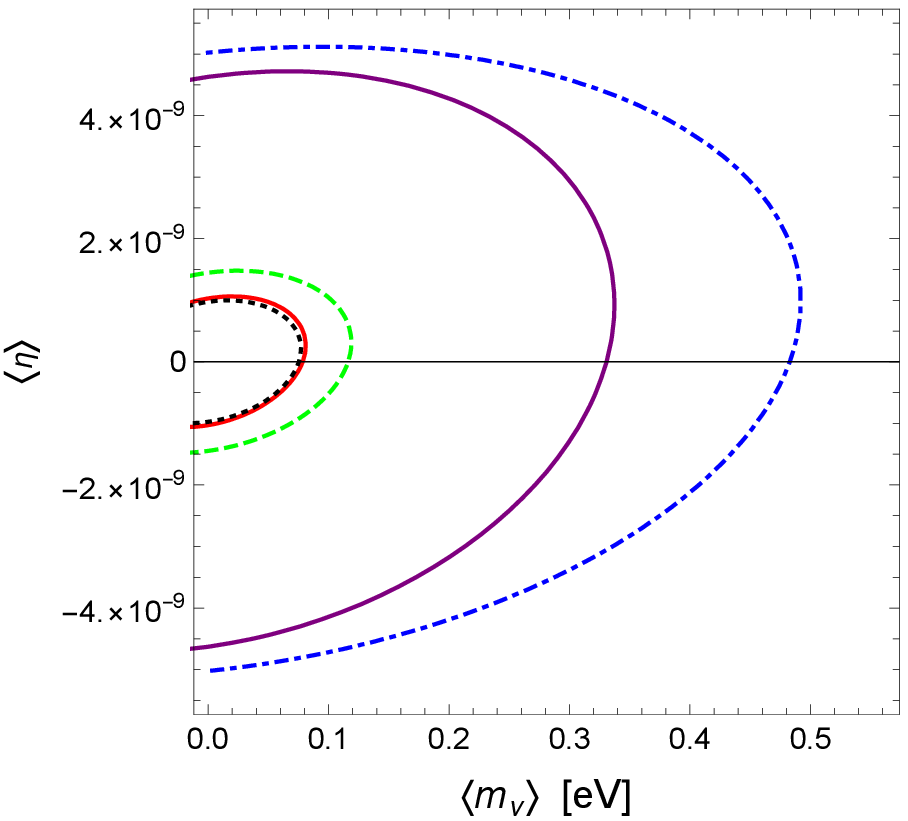} 
\end{minipage}
\caption{Limits on the combination of the mass term, $\langle m_\nu \rangle$ [eV], with the non-standard a) $\bar \epsilon_{V+A}^{V+A} \equiv \langle \lambda \rangle$ term and b)  $\bar \epsilon_{V+A}^{V-A} \equiv \langle \eta \rangle$  in the case of $^{76}$Ge (continuous red line), $^{82}$Se (continuous purple line), $^{100}$Mo (blue dot-dashed line), $^{130}$Te (green dashed line), and $^{136}$Xe (black dotted line).}
\label{fig:limits-half-life}
\end{figure}

From the experimental limit on the half life one can put limits on the three dimensional surface defined by $A^{(0)} = C$.
However, this is not very illuminating and we prefer to put limits on the combinations of the mass term, $\frac{\langle m_\nu \rangle}{m_e}$, and one of the other parameters, $\langle \lambda \rangle$ or $\langle \eta \rangle$. These limits are shown in Fig. \ref{fig:limits-half-life}.
%%%%%%%%%%%%%%%%%%%%%

%%%%%%%%%%%%%%%%%%%%%

For $\langle \lambda \rangle = 0$ and $\langle \eta \rangle = 0$, the limit agrees with our previously published value \cite{Barea:2015kwa}.
An alternative, followed by many authors, is instead to give limits on the parameters considered one at a time. These limits are given in Table \ref{tab:limits-param-LR}.
%%%%%%%%%%%%%%%%%%%%%%%%%%%%%%%%%%%%%%%
\begin{table*}[htbp]
\centering	
\begin{tabular}{ccc|ccc}
\hline
\hline
&$T_{1/2}^{\rm exp}$ [y] & & $\frac{\langle m_\nu \rangle}{m_e}$ & $\langle \lambda \rangle$ & $\langle \eta \rangle$  \\
\hline
%\text{Table XII } \\
$^{76}$Ge &$1.8 \times 10^{26}$ & \cite{Agostini:2013mzu} 	& ~ $1.5\times10^{-7}$ ~ & ~ $2.0\times10^{-7}$ ~ & ~ $1.0\times10^{-9}$ ~     \\
$^{82}$Se &$3.5 \times 10^{24}$ & \cite{Azzolini:2019} 		& ~ $6.5\times10^{-7}$ ~ & ~ $5.7\times10^{-7}$ ~ & ~ $4.6\times10^{-9}$ ~     \\
$^{100}$Mo &$1.1 \times 10^{24}$ & \cite{Arnold:2015wpy}	& ~ $9.4\times10^{-7}$ ~ & ~ $7.9\times10^{-7}$ ~ & ~ $5.0\times10^{-9}$ ~   \\
$^{130}$Te &$3.2 \times 10^{25}$ & \cite{Alfonso:2015wka}	& ~ $2.3\times10^{-7}$ ~ & ~ $2.4\times10^{-7}$ ~ & ~ $1.4\times10^{-9}$ ~   \\
$^{136}$Xe &$1.1 \times 10^{26}$ & \cite{Gando:2012zm} 	& ~ $1.5\times10^{-7}$ ~ & ~ $1.6\times10^{-7}$ ~ & ~ $1.0\times10^{-9}$ ~  \\
\hline
\hline
\end{tabular}
\caption{Upper limits on the absolute values of the parameters of L-R models for $g_A=1.269$.}
\label{tab:limits-param-LR}
\end{table*}
%%%%%%%%%%%%%%%%%%%%%%%%%%%%%%%%%%%%%%%
The limits on $\frac{\langle m_\nu \rangle}{m_e}$ agree with those previously published \cite{Barea:2015kwa}.
We note here that even if the neutrino masses $\langle m_\nu \rangle$ were to be zero, still L-R models would give a finite half-life. We also note that the angular correlation coefficient ${\mathcal K}$, Eq. (\ref{eqn:angular-coefficient}), depends on the values of $\frac{\langle m_\nu \rangle}{m_e}$, $\langle \lambda \rangle$ and $\langle \eta \rangle$.
%%%%%%%%%%%%%%%%%%%%%
\begin{figure}[htbp]
%\[
%\text{Fig. 4 } 
%\]
\centering
\begin{minipage}{17.5pc}
\flushleft{a)}
\includegraphics[width=7cm]{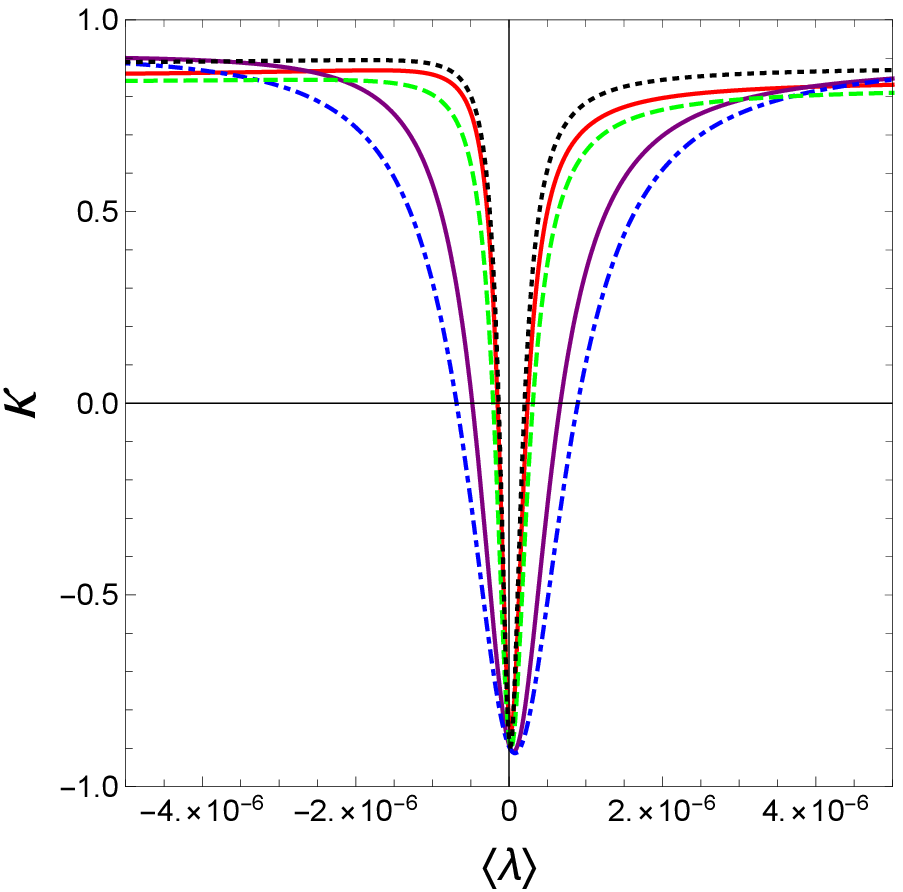} \end{minipage}
\hspace{2pc}
\begin{minipage}{17pc}
\flushleft{b)}
\includegraphics[width=17pc]{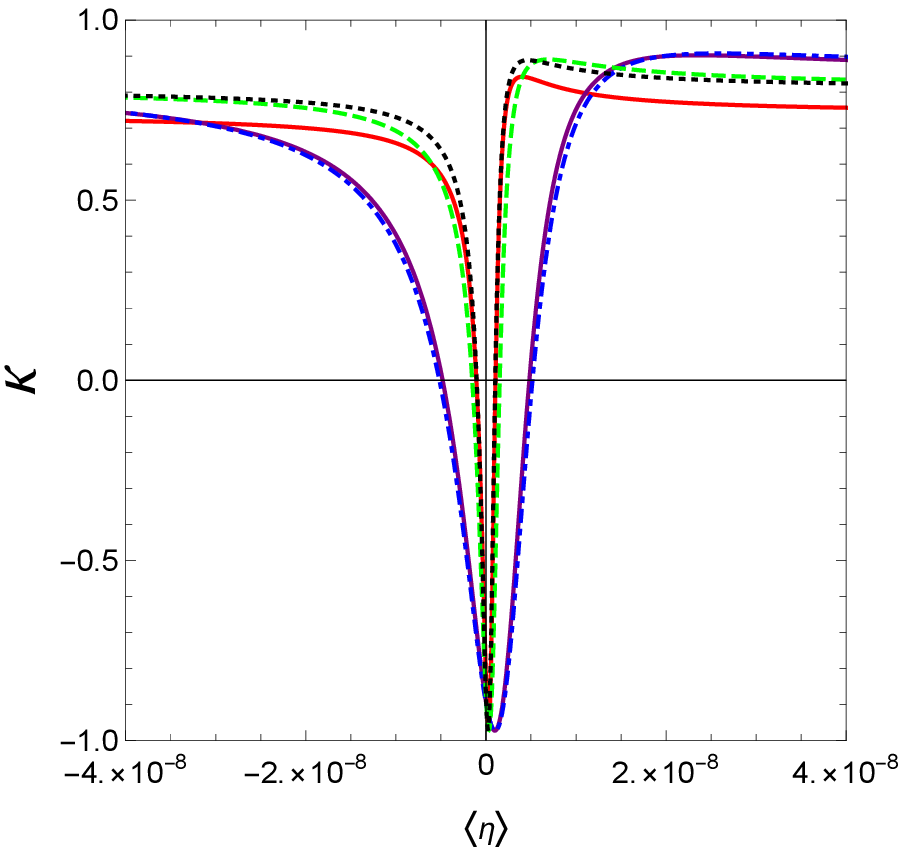} 
\end{minipage}
\caption{Behavior of the coefficient $\mathcal K$, as a function of a) $\langle \lambda \rangle$, and b)  $\langle \eta \rangle$ in the case of $^{76}$Ge (continuous red line), $^{82}$Se (continuous purple line), $^{100}$Mo (blue dot-dashed line), $^{130}$Te (green dashed line), and $^{136}$Xe (black dotted line) for a fixed value of $\frac{\langle m_\nu \rangle}{m_e}$, given in the second column of Table \ref{tab:limits-param-LR}.}
\label{fig:k-VS-lambda}
\end{figure}

\begin{figure*}[htbp]
%\[
%\text{Fig.5 } 
%\]
\small
\begin{minipage}{.48\textwidth}
\centering
\includegraphics[width=0.92\textwidth]{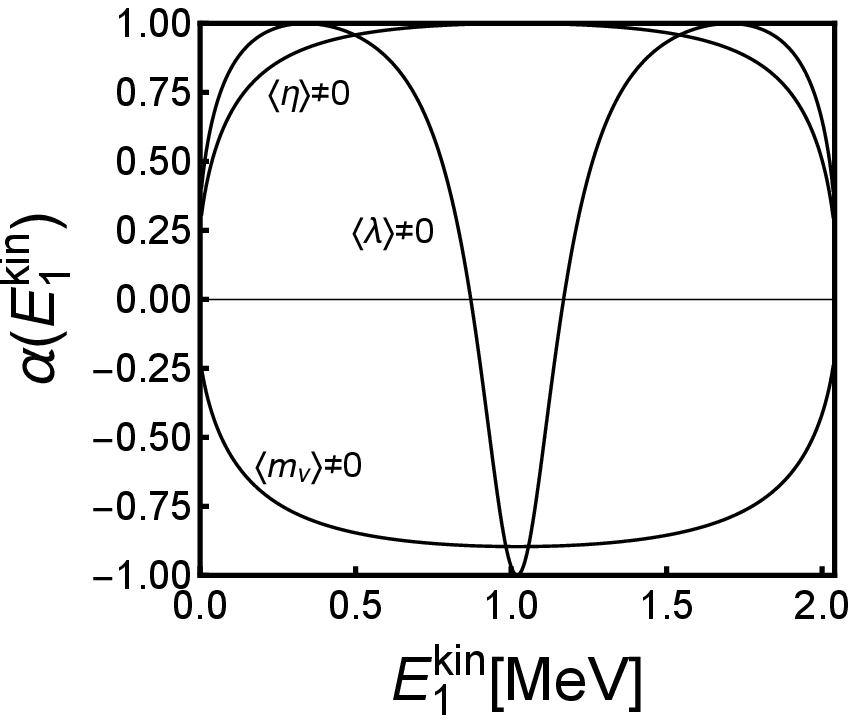}
\end{minipage}
\hfill
\begin{minipage}{.48\textwidth}
\centering
\includegraphics[width=1\textwidth]{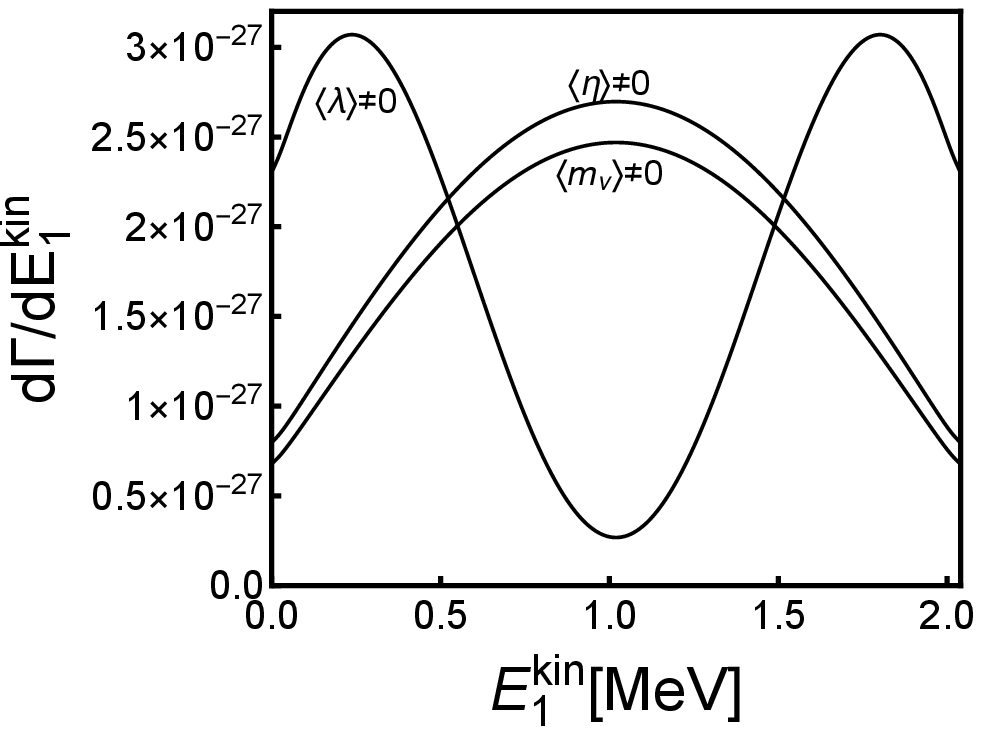}
\end{minipage}
%\begin{minipage}{.48\textwidth}
%\centering
%\includegraphics[width=0.92\textwidth]{fig5-76correps_tomoda.eps}
%\end{minipage}
%\hfill
%\begin{minipage}{.48\textwidth}
%\centering
%\includegraphics[width=1\textwidth]{fig5-76ses_tomoda.eps}
%\end{minipage}
\begin{minipage}{.48\textwidth}
\centering
\includegraphics[width=0.92\textwidth]{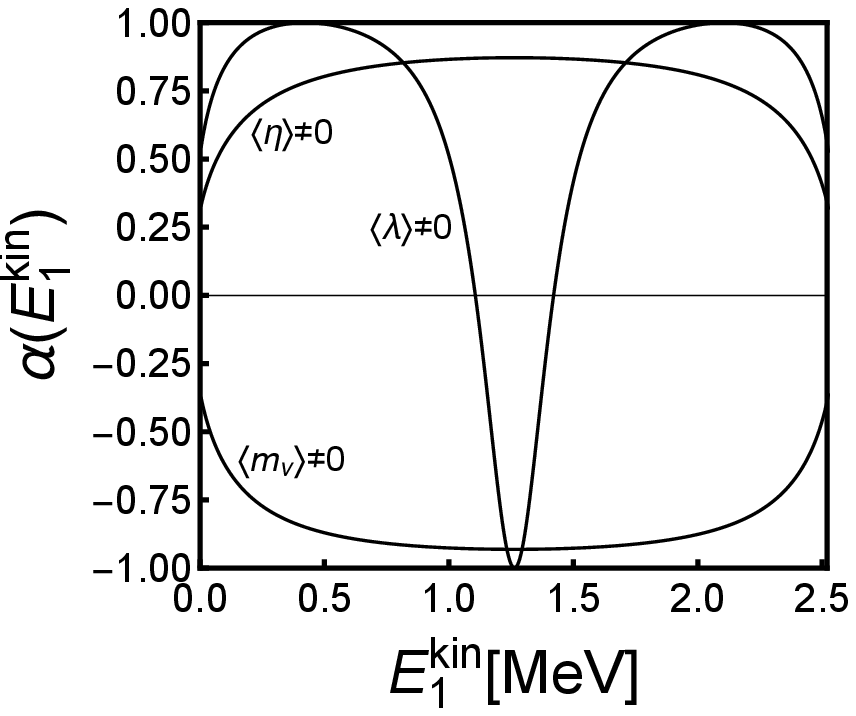}
\end{minipage}
\hfill
\begin{minipage}{.48\textwidth}
\centering
\includegraphics[width=1\textwidth]{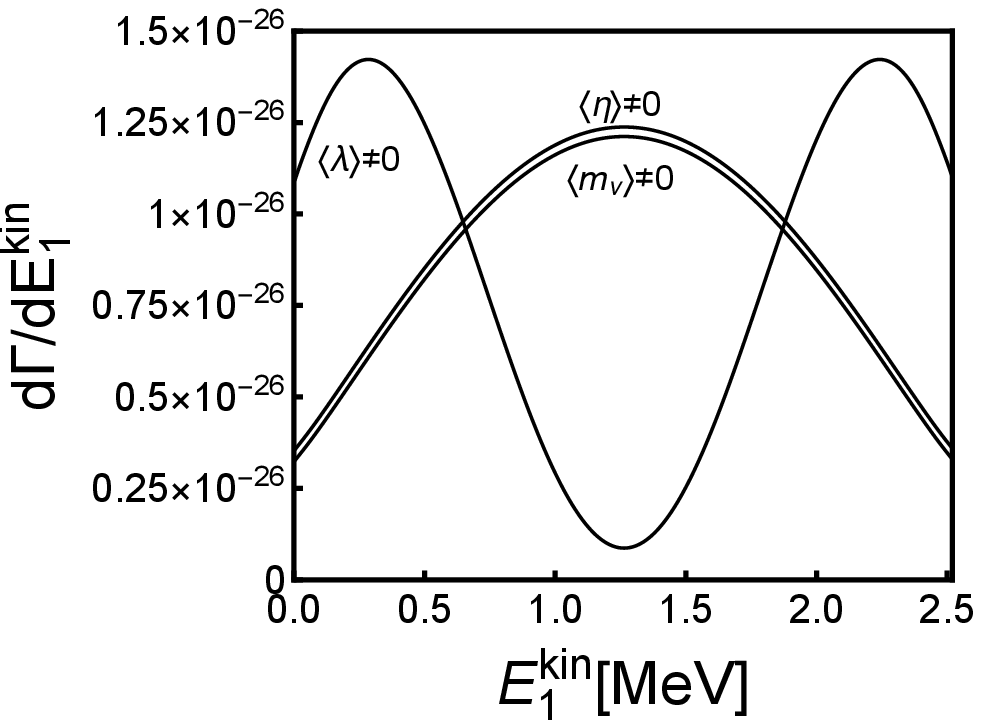}
\end{minipage}
\begin{minipage}{.48\textwidth}
\centering
\includegraphics[width=0.92\textwidth]{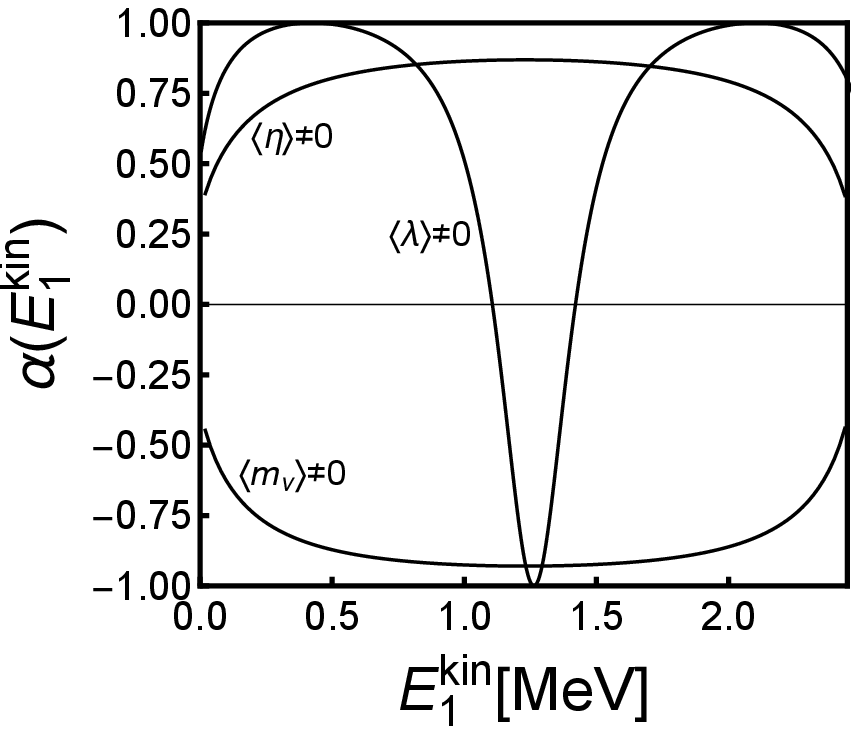}
\end{minipage}
\hfill
\begin{minipage}{.48\textwidth}
\centering
\includegraphics[width=1\textwidth]{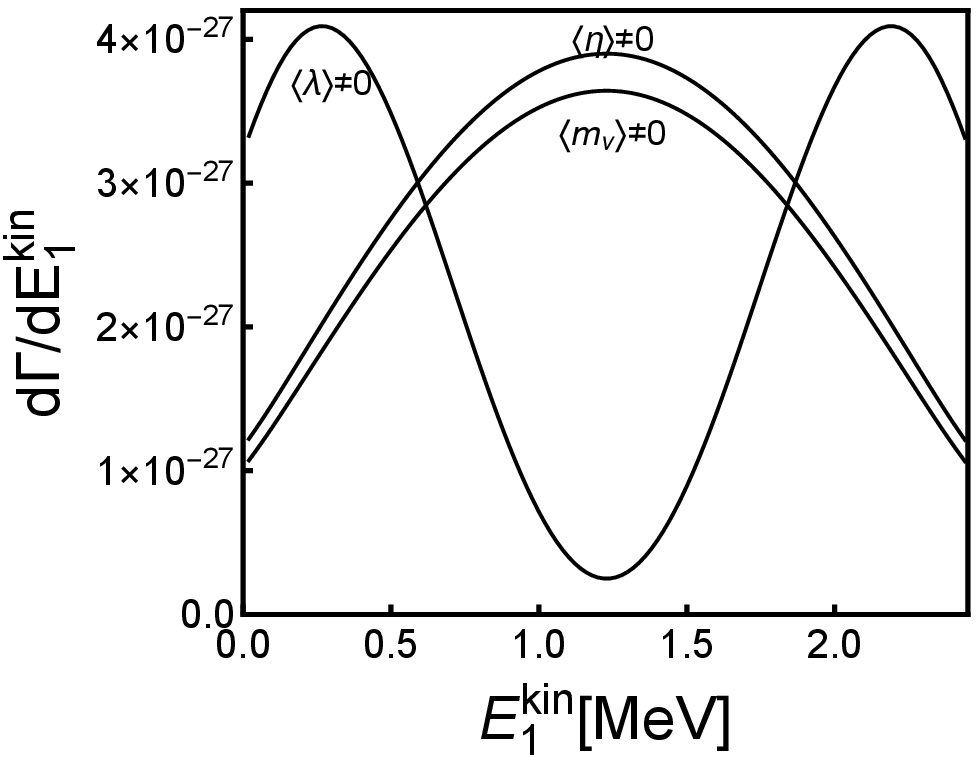}
\end{minipage}
\caption{Right panel: Single electron energy distribution as a function of the kinetic energy $E_1^{\rm kin} = E_1 - m_e$ for the two mechanisms, mass and L-R. Left panel: Energy dependent angular correlation $\alpha(E_1^{\rm kin})$ between the two electrons as a function of the kinetic energy $E_1^{\rm kin}$ for the two mechanisms, mass and L-R. From top to bottom, the first pair of figures corresponds to $^{76}$Ge%with all terms, the second pair to Tomoda's description. }%
, the second pair to $^{130}$Te and, finally, the third pair to $^{136}$Xe.}
\label{fig:e-energy-distribution}
\end{figure*}

%%%%%%%%%%%%%%%%%%%%%
In Fig. \ref{fig:k-VS-lambda}, we show the coefficient $\mathcal K$ for a) $\langle \eta \rangle = 0$ and a fixed value of $\frac{\langle m_\nu \rangle}{m_e}$, as a function of $\langle \lambda \rangle$, and b) $\langle \lambda \rangle = 0$ and a fixed value of $\frac{\langle m_\nu \rangle}{m_e}$, as a function of $\langle \eta \rangle$.
We also show in Table \ref{tab:angular-coeff-LR} the angular correlation coefficients for the different terms, $\frac{\langle m_\nu \rangle}{m_e}$, $\langle \lambda \rangle$ and $\langle \eta \rangle$, considered one at a time.
%%%%%%%%%%%%%%%%%%%%%%%%%%%%%%%%%%%%%%%
\begin{table*}[htbp]
\centering
\begin{tabular}{c|ccc}
\hline
\hline
 & ~ ${\mathcal K} \left(\frac{\langle m_\nu \rangle}{m_e}\right)$ ~ & ~ ${\mathcal K} \left(\langle \lambda \rangle\right)$ ~ 
 & ~ ${\mathcal K} \left(\langle \eta \rangle\right)$ ~ \\
\hline
%\text{Table XIII } 
%\\
$^{76}$Ge & $-0.828$ & $0.841$ & $0.738$     \\
$^{82}$Se & $-0.891$ & $0.896$ & $0.831$ \\
$^{100}$Mo & $-0.896$ & $0.910$ & $0.839$ \\
$^{130}$Te & $-0.876$ & $0.830$ & $0.809$ \\
$^{136}$Xe & $-0.874$ & $0.881$ & $0.806$ \\
\hline
\hline
\end{tabular}
\caption{Angular correlation coefficients for L-R models.}
\label{tab:angular-coeff-LR}
\end{table*}
%%%%%%%%%%%%%%%%%%%%%%%%%%%%%%%%%%%%%%%
These results indicate that, if neutrinoless $0\nu\beta\beta$ decays would be observed, the angular distribution may distinguish between different models. In Fig. \ref{fig:e-energy-distribution} we show the calculated single electron spectrum and the energy-dependent angular correlation 
\begin{equation}
	\alpha(E_1) = \frac{a^{(1)}}{a^{(0)}}
\end{equation}	
for three cases, when the different terms are considered one at a time.
%%%%%%%%%%%%%%%%%%%%%%%%%%%%%%%%%%%%%%%%
%%%%%%%%%%%%%%%%%%%%%%%%%%%%%%%%%%%%%%%%

\subsection{SUSY models}
The matrix elements of interest here are ${\mathcal M}_{4,j}^{LL}$, ${\mathcal M}_{4,j}^{LR}$ and ${\mathcal M}_{5,j}^{LL}$ and ${\mathcal M}_{5,j}^{LR}$, where ${\mathcal M}_4$ and ${\mathcal M}_5$ are given in Eqs. (\ref{eq:nme4}) and (\ref{eq:nme5}). Again, for purpose of calculations, it is convenient to group together various terms and write
\begin{equation}
	\begin{array}{c}
	\tilde h^{LL}_{(4)}(q) \equiv \tilde h^{LL}_{(4), \rm GT}({\bm \sigma}_1 \cdot {\bm \sigma}_2)
	+ \tilde h^{LL}_{(4), \rm T} S_{12} \mbox{ },  \\
	\tilde h_{(4), \rm GT}^{LL} = - i g_A g_{T_1} \tilde h (q^2) + i \frac{g_P g_{T_1}}{12 m_p^2} q^2 \tilde h_{PT_1}(q^2)
	\mbox{ },  \\
	\tilde h_{(4), \rm T}^{LL} = i \frac{g_P g_{T_1}}{12 m_p^2} q^2 \tilde h_{PT_1}(q^2)
	\mbox{ },
	\end{array}
\end{equation}
\begin{equation}
	\begin{array}{c}
	\tilde h^{LR}_{(4)}(q) \equiv \tilde h^{LR}_{(4), \rm GT}({\bm \sigma}_1 \cdot {\bm \sigma}_2)
	+ \tilde h^{LR}_{(4), \rm T} S_{12} \mbox{ },  \\
	\tilde h_{(4), \rm GT}^{LR} = i g_A g_{T_1} \tilde h (q^2) - i \frac{g_P g_{T_1}}{12 m_p^2} q^2 \tilde h_{PT_1}(q^2)
	\mbox{ },  \\
	\tilde h_{(4), \rm T}^{LR} = - i \frac{g_P g_{T_1}}{12 m_p^2} q^2 \tilde h_{PT_1}(q^2)
	\mbox{ },
	\end{array}
\end{equation}
and
\begin{equation}
	\begin{array}{c}
	\tilde h^{LL}_{(5)}(q) \equiv \tilde h_{(5), \rm F} + \tilde h^{LL}_{(5), \rm GT}({\bm \sigma}_1 \cdot {\bm \sigma}_2)
	+ \tilde h^{LL}_{(5), \rm T} S_{12} \mbox{ },  \\
	\tilde h_{(5), \rm F} = g_S g_V \tilde h(q^2)
	\mbox{ },  \\
	\tilde h_{(5), \rm GT}^{LL} = \tilde h_{(5), \rm T}^{LL} = 
	\frac{g_P g_A}{12 m_p^2} \left({\bf q}\cdot{\bf Q}\right)\tilde h_{AP}(q^2)
	- \frac{g_P^2}{24 m_p^2} \left(\frac{q^0}{m_p}\right) q^2 \tilde h_{PP}(q^2)	\mbox{ },  
	\end{array}
\end{equation}
and
\begin{equation}
	\begin{array}{c}
	\tilde h^{LR}_{(5)}(q) \equiv \tilde h_{(5), \rm F} + \tilde h^{LR}_{(5), \rm GT}({\bm \sigma}_1 
	\cdot {\bm \sigma}_2) + \tilde  h^{LR}_{(5), \rm T} S_{12} \mbox{ },  \\
	\tilde h_{(5), \rm F} = g_S g_V \tilde h(q^2) \mbox{ },  \\
	\tilde h_{(5), \rm GT}^{LR} = \tilde h_{(5), \rm T}^{LR} = 
	- \frac{g_P g_A}{12 m_p^2} \left({\bf q}\cdot{\bf Q}\right)\tilde h_{AP}(q^2)
	+ \frac{g_P^2}{24 m_p^2} \left(\frac{q^0}{m_p}\right) q^2 \tilde h_{PP}(q^2)	\mbox{ }.  
	\end{array}
\end{equation}
In this case
\begin{equation}
	\begin{array}{c}
	\tilde h^{LR}_{(4)}(q) = - \tilde h^{LL}_{(4)}(q) \mbox{ },  \\
	\tilde h^{LR}_{(5), \rm GT}(q) = - \tilde h^{LL}_{(5), \rm GT}(q) \mbox{ },  \\
	\tilde h^{LR}_{(5), \rm T}(q) = - \tilde h^{LL}_{(5), \rm T}(q)  \mbox{ }.  
	\end{array}
\end{equation}
%\color{red}
As discussed in Sec. \ref{Nuclear matrix elements}, one then multiplies $\tilde{h}_{(4)}$ and $\tilde{%
h}_{(5)}$ by the neutrino potentials $v_{1},v_{3}$ and $v_{4},$%
\begin{equation}
	\begin{array}{rcl}
h_{(4),j}^{LL} &=&v_{j}(q)\tilde{h}_{(4)}^{LL}(q),\text{ \ \ }%
h_{(4),j}^{LR}=v_{j}(q)\tilde{h}_{(4)}^{LR}(q),   \\
h_{(5),j}^{LL} &=&v_{j}(q)\tilde{h}_{(5)}^{LL}(q),\text{ \ \ }%
h_{(5),j}^{LR}=v_{j}(q)\tilde{h}_{(5)}^{LR}(q),
\end{array}
\end{equation}%
with $j=1,3,4$. From these one can calculate the matrix elements for $j=1,3$%
\begin{equation}
	\begin{array}{rcl}
\mathcal{M}_{(4),j}^{LL} &=&\left\langle h_{(4),j}^{LL}(q)\right\rangle ,%
\text{ \ \ }\mathcal{M}_{(4),j}^{LR}=\left\langle
h_{(4),j}^{LR}(q)\right\rangle ,   \\
\mathcal{M}_{(5),j}^{LL} &=&\left\langle h_{(5),j}^{LL}(q)\right\rangle ,%
\text{ \ \ }\mathcal{M}_{(5),j}^{LR}=\left\langle
h_{(5),j}^{LR}(q)\right\rangle .
\end{array}
\end{equation}%
For $j=4$, as discussed in Sec. \ref{Hadronic matrix elements}, one has to consider the modification
Eq. (\ref{modification}) and calculate%
\begin{equation}
	\begin{array}{rcl}
\mathcal{M}_{(4),4}^{\prime LL} &=&\left\langle
h_{(4),4}^{LL}(q)\right\rangle ,\text{ \ \ }\mathcal{M}_{(4),4}^{\prime
LR}=\left\langle h_{(4),4}^{LR}(q)\right\rangle ,   \\
\mathcal{M}_{(5),4}^{\prime LL} &=&\left\langle
h_{(5),4}^{LL}(q)\right\rangle ,\text{ \ \ }\mathcal{M}_{(5),4}^{\prime
LR}=\left\langle h_{(5),4}^{LR}(q)\right\rangle .
\end{array}%
\end{equation}
The values of the matrix elements for $g_A=1.269$ are given in Table \ref{tab:NMEs-SUSY}. From these, one
can calculate the quantities $Y_{j}$ of Eq. (\ref{eqn:Xi-Yj}) and then $a^{(0)}$ and $%
a^{(1)}$ of Eq. (\ref{eqn:a0-a1}) and the half-life and angular coefficients.

%%%%%%%%%%%%%%%%%%%%%%%%%%%%%%%%%%%%%%%
\begin{table}[htbp]
\centering
\begin{tabular}{l|ccc|ccc}
\hline 
\hline
		&${\cal M}^{LL}_{4,1}$		&${\cal M}^{LL}_{4,3}$		&${\cal M'}^{LL}_{4,4}$		&${\cal M}^{LR}_{4,1}$		&${\cal M}^{LR}_{4,3}$		&${\cal M'}^{LR}_{4,4}$\\
\hline
%		\text{Table XIV } \\

$^{76}$Ge	&$-6.98i$		&$-5.72i$		&$-4.27i$		&$6.98i$		&$5.72i$		&$4.27i$\\
$^{82}$Se		&$-5.69i$		&$-4.63i$		&$-3.58i$		&$5.69i$		&$4.63i$		&$3.58i$\\
$^{100}$Mo	&$-5.88i$		&$-4.70i$		&$-2.72i$		&$5.88i$		&$4.70i$		&$2.72i$\\
$^{130}$Te	&$-4.44i$		&$-3.44i$		&$-2.82i$		&$4.44i$		&$3.44i$		&$2.82i$\\
$^{136}$Xe	&$-3.65i$		&$-2.85i$		&$-2.29i$		&$3.65i$		&$2.85i$		&$2.29i$\\
\hline								
		&${\cal M}^{LL}_{5,1}$		&${\cal M}^{LL}_{5,3}$		&${\cal M'}^{LL}_{5,4}$		&${\cal M}^{LR}_{5,1}$		&${\cal M}^{LR}_{5,3}$		&${\cal M'}^{LR}_{5,4}$\\
		\hline
%		\text{Table XIV } \\
$^{76}$Ge	&$-0.28$		&$-0.26$		&$-0.47$		&$-1.28$		&$-1.20$		&$-1.19$\\
$^{82}$Se		&$-0.27$		&$-0.30$		&$-0.43$		&$-1.06$		&$-0.95$		&$-1.00$\\
$^{100}$Mo	&$-0.61$		&$-0.58$		&$ 0.26$		&$-1.64$		&$-1.54$		&$-1.34$\\
$^{130}$Te	&$-0.28$		&$-0.25$		&$-0.44$		&$-1.02$		&$-0.93$		&$-0.98$\\
$^{136}$Xe	&$-0.23$		&$-0.18$		&$-0.36$		&$-0.81$		&$-0.76$		&$-0.78$\\		
\hline
\hline
\end{tabular}
\caption{Nuclear matrix elements for SUSY models.}
\label{tab:NMEs-SUSY}
\end{table}
%%%%%%%%%%%%%%%%%%%%%%%%%%%%%%%%%%%%%%%
\color{black}

\begin{figure}[htbp]
%\[
%\text{Fig. 6 new} 
%\]
\centering
\begin{minipage}{12.7pc}
\flushleft{a)}
\includegraphics[width=12.7pc]{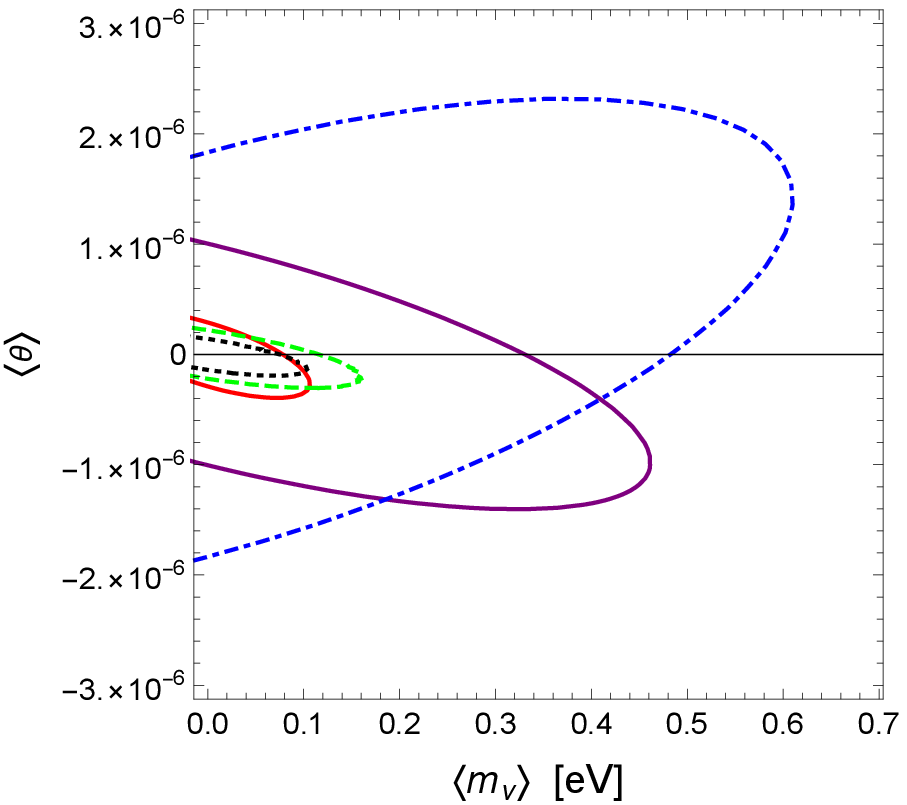} 
\end{minipage}
\hspace{0pc}
\begin{minipage}{12.5pc}
\flushleft{b)}
\includegraphics[width=12.5pc]{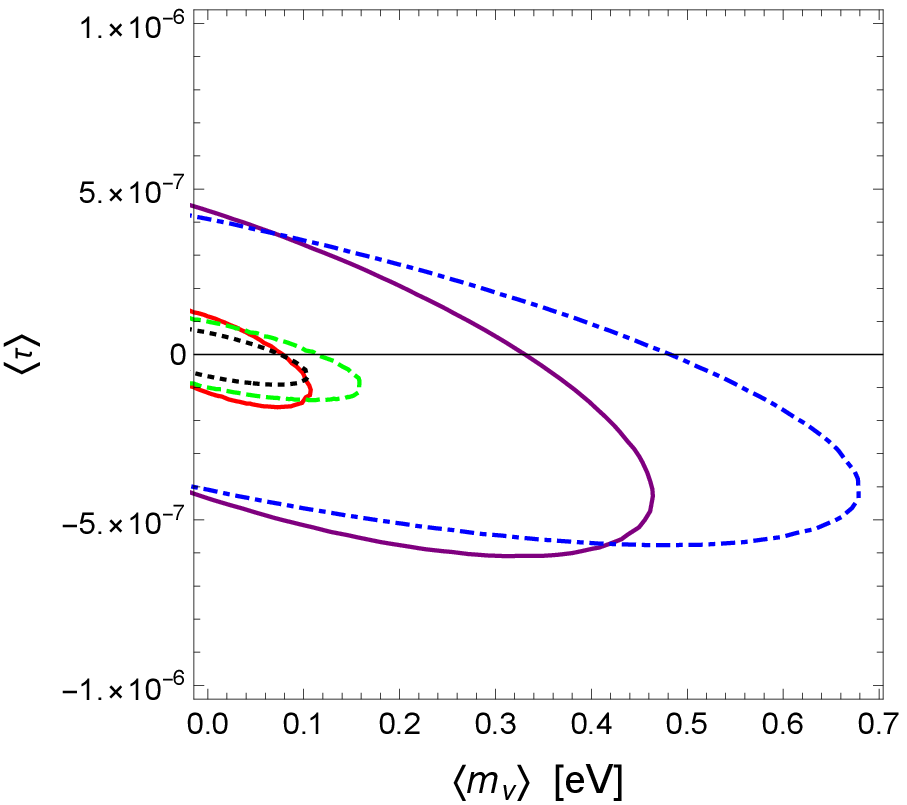} 
\end{minipage}
\hspace{0pc}
\begin{minipage}{12.7pc}
\flushleft{c)}
\includegraphics[width=12.7pc]{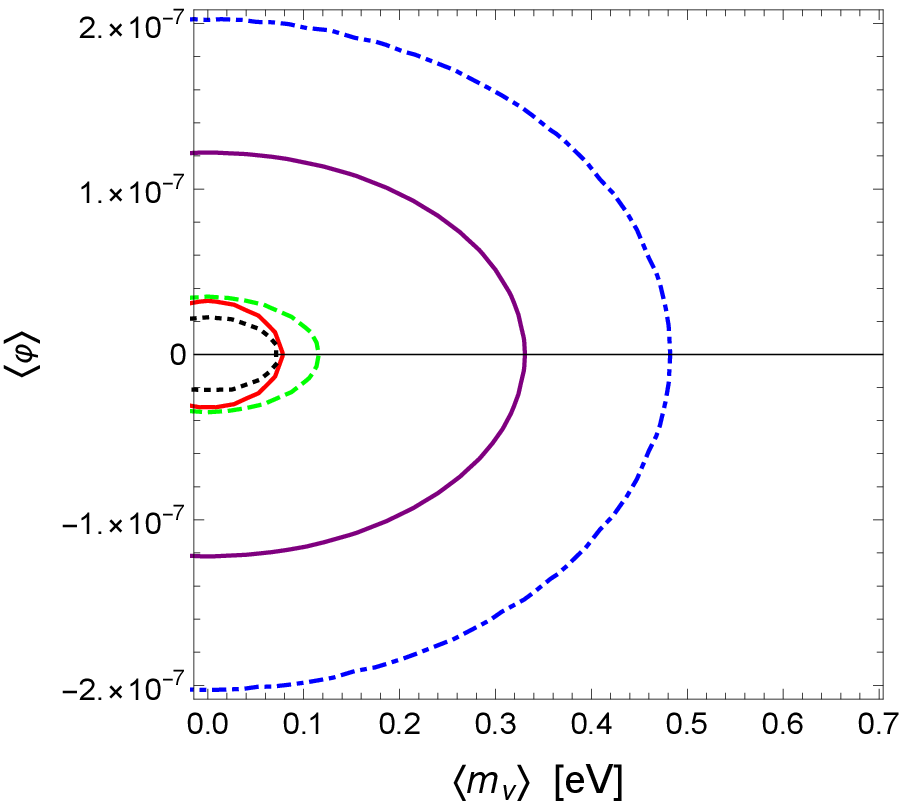} 
\end{minipage}
\caption{Limits on the combination of the mass term, $\langle m_\nu \rangle$ [eV], with the SUSY a) $\bar \epsilon_{S+P}^{S+P} \equiv \langle \theta \rangle$ term, b)  $\bar \epsilon_{S+P}^{S-P} \equiv \langle \tau \rangle$, and c)  $\bar \epsilon_{T+T_5}^{T+T_5} \equiv \langle \varphi \rangle$  in the case of $^{76}$Ge (continuous red line), $^{82}$Se (continuous purple line), $^{100}$Mo (blue dot-dashed line), $^{130}$Te (green dashed line), and $^{136}$Xe (black dotted line). The anomalous behavior of $^{100}$Mo in a) is due to the opposite sign of the tensor term for $^{100}$Mo.}
\label{fig:limits-half-life-SUSY}
\end{figure}

\begin{figure}[htbp]
%\[
%\text{Fig. 7 new} 
%\]
\centering
\begin{minipage}{12.7pc}
\flushleft{a)}
\includegraphics[width=12.7pc]{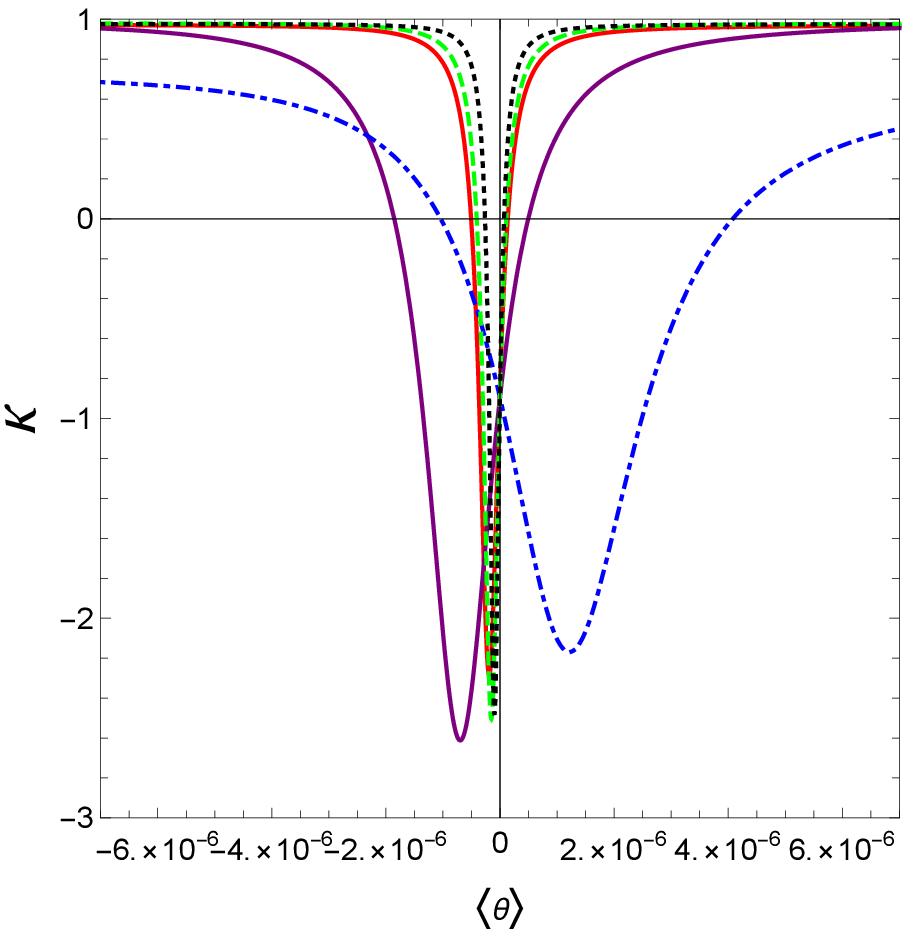} 
\end{minipage}
\hspace{0pc}
\begin{minipage}{12.5pc}
\flushleft{b)}
\includegraphics[width=12.5pc]{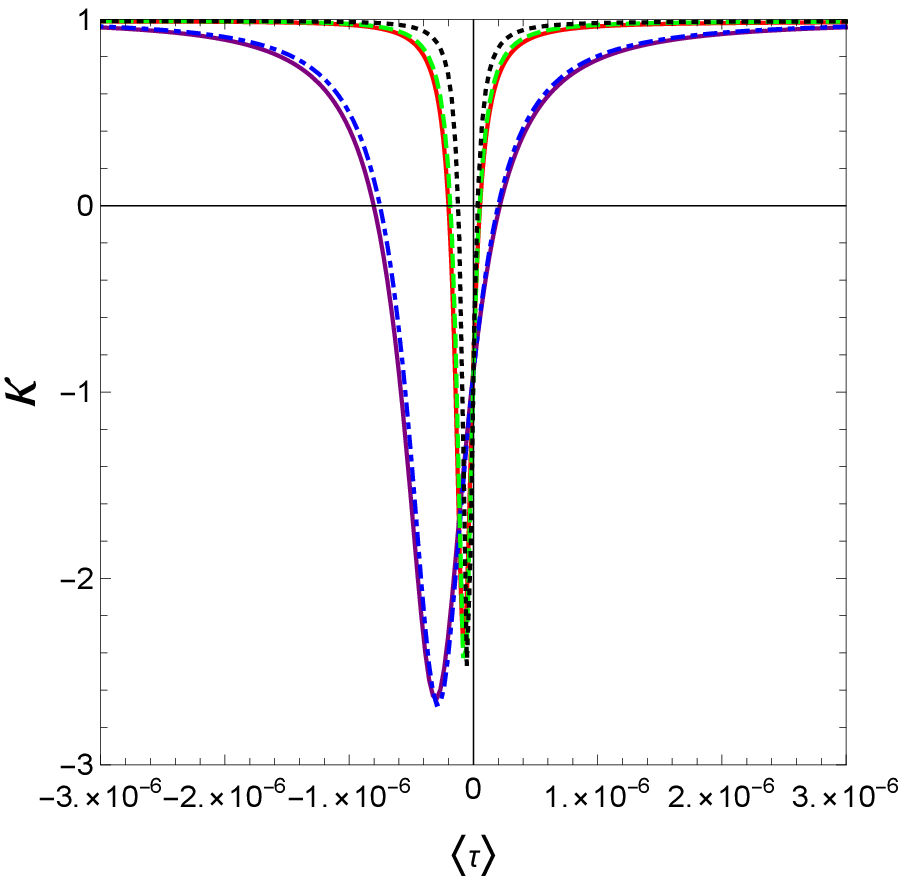} 
\end{minipage}
\hspace{0pc}
\begin{minipage}{12.7pc}
\flushleft{c)}
\includegraphics[width=12.7pc]{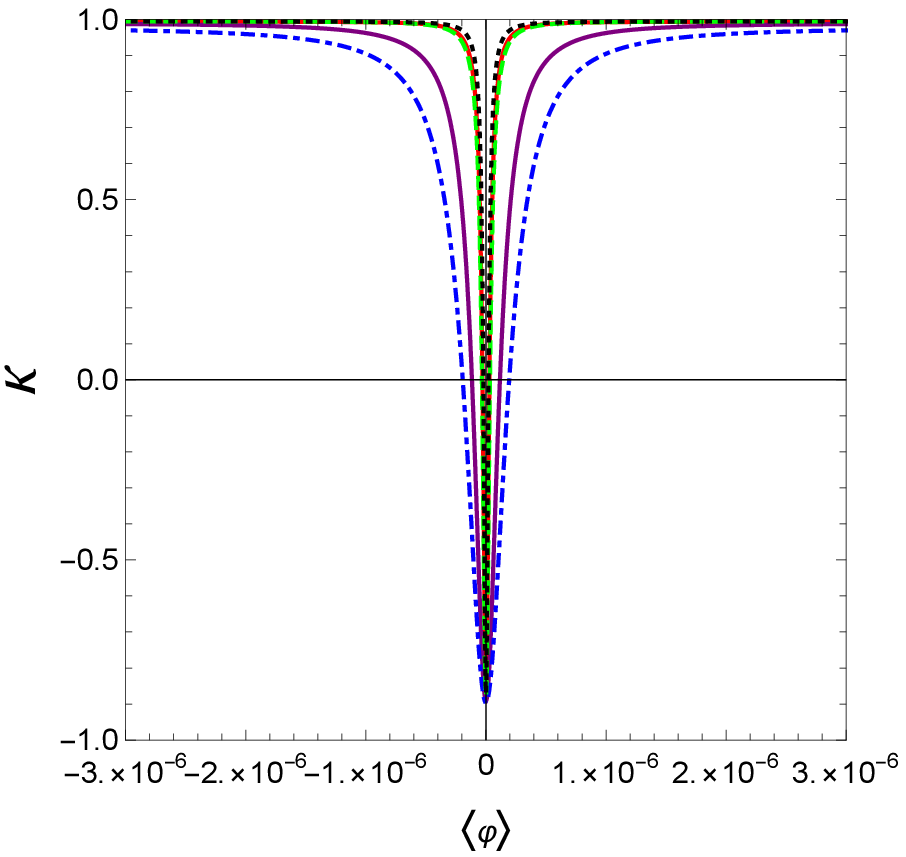} 
\end{minipage}
\caption{Behavior of the coefficient $\mathcal K$, as a function of a) $\langle \theta \rangle$, b)  $\langle \tau \rangle$, and b)  $\langle \varphi \rangle$ in the case of $^{76}$Ge (continuous red line), $^{82}$Se (continuous purple line), $^{100}$Mo (blue dot-dashed line), $^{130}$Te (green dashed line), and $^{136}$Xe (black dotted line) for a fixed value of $\frac{\langle m_\nu \rangle}{m_e}$, given in the second column of Table \ref{tab:limits-param-LR}. The anomalous behavior of $^{100}$Mo in a) is due to the opposite sign of the tensor term for $^{100}$Mo.}
\label{fig:k-VS-SUSY}
\end{figure}

\begin{figure*}[htbp]
%\[
%\text{Fig.8 new} 
%\]
\small
\begin{minipage}{.48\textwidth}
\centering
\includegraphics[width=0.92\textwidth]{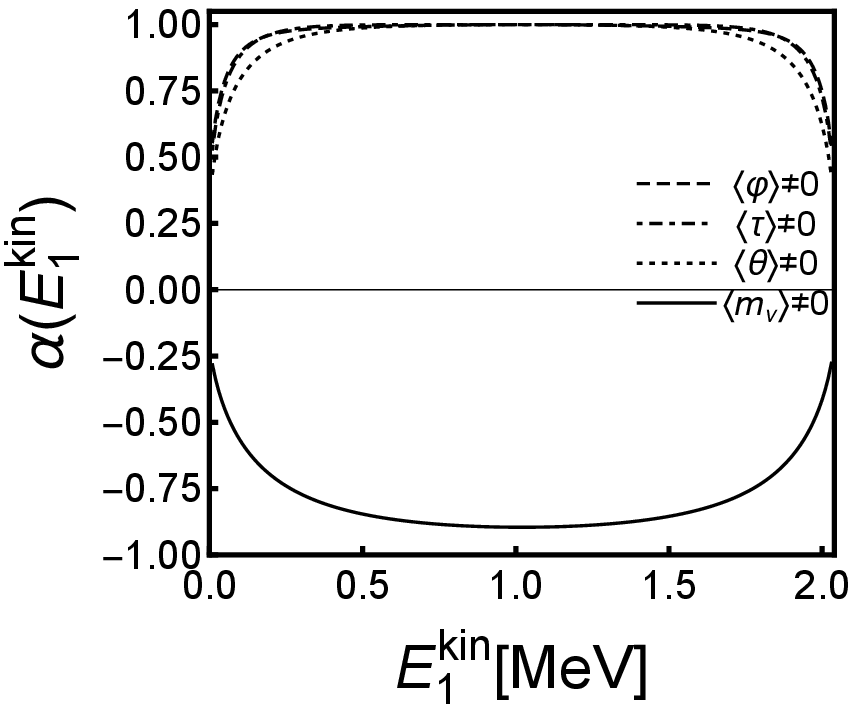}
\end{minipage}
\hfill
\begin{minipage}{.48\textwidth}
\centering
\includegraphics[width=1\textwidth]{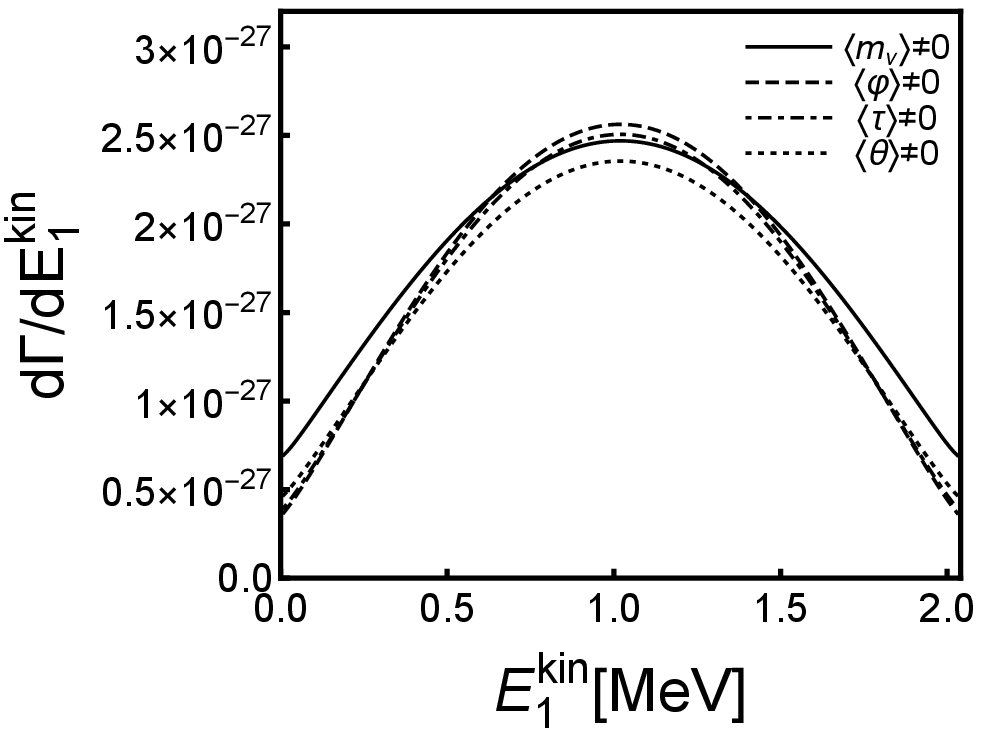}
\end{minipage}
%\begin{minipage}{.48\textwidth}
%\centering
%\includegraphics[width=0.92\textwidth]{fig5-76correps_tomoda.eps}
%\end{minipage}
%\hfill
%\begin{minipage}{.48\textwidth}
%\centering
%\includegraphics[width=1\textwidth]{fig5-76ses_tomoda.eps}
%\end{minipage}
\begin{minipage}{.48\textwidth}
\centering
\includegraphics[width=0.92\textwidth]{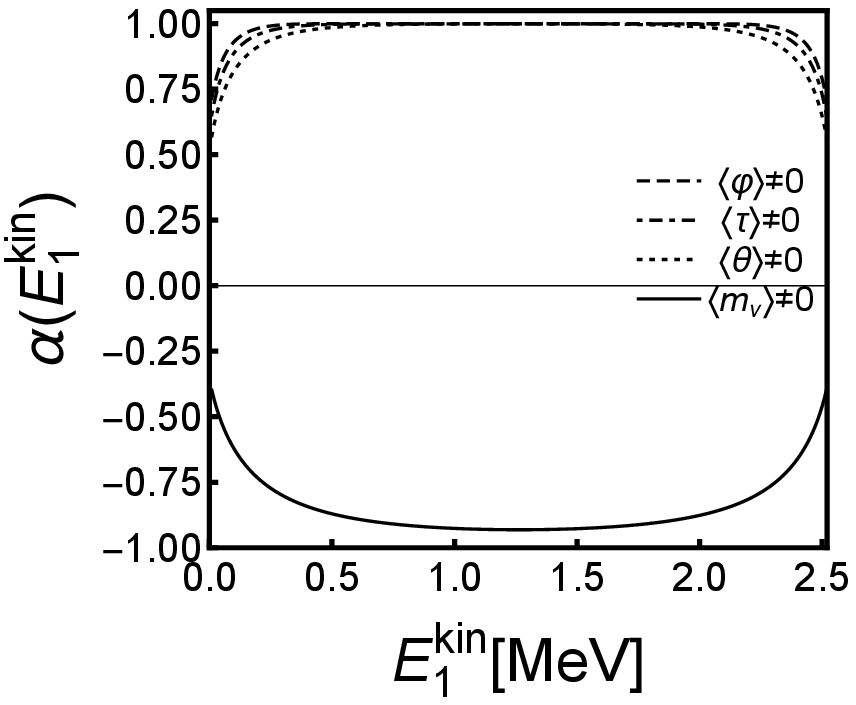}
\end{minipage}
\hfill
\begin{minipage}{.48\textwidth}
\centering
\includegraphics[width=1\textwidth]{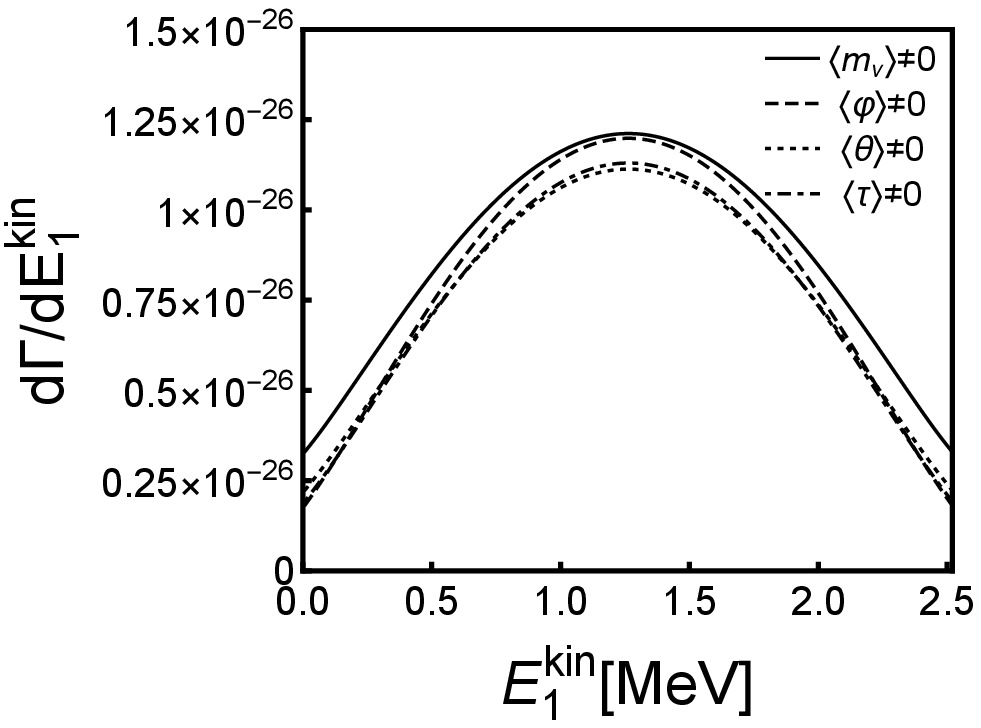}
\end{minipage}
\begin{minipage}{.48\textwidth}
\centering
\includegraphics[width=0.92\textwidth]{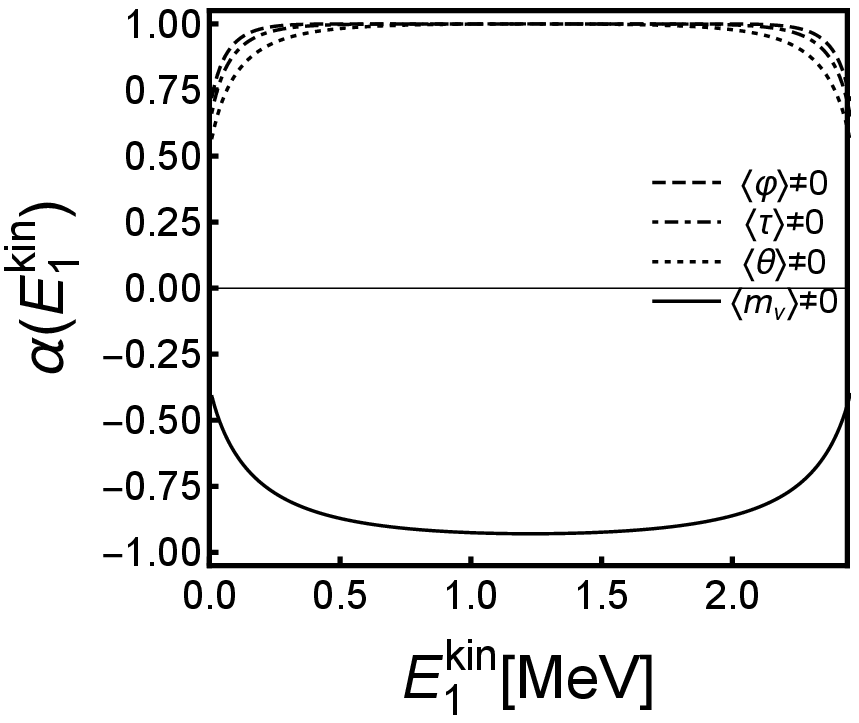}
\end{minipage}
\hfill
\begin{minipage}{.48\textwidth}
\centering
\includegraphics[width=1\textwidth]{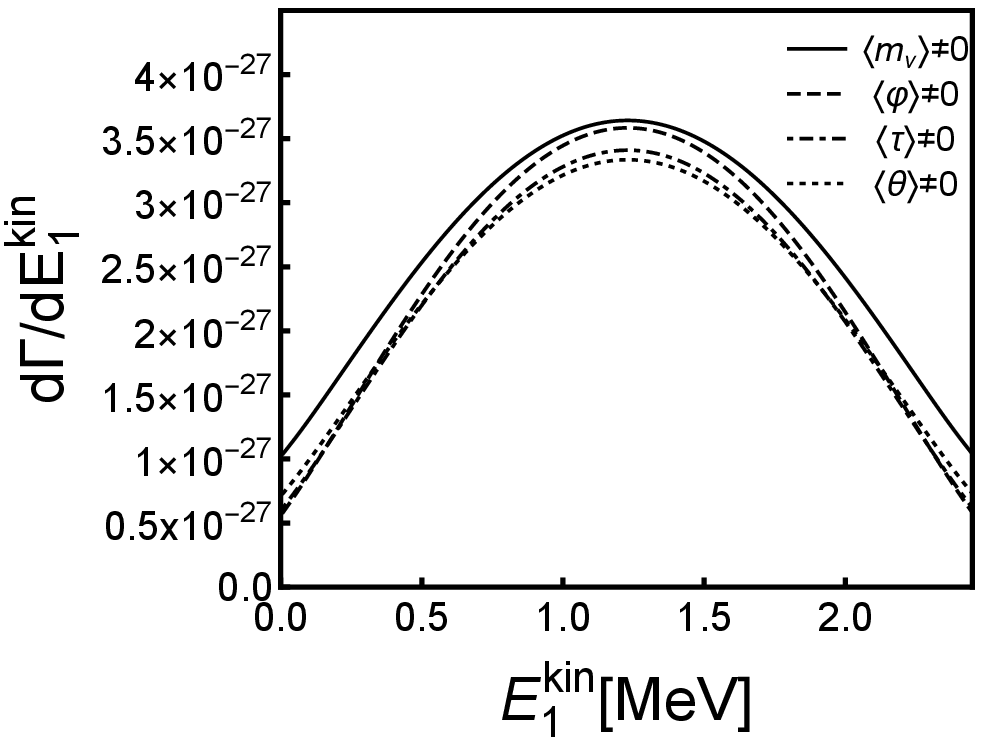}
\end{minipage}
\caption{Right panel: Single electron energy distribution as a function of the kinetic energy $E_1^{\rm kin} = E_1 - m_e$ for the SUSY mechanism. Left panel: Energy dependent angular correlation $\alpha(E_1^{\rm kin})$ between the two electrons as a function of the kinetic energy $E_1^{\rm kin}$ for the SUSY mechanisms. From top to bottom, the first pair of figures corresponds to $^{76}$Ge%with all terms, the second pair to Tomoda's description. }%
, the second pair to $^{130}$Te and, finally, the third pair to $^{136}$Xe.}
\label{fig:e-energy-distribution-SUSY}
\end{figure*}

We set here $\langle \gamma \rangle = 0$, since this parameter always appears in combination with the mass term. The half-life depends on the four parameters $\frac{\langle m_\nu \rangle}{m_e}$, $\langle \theta \rangle$, $\langle \tau \rangle$ and $\langle \varphi \rangle$. One can repeat for this class of models the previous analysis. The results are shown in Figs. \ref{fig:limits-half-life-SUSY}, \ref{fig:k-VS-SUSY}, and \ref{fig:e-energy-distribution-SUSY}.  Note that these figures, especially Fig. \ref{fig:k-VS-SUSY}, depend crucially on the sign of the interference terms between the mass term and the SUSY terms, especially on the signs of $G_{14}^{\prime\prime (0)}$ and $G_{14}^{\prime\prime (1)}$ in Table \ref{tab:Gij-c}. We have adopted in this paper a positive sign. For other signs one would obtain a different behavior of the coefficient $\mathcal K$.

%%%%%%%%%%%%%%%%%%%%%

%For the sake of conciseness, we present here only the upper limits on the parameters considered one at a time. These limits are given in Table \ref{tab:limits-param-SUSY}.
The upper limits on the parameters considered one at a time are given in Table \ref{tab:limits-param-SUSY} and the angular correlation coefficients in Table \ref{tab:angular-coeff-SUSY}.

%%%%%%%%%%%%%%%%%%%%%%%%%%%%%%%%%%%%%%%
\begin{table*}[htbp]
\centering
\begin{tabular}{ccc|cccc}
\hline
\hline
& $T_{1/2}^{\rm exp}$ [y] & & $\frac{\langle m_\nu \rangle}{m_e}$ & $\langle \theta \rangle$ & 
	$\langle \tau \rangle$ & $\langle \varphi \rangle$ \\
\hline
%\text{Table XV } 
$^{76}$Ge &$1.8 \times 10^{26}$ & \cite{Agostini:2013mzu}  	& ~ $1.5\times10^{-7}$ ~ & ~ $2.9\times10^{-7}$ ~ & ~ $1.2\times10^{-7}$ ~ & ~ $3.3\times10^{-8}$ ~   \\
$^{82}$Se &$3.5 \times 10^{24}$ & \cite{Azzolini:2019} 		& ~ $6.5\times10^{-7}$ ~ & ~ $1.0\times10^{-6}$ ~ & ~ $4.4\times10^{-7}$ ~ & ~ $1.2\times10^{-7}$ ~   \\
$^{100}$Mo &$1.1 \times 10^{24}$ & \cite{Arnold:2015wpy} 	& ~ $9.4\times10^{-7}$ ~ & ~ $2.2\times10^{-6}$ ~& ~ $4.1\times10^{-7}$ ~ & ~ $2.0\times10^{-7}$ ~   \\
$^{130}$Te &$3.2 \times 10^{25}$ & \cite{Alfonso:2015wka}	& ~ $2.3\times10^{-7}$ ~ & ~ $2.2\times10^{-7}$ ~ & ~ $1.0\times10^{-7}$ ~ & ~ $3.5\times10^{-8}$ ~   \\
$^{136}$Xe &$1.1 \times 10^{26}$ & \cite{Gando:2012zm} 	& ~ $1.5\times10^{-7}$ ~ & ~ $1.4\times10^{-7}$ ~ & ~ $6.6\times10^{-8}$ ~ & ~ $2.3\times10^{-8}$ ~   \\
\hline
\hline
\end{tabular}
\caption{Upper limits on the absolute values of the parameters of SUSY models for $g_A=1.269$.}
\label{tab:limits-param-SUSY}
\end{table*}
%%%%%%%%%%%%%%%%%%%%%%%%%%%%%%%%%%%%%%%

\begin{table*}[htbp]
\centering
\begin{tabular}{c|cccc}
\hline
\hline
 & ~ ${\mathcal K} \left(\frac{\langle m_\nu \rangle}{m_e}\right)$ ~ & ~ ${\mathcal K} \left(\langle \theta \rangle\right)$ ~ 
 & ~ ${\mathcal K} \left(\langle \tau \rangle\right)$ ~ & ~ ${\mathcal K} \left(\langle \varphi \rangle\right)$ ~ \\
\hline
%\text{Table XIII } 
%\\
$^{76}$Ge & $-0.828$ & $0.971$ & $0.988$ & $0.995$     \\
$^{82}$Se & $-0.891$ & $0.993$ & $0.996$  & $0.991$\\
$^{100}$Mo & $-0.896$ & $0.941$ & $0.995$  & $0.979$\\
$^{130}$Te & $-0.876$ & $0.980$ & $0.991$  & $$0.995\\
$^{136}$Xe & $-0.874$ & $0.976$ & $0.991$  & $0.995$\\
\hline
\hline
\end{tabular}
\caption{Angular correlation coefficients for SUSY models.}
\label{tab:angular-coeff-SUSY}
\end{table*}
%%%%%%%%%%%%%%%%%%%%%%%%%%%%%%%%%%%%%%%

\section{Summary and conclusions}
In this article, we have developed a general formalism for long-range neutrinoless double beta decay.
In the first part, we have derived the expression for the hadronic matrix elements up to order $\frac{q^2}{m_p^2}$ in the nonrelativistic expansion of the currents.
This part is similar to the derivation of the hadronic matrix elements for short-range mechanisms reported in \cite{Graf:2018ozy}, but more complex than it due to the presence here of three types of neutrino potentials instead of one.
In the second part, we have derived the expressions for the leptonic matrix elements. This part is also similar to that of \cite{Graf:2018ozy}, but much more complex than that since there are here 36 matrix elements instead of three.

Numerical values of the nucleon matrix elements (NME) and phase space factors (PSF) have been given in Secs. \ref{Numerical values of the nuclear matrix elements} and \ref{Numerical values of the PSF}, respectively. 
The NMEs have been calculated by making use of the microscopic interacting boson model of the nucleus (IBM-2) \cite{Barea:2009zza,Barea:2013bz,Barea:2015kwa} with $g_A=1.269$. For quenched values of $g_A$ one needs to use an effective value $g_A^{eff}=g_A/1.269$.
The PSF have been calculated by making use of exact Dirac wave functions \cite{Kotila:2012zza}.
Explicit calculations have been done for two classes of models, Left-Right (L-R) and SUSY models, and limits on the coupling constants, $\bar \epsilon$, of these models have been obtained.
Energy-dependent angular correlation between the two emitted electrons and single electron spectra for L-R models have been calculated.

Our conclusions are: (I) Non-standard L-R and SUSY models give rise to lepton number violation even if the neutrino masses are very small. This conclusion is of extreme importance for experiments, since current and projected experimental sensitivities are such that it is quite unlikely that the so-called normal hierarchy (see Fig. \ref{fig:EXP-LIMITS}) can be reached, especially in view of the quenching of the axial-vector coupling constant in nuclei.
Non-standard models provide a way out of the impasse, which is consistent with the limits on the sum of the neutrino masses coming from cosmology. (II) Different models have very different angular correlation between the two emitted electrons (see Fig. \ref{fig:e-energy-distribution} and Fig. \ref{fig:e-energy-distribution-SUSY}).
Therefore, if sufficient number of events will be observed allowing a measurement of the angular correlation, it may be possible to discriminate between different models of lepton number violation \cite{supernemo2010}.

\section*{Acknowledgements}
This work was supported in part by the  Academy of Finland (Grant Nos. 314733 and 320062). We wish to thank Frank Deppisch and Lukas Graf for many useful discussions and for checking some of the formulas.

\begin{appendix}

\section{Notation}
In all calculations of hadronic and leptonic matrix elements, we use the standard notation \cite{Bjorken-Drell} for covariant, $x_\mu$, and contravariant, $x^\mu$, vectors with metric 
\begin{equation}
	\label{eqn:gmunu}
	g_{\mu\nu} = 
	\left( \begin{array}{cccc} 1 & 0 & 0 & 0 \\ 0 & -1 & 0 & 0 \\ 0 & 0 & -1 & 0 \\ 0 & 0 & 0 & -1\end{array}\right)  \mbox{ },
\end{equation}
and
\begin{equation}
	x_\mu = g_{\mu\nu} x^\nu  \mbox{ }.
\end{equation}	
The inner product is
\begin{equation}
	x^2 = x_\mu x^\mu = (x^0)^2 - ({\bf x})^2  \mbox{ }.
\end{equation}	
We also use the standard representation of the $\gamma$-matrices
\begin{equation}
	\begin{array}{c}
	\gamma^\mu \equiv (\gamma^0,\gamma^1,\gamma^2,\gamma^3) \equiv (\gamma^0,{\bm \gamma}) \equiv
	(\gamma^0,\gamma^j)  \mbox{ , } \mbox{ } \gamma^\mu\gamma_\mu = 4  \mbox{ }, \\
	\gamma^0 \equiv \beta \mbox{ , } \mbox{ } {\bm \gamma} = \beta {\bm \alpha} \mbox{ , } \mbox{ } 
	\gamma^j = \beta \alpha^j \mbox{ }, \\
	\gamma_0 = 
	\left(\begin{array}{cc}
		1 & 0 \\
		0 & -1
	\end{array} \right), \mbox{ }
	\gamma^j = 
	\left( \begin{array}{cc}
		0 & \sigma_j \\
		- \sigma_j & 0
	\end{array} \right), \mbox{ }
	\gamma_5 \equiv i \gamma^0 \gamma^1 \gamma^2 \gamma^3 = \Bigg( \begin{array}{cc}
		0 & 1 \\
		1 & 0
	\end{array} \Bigg)  \mbox{ },  \\
	\sigma_{\mu\nu} = \frac{i}{2} [\gamma_\mu,\gamma_\nu] = 
	\frac{i}{2} [\gamma_\mu\gamma_\nu - \gamma_\nu\gamma_\mu]  \mbox{ , } \mbox{ }
	\sigma^{\mu\nu}\sigma_{\mu\nu}  = 12 \mbox{ }.
	\end{array}
\end{equation}
We denote the relativistic wave functions of electrons by Dirac four component spinors $e_{{\bf p},s}$, with momentum $\bf p$ and spin component $s$. We use the standard definition of adjoint and charge conjugate \cite{Bjorken-Drell}
\begin{equation}
	\label{eqn:Bjorken-definitions}  
	\bar e_{{\bf p},s} = e_{{\bf p},s}^\dag \gamma_0 \mbox{ , } \mbox{ } 
	e_{{\bf p},s}^{\rm c} = i \gamma_2 e_{{\bf p},s}^* \mbox{ }.
\end{equation}

\section{Non-relativistic expansion of nuclear currents}
\label{Nuclear currents}
The non-relativistic expansion of the nuclear currents is obtained by means of a Foldy-Wouthuysen (FW) transformation \cite{Bhalla,Rose,Tomoda:1990rs}. In Table \ref{tab:nuclear-currents-FWs}, we give the FW expansion up to the order $k = 2$ in $\frac{1}{m_p}$ of the terms relevant to this article.
%%%%%%%%%%%%%%%%%%%%%%%%%%%%%%%%%%%%%%%
\begin{table*}[htbp]
\scriptsize
%\centering
\begin{tabular}{llllllll}
\hline
\hline
                        & & & $k = 0$ & & $k = 1$ & & $k = 2$ \\
\hline
1 & & $S^{(k)}$       & 1           && 0            && $- \frac{1}{8 m_p^2} {\bf q}^2$ \\
$\gamma_5$ & & $P^{(k)}$      & 0           && $-\frac{1}{2m_p} {\bm \sigma} \cdot {\bf q}$ && 0 \\ 
$\gamma_5 q_0$ & & $P'^{(k)0}$ & 0           && $-\frac{1}{2m_p} q_0 {\bm \sigma} \cdot {\bf q}$ && 0 \\
$\gamma_5 q_i$ & & ${\bf P}'^{(k)}$ & 0           && $-\frac{1}{2m_p} {\bf q}_i {\bm \sigma} \cdot {\bf q}$ && 0 \\ \\
$\gamma_0$ & & $V^{(k)0}$ & 1           && 0            && $- \frac{1}{8 m_p^2} {\bf q} \cdot \left({\bf q} - i {\bm \sigma} \times {\bf Q} \right)$ \\
$\gamma_i$ & & ${\bf V}^{(k)}$ & 0           && $\frac{1}{2m_p}\left({\bf Q} - i {\bm \sigma} \times {\bf q}\right)_i$ && 0 \\ 
$\gamma_0 \gamma_5$ & & $A^{(k)0}$ & 0           && $\frac{1}{2m_p} {\bm \sigma} \cdot {\bf Q}$ && 0 \\
$\gamma_i \gamma_5$ & & ${\bf A}^{(k)}$ & ${\bm \sigma}_i$ && 0 && $-\frac{1}{8 m_p^2} \left[ {\bm \sigma} {\bf Q}^2 - {\bf Q} {\bf Q} \cdot {\bm \sigma} + {\bf q} {\bf q} \cdot {\bm \sigma} - i {\bf q} \times {\bf Q} \right]_i$ \\ \\
$-i \sigma_{0\nu}q^\nu$ & & $W^{(k)0}$ & 0           && $-\frac{1}{2m_p} {\bf q} \cdot \left({\bf q} - i {\bm \sigma} \times {\bf Q}\right)$ && 0 \\
$-i \sigma_{i\nu}q^\nu$ & & ${\bf W}^{(k)}$ & $-i ({\bm \sigma} \times {\bf q})_i$ && $-\frac{1}{2m_p}q^0 \left({\bf q} - i {\bm \sigma} \times {\bf Q}\right)_i$ && $\frac{1}{8 m_p^2} \left[ i {\bf q}^2 {\bm \sigma} \times {\bf q} + i ({\bm \sigma} \cdot {\bf Q}) ({\bf Q} \times {\bf q}) - {\bf q}^2 {\bf Q} + {\bf q} \cdot {\bf Q q} \right]_i$ \\
$\sigma_{00}$ & & $T_1^{(k)00}$ & 0           && 0 && 0 \\
$\sigma_{0i}$ & & $T_1^{(k)0i}$ & 0 && $-\frac{i}{2m_p} \left({\bf q} - i {\bm \sigma} \times {\bf Q}\right)_i$ && 0 \\
$\sigma_{ij}$ & & $T_1^{(k)ij}$ & $i \epsilon_{kij} \sigma_k$ && 0 && $\frac{1}{8 m_p^2} \left[ i {\bf q}_i {\bf Q}_j - {\bf q}_i ({\bm \sigma} \times {\bf q})_j - ({\bm \sigma} \times {\bf Q})_i {\bf Q}_j - i ({\bm \sigma} \times {\bf Q})_i ({\bm \sigma} \times {\bf q})_j \right]$ \\ 
$\sigma_{00} \gamma^5$ & & $T_{1,5}^{(k)00}$ & 0           && 0 && 0 \\
$\sigma_{0i} \gamma^5$ & & $T_{1,5}^{(k)0i}$ & $- i {\bm \sigma}_i$ && 0 && $\frac{i}{16m_p^2} q_0 ({\bf q} + i {\bm \sigma} \times {\bf Q})_i$ \\
$\sigma_{ij} \gamma^5$ & & $T_{1,5}^{(k)ij}$ & 0 && $- \frac{i}{2m_p} {\bm \sigma}_i ({\bf q} - i {\bm \sigma} \times {\bf Q})_j$ && 0 \\ 
\hline
\hline
\end{tabular}
\caption{Nonrelativistic reduction of the nuclear currents by means of Foldy-Wouthuysen transformations \cite{Foldy:1949wa,Bhalla,Rose,Tomoda:1990rs}. In this table, the time and spatial components of the terms are listed according to the order $k$ in $1/m_p$. We also define $q^0 = p^0 - p'^0$, ${\bf q} = {\bf p} - {\bf p}'$, and ${\bf Q} = {\bf p} + {\bf p}'$, where $p$ and $p'$ are the initial and final 4-momenta of the nucleon, and $\bm \sigma$ Pauli spin matrices. 
Some of the results listed here were also given in \cite[Table 2]{Tomoda:1990rs} and \cite{Ali:2007ec}, with notation as in column 2.}
\label{tab:nuclear-currents-FWs}
\end{table*}
%%%%%%%%%%%%%%%%%%%%%%%%%%%%%%%%%%%%%%%

\section{Nucleon current products}
\label{Nucleon current products}
In this appendix, we explicitly show the products of the non-relativistic hadronic currents for each of the five terms $\Pi_i$ ($i = 1, ..., 5$) which contribute to transitions $0^+ \rightarrow 0^+$.
These products are symmetrized in the nucleon indices $a \leftrightarrow b$.
Following the notation of \cite{Graf:2018ozy}, we also provide the signs corresponding to all possible combinations of $L$ and $R$ chiralities and show them in front of each term of the expression.
For $\Pi_{1,2,3}$ three different combinations are possible, shown in the order $RR$, $LL$, $\frac{1}{2}(RL+LR)$. For $\Pi_{4,5}$, since the two factors in the product have different Lorentz structure, all four combinations are possible, shown in the order $RR$, $LL$, $RL$, and $LR$.
%\color{red}
\begin{enumerate}
\item $JJ$ {\bf product.} The product of the $JJ$ scalar/pseudoscalar nuclear currents is given by
\begin{equation}
	\begin{array}{rcl}
\Pi _{1} &\equiv &\frac{1}{2}\left[ J_{a}J_{b}+J_{b}J_{a}\right]  
\\
&=&\left( +++\right) F_{S}^{2}(q^{2})I_{a}I_{b}   \\
&&\left( ++-\right) \frac{F_{P}^{2}}{4m_{p}^{2}}\left( {\bm\sigma_{a}}%
\mathbf{\cdot q}\right) \left( {\bm\sigma_{b}}\mathbf{\cdot q}\right)
+...,
	\end{array}
\end{equation}
where the term proportional to $F_{P}^{2}(q^{2})$ can be recoupled by means
of 
\begin{equation}
\label{eqn:recoupling01}
\left( {\bm\sigma_{a}}\mathbf{\cdot q}\right) \left( {\bm\sigma%
_{b}}\mathbf{\cdot q}\right) =\frac{1}{3}\left( {\bm\sigma_{a}}\mathbf{%
\cdot}{\bm\sigma_{b}}\right) \mathbf{q}^{2}+\frac{1}{3}\mathbf{q}^{2}S_{ab},
\end{equation}%
with $S_{ab}=3\left( {\bm\sigma_{a}}\mathbf{\cdot \hat{q}}\right)
\left( {\bm\sigma_{b}}\mathbf{\cdot \hat{q}}\right) -\left( {\bm%
\sigma_{a}}\mathbf{\cdot} {\bm\sigma_{b}}\right) $. For this product there are
only rank-0 terms.

\item $J^{\mu\nu}J_{\rho\sigma}$ {\bf product.} The rank-0 product $J^{\mu\nu}J_{\mu\nu}$ of currents is given by:
\begin{equation}
	\begin{array}{rcl}
\Pi _{2} &\equiv &\frac{1}{2}\left[ J_{a}^{\mu \nu }J_{\mu \nu
,b}+J_{b}^{\mu \nu }J_{\mu \nu ,a}\right]   \\
&=&\left( ---\right) 2F_{T_{1}}^{2}(q^{2})\left( {\bm\sigma_{a}}\mathbf{%
\cdot} {\bm\sigma_{b}}\right) +...\text{ \ \ .}
	\end{array}
\end{equation}
We do not consider here rank-1 product of currents since we do not use them
in this paper.

\item $J^{\mu} J_{\nu}$ {\bf product.} The rank-0 product $J^{\mu} J_{\mu}$ of currents can be expressed as:
\begin{equation}
	\begin{array}{rcl}
\Pi _{3} &\equiv &\frac{1}{2}\left[ J_{a}^{\mu }J_{\mu ,b}+J_{b}^{\mu
}J_{\mu ,a}\right]   \\
&=&\left( +++\right) F_{V}^{2}(q^{2})I_{a}I_{b}   \\
&&\left( ++-\right) \frac{F_{A}^{2}}{4m_{p}^{2}}\left( {\bm\sigma_{a}}%
\mathbf{\cdot Q}\right) \left( {\bm\sigma_{b}}\mathbf{\cdot Q}\right)  \\
&&\left( --+\right) F_{A}^{2}(q^{2})\left( {\bm\sigma_{a}}\mathbf{\cdot}
{\bm\sigma_{b}}\right)   \\
&&\left( ++-\right) 2\frac{F_{A}(q)F_{P^{\prime }}(q)}{4m_{p}^{2}}\left( 
{\bm\sigma_{a}}\mathbf{\cdot q}\right) \left( {\bm\sigma_{b}}%
\mathbf{\cdot q}\right)   \\
&&\left( +++\right) \frac{\left( F_{V}(q^{2})+F_{W}(q^{2})\right) ^{2}}{%
4m_{p}^{2}}\left( {\bm\sigma_{a}}\mathbf{\times q}\right) \cdot \left( 
{\bm\sigma_{b}}\mathbf{\times q}\right)   \\
&&\left( ---\right) \frac{F_{V}^{2}}{4m_{p}^{2}}\mathbf{Q}^{2}  \\
&&\left( ++-\right) \frac{F_{V}(q^{2})F_{A}(q^{2})}{2m_{p}}\left[
I_{a}\left( {\bm\sigma_{b}}\mathbf{\cdot Q}\right) +\left( {\bm%
\sigma_{a}}\mathbf{\cdot Q}\right) I_{b}\right]   \\
&&\left( --+\right) \frac{F_{P^{\prime }}^{2}(q^{2})}{16m_{p}^{4}}\mathbf{q}%
^{2}\left( {\bm\sigma_{a}}\mathbf{\cdot q}\right) \left( {\bm\sigma_{b}}\mathbf{\cdot q}\right) +...\text{ \ \ ,}
	\end{array}
\end{equation}
where the term proportional to $\left( F_{V}(q^{2})+F_{W}(q^{2})\right) ^{2}$
can be re-coupled as follows%
\begin{equation}
	\label{eqn:recoupling02}
\left( {\bm\sigma }_{a}\mathbf{\times q}\right) \cdot \left( {\bm%
\sigma_{b}}\mathbf{\times q}\right) =-\frac{2}{3}\left( {\bm\sigma_{a} }%
\mathbf{\cdot}{\bm\sigma_{b}}\right) \mathbf{q}^{2}+\frac{1}{3}\mathbf{q}%
^{2}S_{ab},
\end{equation}%
and we have kept only terms up to $\frac{1}{m_{p}^{2}}$ except for the term
proportional to $F_{P^{\prime }}^{2}(q^{2})$ which is enhanced by the pion
pole.

The rank-1 product $J^{\mu }J_{\nu }\left( \nu \neq \mu \right) $ can be
written as:%
\begin{equation}
	\begin{array}{rcl}
\label{eq:rank1}
\bar{\Pi}_{3} &\equiv &J^{i}J^{k}-J^{k}J^{i}   \\
&=&\left( ++-\right) F_{A}^{2}(q^{2})\left( {\bm\sigma_{a} }\mathbf{%
\times}{\bm\sigma_{b }}\right)   \\
&&\left( +--\right) \frac{F_{A}(q^{2})F_{V}(q^{2})}{2m_{p}}\left( {\bm%
\sigma_{a }}I_{b}-{\bm\sigma_{b} }I_{a}\right) \times \mathbf{Q} 
 \\
&&\left( +--\right) (-i)\frac{F_{A}(q^{2})F_{V}(q^{2})}{2m_{p}}\left( 1+%
\frac{F_{W}(q^{2})}{F_{V}(q^{2})}\right) \left( {\bm\sigma_{a }}\mathbf{%
\times }\left( {\bm\sigma_{b }}\mathbf{\times q}\right) -{\bm\sigma%
_{b }}\mathbf{\times }\left( {\bm\sigma_{a }}\mathbf{\times q}\right)
\right)   \\
&&\left( --+\right) (-)\frac{F_{A}(q^{2})F_{P^{\prime }}(q^{2})}{4m_{p}^{2}}%
\left[ \left( {\bm\sigma_{a }}\mathbf{\times q}\right) \left( {\bm%
\sigma_{b }}\mathbf{\cdot q}\right) -\left( {\bm\sigma_{b }}\mathbf{%
\times q}\right) \left( {\bm\sigma_{a}}\mathbf{\cdot q}\right) \right] 
 \\
&&\left( ---\right) (-i)\frac{F_{V}^{2}(q^{2})\left(
F_{V}(q^{2})+F_{W}(q^{2})\right) }{4m_{p}^{2}}\left[ \left( {\bm\sigma%
_{a}}\mathbf{\times q}\right) \times \mathbf{Q-}\left( {\bm\sigma_{b }}%
\mathbf{\times q}\right) \times \mathbf{Q}\right]
	\end{array}
\end{equation}
and%
\begin{equation}
	\begin{array}{rcl}
\bar{\Pi}_{3} &\equiv &J^{0}J^{k}-J^{k}J^{0}   \\
&=&\left( +--\right) F_{V}(q^{2})F_{A}(q^{2})\left( I_{a}{\bm\sigma_{b} }%
-{\bm\sigma_{a} }I_{b}\right) .
	\end{array}
\end{equation}
Note that in constructing the matrix elements the rank-1 terms must be
dotted with $\mathbf{\hat{q}}$ to make them scalars. They contribute only to 
$\mathbf{k}$-terms.

\item $J^{\mu }J_{\rho \nu }$ {\bf product.}. The Lorentz rank-1 product $J^{\mu
}J_{\mu \nu }$ can be relativistically approximated by%
\begin{equation}
	\begin{array}{rcl}
	\Pi_{4\nu,ab} & \equiv &  \frac{1}{2}\sqb{J^{\mu}_a J_{\mu\nu,b} + J^{\mu}_b J_{\mu\nu,a} } \\
	& =& {g_{\nu}}^{0}\Big\{ \foursigns{-}{-}{+}{+} iF_A(q^2)F_{T_1}(q^2) 
	\nba{\vecs{\sigma}_a\cdot\vecs{\sigma}_b}  \\
	& &\foursigns{+}{+}{-}{-} i\frac{F_{P'}(q^2)F_{T_1}(q^2)}{4 m_p^2} \nba{\vecs{\sigma}_a\cdot\vecl{q}} 
	\nba{\vecs{\sigma}_b\cdot\vecl{q}} \Big\} \\
	& + & {g_{\nu}}^{i} \Big\{\foursigns{+}{-}{-}{+} \frac{i}{2} F_V(q^2)F_{T_1}(q^2) \nba{ I_a\sigma_{b i} 
	+ I_b\sigma_{a i} }  \\
	& & \foursigns{-}{-}{-}{-} i\frac{F_V(q^2)\sqb{F_{T_1}(q^2)-2F_{T_2}(q^2)}}{2m_p}q_iI_aI_b \\
	& & \foursigns{-}{-}{-}{-} 2 \frac{F_V(q^2)F_{T_1}(q^2)}{4m_p} 
	 \sqb{ I_a(\vecs{\sigma}_b\times\vecl{Q})_i + I_b(\vecs{\sigma}_a\times\vecl{Q})_i }  \\
	& & \foursigns{-}{-}{-}{-} i \frac{\sqb{F_V(q^2)+F_W(q^2)}F_{T_1}(q^2)}{4m_p} \\
	& & \times \sqb{ 2 q_i(\vecs{\sigma}_a\cdot\vecs{\sigma}_b) - \sigma_{ai}(\vecl{q}\cdot\vecs{\sigma}_b) 
	- \sigma_{bi} (\vecl{q} \cdot\vecs{\sigma}_a) }  \\
	& & \foursigns{-}{-}{+}{+} \frac{F_A(q^2)F_{T_1}(q^2)}{4m_p} \sqb{ (\vecs{\sigma}_a\cdot\vecl{Q})\sigma_{b i} 
	+ (\vecs{\sigma}_b\cdot\vecl{Q})\sigma_{a i} }  \\
	& & \foursigns{+}{+}{-}{-} \frac{F_A(q^2)\sqb{F_{T_1}(q^2)-2F_{T_2}(q^2)}}{4m_p} 
	 \sqb{ (\vecs{\sigma}_a\times\vecl{q})_i I_b + (\vecs{\sigma}_b\times\vecl{q})_i I_a }  \\
	& & \foursigns{-}{-}{+}{+} i \frac{F_A(q^2) F_{T_1}(q^2)}{4m_p} 
	 \sqb{ \sigma_{ai}(\vecl{Q}\cdot\vecs{\sigma}_b) + \sigma_{bi}(\vecl{Q} \cdot\vecs{\sigma}_a) 
	- 2 Q_i(\vecs{\sigma}_a\cdot\vecs{\sigma}_b) }	\\
	& & \foursigns{-}{-}{+}{+} i \frac{F_{P'}(q^2) F_{T_1}(q^2)}{8 m_p^2} q^0 \times \sqb{ (\vecs{\sigma}_a\cdot\vecl{q})
	\sigma_{b i} + (\vecs{\sigma}_b\cdot\vecl{q})\sigma_{a i} } 	 \\
	& & \foursigns{+}{+}{-}{-} i \frac{F_{P'}(q^2) F_{T_1}(q^2)}{16 m_p^3} 
	\left[ \sigma_{ai}(\vecl{q} \cdot\vecl{Q})(\vecl{q} \cdot\vecs{\sigma}_b) \right. \\
	& & + \left. \sigma_{bi}(\vecl{q} \cdot\vecl{Q})(\vecl{q} \cdot\vecs{\sigma}_a) - 2 Q_i(\vecl{q}\cdot\vecs{\sigma}_b)
	(\vecl{q}\cdot\vecs{\sigma}_b) \right]	 \Big\} + \dots 
	\end{array}
\end{equation}
where the zero-th compoment is rank-0 under rotations and the i-th component
is rank-1.

\item $J^{\mu }J$ {\bf product.}. This product is Lorentz rank-1 and we have 
\begin{equation}
	\begin{array}{rcl}
	\Pi^\mu_5 & \equiv & \frac{1}{2}\sqb{J^\mu_a J_b + J^\mu_b J_a} \\
	& = & {g^{\mu}}_{0} \Big\{ \foursigns{+}{+}{+}{+} F_S(q^2)F_V(q^2)I_a I_b \\
	& & \foursigns{+}{+}{-}{-} \frac{F_P(q^2) F_A(q^2)}{8m_p^2} \sqb{ (\vecs{\sigma}_a\cdot\vecl{Q})
	(\vecs{\sigma}_b\cdot\vecl{q}) + (\vecs{\sigma}_a\cdot\vecl{q})(\vecs{\sigma}_b\cdot\vecl{Q}) } \\
	& & \foursigns{-}{-}{+}{+} \frac{F_P(q^2) F_{P'}(q^2)}{8m_p^3} q^0 (\vecs{\sigma}_a\cdot\vecl{q})
	(\vecs{\sigma}_b\cdot\vecl{q}) \Big\}  \\
	& + & {g^{\mu}}_{i}\Big\{ \foursigns{-}{+}{-}{+} \frac{F_S(q^2) F_A(q^2)}{2} \nba{\sigma_a^i I_b + \sigma_b^i I_a} \\
	& & \foursigns{-}{-}{-}{-} \frac{F_S(q^2) F_V(q^2)}{2m_p} Q^i I_a I_b \\
	& & \foursigns{+}{+}{+}{+} i \frac{F_S(q^2) \sqb{F_V(q^2) + F_W(q^2)}}{4m_p}  
	\sqb{ (\vecs{\sigma}_a\times\vecl{q})^i I_b + (\vecs{\sigma}_b\times\vecl{q})^i I_a } \\
	& & \foursigns{-}{-}{+}{+} \frac{F_P(q^2) F_A(q^2)}{4m_p} \sqb{ \sigma_a^i(\vecs{\sigma}_b\cdot\vecl{q}) 
	+ \sigma_b^i(\vecs{\sigma}_a\cdot\vecl{q}) } \\
	& & \foursigns{-}{+}{+}{-} \frac{F_P(q^2) F_V(q^2)}{8m_p^2} Q^i \sqb{ I_a (\vecs{\sigma}_b\cdot\vecl{q}) 
	+ I_b (\vecs{\sigma}_a\cdot\vecl{q}) } \\
	& & \foursigns{+}{-}{+}{-} \frac{F_{S}(q^2) F_{P'}(q^2)}{8m_p^2} q^i \sqb{ (\vecs{\sigma}_a\cdot\vecl{q}) I_b 
	+ (\vecs{\sigma}_b\cdot\vecl{q}) I_a}  \\
	& & \foursigns{+}{+}{-}{-} \frac{F_P(q^2) F_{P'}(q^2)}{8m_p^3} q^i (\vecs{\sigma}_a\cdot\vecl{q}) 
	(\vecs{\sigma}_b\cdot\vecl{q}) \Big\} + \dots
	\end{array}		
\end{equation}
where again the zeroth component is rank-0 under rotations and the i-th
component is rank-1.
\end{enumerate}

\color{black}
The rank-0 products $\Pi_i$ were given in Ref. \cite{Graf:2018ozy}. 
Here, we have used the index $P$ for the intrinsic pseudoscalar form factor, denoted by $PS$ in \cite{Graf:2018ozy}, and the index $P'$ for the induced pseudoscalar form factor, denoted by $P$ in \cite{Graf:2018ozy}.
The matrix elements ${\mathcal M}_k$ of Eqs. (\ref{eq:nme1})-(\ref{eq:nme5}) are obtained from the five products of currents $\Pi_i$ by re-coupling the terms $({\bm \sigma}_a \cdot {\bf q}) ({\bm \sigma}_b \cdot {\bf q})$ and $({\bm \sigma}_a \times {\bf q}) \cdot ({\bm \sigma}_b \times {\bf q})$, as in Eqs. (\ref{eqn:recoupling01}),(\ref{eqn:recoupling02}).
These re-couplings are used here to conform with the standard notation \cite{Ericson}.

\section{Leptonic matrix elements and phase space factors}
\label{Leptonic matrix elements and phase space factors}
The matrix elements that need to be calculated are those of Eqs. (\ref{eqn:LME1})-(\ref{eqn:LME3}).
In the derivation, we use the property of the helicity operators $P_\alpha P_\beta = P_\alpha \delta_{\alpha\beta}$. 
The $\bf k$ terms require the use of $P^{1/2}$-wave electron wave functions, since they are multiplied by the vector $\bf k$ in Eq. (\ref{eqn:rate-Tomoda}).

\begin{itemize}
\item $\Pi_1: jj$
\begin{itemize}
\item[$\rhd$] $m_i$ terms $\left[LL\text{ and }RR\right]$
\begin{equation}
	\label{eqn:D3}
	\bar e_1 \left( \frac{1 \mp \gamma_5}{2} \right) \left( \frac{1 \mp \gamma_5}{2} \right) e_2^{\rm c} 
	= \bar e_1 \left( \frac{1 \mp \gamma_5}{2} \right) e_2^{\rm c}
\end{equation}
\item[$\rhd$] $\omega$ terms $\left[LR + RL\right]$
\begin{equation}
	\bar e_1 \left( \frac{1 - \gamma_5}{2} \right) \gamma^0\left( \frac{1 + \gamma_5}{2} \right) e_2^{\rm c} 
	+ \bar e_1 \left( \frac{1 + \gamma_5}{2} \right) \gamma^0\left( \frac{1 - \gamma_5}{2} \right) e_2^{\rm c} 
	= \bar e_1 \gamma^0 e_2^{\rm c}
\end{equation}
\item[$\rhd$] ${\bf k}$ terms $\left[LR + RL\text{; }S^{1/2}-P^{1/2} \text{ waves}\right]$
\begin{equation}
	\bar e_1 \left( \frac{1 - \gamma_5}{2} \right) \gamma^j\left( \frac{1 + \gamma_5}{2} \right) e_2^{\rm c} 
	+ \bar e_1 \left( \frac{1 + \gamma_5}{2} \right) \gamma^j\left( \frac{1 - \gamma_5}{2} \right) e_2^{\rm c} 
	= \bar e_1 \gamma^j e_2^{\rm c}
\end{equation}
\end{itemize}
\item $\Pi_2: j^{\mu\nu}j_{\mu\nu}$
%\color{red}
\begin{itemize}
\item[$\rhd$] $m_{i}$ terms $\left[ LL\text{ and }RR\right] $%
\begin{equation}
\bar{e}_{1}\sigma ^{\mu \nu }\left( \frac{1\mp \gamma _{5}}{2}\right) \left( 
\frac{1\mp \gamma _{5}}{2}\right) \sigma _{\mu \nu }e_{2}^{c}=12\bar{e}%
_{1}\left( \frac{1\mp \gamma _{5}}{2}\right) e_{2}^{c}
\end{equation}
\item[$\rhd$] $\omega $ terms $\left[ (LR+RL)\right] $%
\begin{equation}
\bar{e}_{1}\sigma ^{\mu \nu }\left( \frac{1-\gamma _{5}}{2}\right) \gamma
^{0}\left( \frac{1+\gamma _{5}}{2}\right) \sigma _{\mu \nu }e_{2}^{c}+\bar{e}%
_{1}\sigma ^{\mu \nu }\left( \frac{1+\gamma _{5}}{2}\right) \gamma
^{0}\left( \frac{1-\gamma _{5}}{2}\right) \sigma _{\mu \nu }e_{2}^{c}=0
\end{equation}
\item[$\rhd$] $\mathbf{k}$ terms $\left[ (LR+RL)\right] $%
\begin{equation}
\bar{e}_{1}\sigma ^{\mu \nu }\left( \frac{1-\gamma _{5}}{2}\right) \gamma
^{j}\left( \frac{1+\gamma _{5}}{2}\right) \sigma _{\mu \nu }e_{2}^{c}+\bar{e}%
_{1}\sigma ^{\mu \nu }\left( \frac{1+\gamma _{5}}{2}\right) \gamma
^{j}\left( \frac{1-\gamma _{5}}{2}\right) \sigma _{\mu \nu }e_{2}^{c}=0
\end{equation}
\end{itemize}

\item $\Pi _{3}:j^{\mu }j_{\mu }$ rank-0
\begin{itemize}
\item[$\rhd$]  $m_{i}$ terms $\left[ LL\text{ and }RR\right] $%
\begin{equation}
\bar{e}_{1}\gamma ^{\mu }\left( \frac{1\mp \gamma _{5}}{2}\right) \left( 
\frac{1\mp \gamma _{5}}{2}\right) \gamma _{\mu }e_{2}^{c}=4\bar{e}_{1}\left( 
\frac{1\pm \gamma _{5}}{2}\right) e_{2}^{c}
\end{equation}

\item[$\rhd$] $\omega $ terms $\left[ LR+RL\right] $%
\begin{equation}
\bar{e}_{1}\gamma ^{\mu }\left( \frac{1-\gamma _{5}}{2}\right) \gamma
^{0}\left( \frac{1+\gamma _{5}}{2}\right) \gamma _{\mu }e_{2}^{c}+\bar{e}%
_{1}\gamma ^{\mu }\left( \frac{1+\gamma _{5}}{2}\right) \gamma ^{0}\left( 
\frac{1-\gamma _{5}}{2}\right) \gamma _{\mu }e_{2}^{c}=-2\bar{e}_{1}\gamma
^{0}e_{2}^{c}
\end{equation}

\item[$\rhd$] $\mathbf{k}$ terms $\left[ LR+RL\right] $%
\begin{equation}
\bar{e}_{1}\gamma ^{\mu }\left( \frac{1-\gamma _{5}}{2}\right) \gamma
^{j}\left( \frac{1+\gamma _{5}}{2}\right) \gamma _{\mu }e_{2}^{c}+\bar{e}%
_{1}\gamma ^{\mu }\left( \frac{1+\gamma _{5}}{2}\right) \gamma ^{j}\left( 
\frac{1-\gamma _{5}}{2}\right) \gamma _{\mu }e_{2}^{c}=-2\bar{e}_{1}\gamma
^{j}e_{2}^{c}
\end{equation}
\end{itemize}
\item $\bar{\Pi}_{3}:j^{0}j^{k}-j^{k}j^{0}$ rank-1
\begin{itemize}
\item[$\rhd$] $\mathbf{k}$ terms $\left[ LR+RL\right] $%
\begin{equation}
\begin{array}{rcl}
&&\left[ \bar{e}_{1}\gamma ^{0}\left( \frac{1-\gamma _{5}}{2}\right) \gamma^{j}\left( \frac{1+\gamma _{5}}{2}\right) \gamma ^{k}e_{2}^{c}-\bar{e}%
_{1}\gamma ^{k}\left( \frac{1-\gamma _{5}}{2}\right) \gamma ^{j}\left( \frac{%
1+\gamma _{5}}{2}\right) \gamma ^{0}e_{2}^{c}\right]   \\
&&+\left[ \bar{e}_{1}\gamma ^{0}\left( \frac{1+\gamma _{5}}{2}\right) \gamma^{j}\left( \frac{1-\gamma _{5}}{2}\right) \gamma ^{k}e_{2}^{c}-\bar{e}%
_{1}\gamma ^{k}\left( \frac{1+\gamma _{5}}{2}\right) \gamma ^{j}\left( \frac{%
1-\gamma _{5}}{2}\right) \gamma ^{0}e_{2}^{c}\right]   \\
&=&2\bar{e}_{1}\gamma ^{0}\gamma ^{j}\gamma ^{k}e_{2}^{c}
\end{array}
\end{equation}
\end{itemize}
\item $\bar{\Pi}_{3}:j^{i}j^{k}-j^{k}j^{i},i\neq k$ rank-1
\begin{itemize}
\item[$\rhd$] $\mathbf{k}$ terms $\left[ LR+RL\right] $%
\begin{equation}
\begin{array}{rcl}
&&\left[ \bar{e}_{1}\gamma ^{i}\left( \frac{1-\gamma _{5}}{2}\right) \gamma
^{j}\left( \frac{1+\gamma _{5}}{2}\right) \gamma ^{k}e_{2}^{c}-\bar{e}%
_{1}\gamma ^{k}\left( \frac{1-\gamma _{5}}{2}\right) \gamma ^{j}\left( \frac{%
1+\gamma _{5}}{2}\right) \gamma ^{i}e_{2}^{c}\right]   \\
&&+\left[ \bar{e}_{1}\gamma ^{i}\left( \frac{1+\gamma _{5}}{2}\right) \gamma
^{j}\left( \frac{1-\gamma _{5}}{2}\right) \gamma ^{k}e_{2}^{c}-\bar{e}%
_{1}\gamma ^{k}\left( \frac{1+\gamma _{5}}{2}\right) \gamma ^{j}\left( \frac{%
1-\gamma _{5}}{2}\right) \gamma ^{i}e_{2}^{c}\right]   \\
&=&-4\bar{e}_{1}\gamma ^{i}e_{2}^{c},\text{ \ \ }j=k\text{ or }i.
\end{array}
\end{equation}%
The matrix elements $RR$ and $LL$ vanish for these terms.
\end{itemize}
\item $\Pi _{4}:j^{\mu }j_{\mu \nu }$
\begin{itemize}
\item[$\rhd$] $m_{i}$ terms $\left[ LL\text{ and }RR\right] $%
\begin{equation}
\bar{e}_{1}\gamma ^{\mu }\left( \frac{1\mp \gamma _{5}}{2}\right) \left( 
\frac{1\mp \gamma _{5}}{2}\right) \sigma _{\mu \nu }e_{2}^{c}=4i\bar{e}%
_{1}\gamma _{\nu }\left( \frac{1\mp \gamma _{5}}{2}\right) e_{2}^{c}
\end{equation}

\item[$\rhd$] $\omega $ terms $\left[ LR+RL\right] $%
\begin{equation}
\bar{e}_{1}\gamma ^{\mu }\left( \frac{1-\gamma _{5}}{2}\right) \gamma
^{0}\left( \frac{1+\gamma _{5}}{2}\right) \sigma _{\mu \nu }e_{2}^{c}+\bar{e}%
_{1}\gamma ^{\mu }\left( \frac{1+\gamma _{5}}{2}\right) \gamma ^{0}\left( 
\frac{1-\gamma _{5}}{2}\right) \sigma _{\mu \nu }e_{2}^{c}=i\bar{e}%
_{1}\gamma _{\nu }\gamma ^{0}e_{2}^{c}
\end{equation}

\item[$\rhd$] $\mathbf{k}$ terms $\left[ LR+RL\right] $%
\begin{equation}
\bar{e}_{1}\gamma ^{\mu }\left( \frac{1-\gamma _{5}}{2}\right) \gamma
^{j}\left( \frac{1+\gamma _{5}}{2}\right) \sigma _{\mu \nu }e_{2}^{c}+\bar{e}%
_{1}\gamma ^{\mu }\left( \frac{1+\gamma _{5}}{2}\right) \gamma ^{j}\left( 
\frac{1-\gamma _{5}}{2}\right) \sigma _{\mu \nu }e_{2}^{c}=i\bar{e}%
_{1}\gamma _{\nu }\gamma ^{j}e_{2}^{c}
\end{equation}
\end{itemize}
\item $\Pi _{5}:j^{\mu }j$
\begin{itemize}
\item[$\rhd$] $m_{i}$ terms $\left[ LL\text{ and }RR\right] $%
\begin{equation}
\bar{e}_{1}\gamma ^{\mu }\left( \frac{1\mp \gamma _{5}}{2}\right) \left( 
\frac{1\mp \gamma _{5}}{2}\right) e_{2}^{c}=\bar{e}_{1}\gamma ^{\mu }\left( 
\frac{1\mp \gamma _{5}}{2}\right) e_{2}^{c}
\end{equation}

\item[$\rhd$] $\omega $ terms $[LR+RL]$%
\begin{equation}
\bar{e}_{1}\gamma ^{\mu }\left( \frac{1-\gamma _{5}}{2}\right) \gamma
^{0}\left( \frac{1+\gamma _{5}}{2}\right) e_{2}^{c}+\bar{e}_{1}\gamma ^{\mu
}\left( \frac{1+\gamma _{5}}{2}\right) \gamma ^{0}\left( \frac{1-\gamma _{5}%
}{2}\right) e_{2}^{c}=\bar{e}_{1}\gamma ^{\mu }\gamma ^{0}e_{2}^{c}
\end{equation}

\item[$\rhd$] $\mathbf{k}$ terms $\left[ LR+RL\right] $%
\begin{equation}
	\label{eqn:D15}
\bar{e}_{1}\gamma ^{\mu }\left( \frac{1-\gamma _{5}}{2}\right) \gamma
^{j}\left( \frac{1+\gamma _{5}}{2}\right) e_{2}^{c}+\bar{e}_{1}\gamma ^{\mu
}\left( \frac{1+\gamma _{5}}{2}\right) \gamma ^{j}\left( \frac{1-\gamma _{5}%
}{2}\right) e_{2}^{c}=\bar{e}_{1}\gamma ^{\mu }\gamma ^{j}e_{2}^{c}
\end{equation}
\end{itemize}
\end{itemize}
\color{black}
In the explicit evaluation of the matrix elements of Eqs. (\ref{eqn:D3})-(\ref{eqn:D15}),
\begin{equation}
	\bar e_1 {\mathcal O}_\alpha \cdots {\mathcal O}_\beta e_2^{\rm c}  \mbox{ },
\end{equation}
and their squares and products summed over spins,
\begin{equation}
	\displaystyle \sum_{s,s'} \left|\bar e_1 {\mathcal O}_\alpha \cdots {\mathcal O}_\beta e_2^{\rm c}\right|^2  \mbox{ },
\end{equation}	
we use the electron wave functions of Eqs. (\ref{eq:radwaves-S}),(\ref{eq:radwaves-P}), the notation of Tomoda \cite{Tomoda:1990rs} with spinors $\chi_s$ normalized to 1, and the standard definitions for the adjoint, $\bar e_{{\bf p},s}$, and charge conjugate, $e_{{\bf p},s}^{\rm c}$, wave functions; see Eq. (\ref{eqn:Bjorken-definitions}) and Ref. \cite{Bjorken-Drell}.

\begin{itemize}
\item $\Pi_1,\Pi_2,\Pi_3$ {\bf currents. Rank-0.}
\begin{itemize}
\item[$\rhd$] $m_{i}$ terms $\left[ LL\text{ and }RR\right] \left[
S_{1}^{1/2}S_{2}^{1/2}\right] $ 
\\The matrix elements of Eq. (\ref{eqn:D3}) can be expressed as
\begin{equation}
\label{eq:PSFder1}
\begin{array}{rcl}
	\bar{e}_1 \left(\frac{1 \mp \gamma_5}{2}\right) e_2^{\rm c} & \simeq & (\bar{e}_{\vecl{p}_1 s})^{S_{1/2}}
	\left(\frac{1 \mp \gamma_5}{2}\right) (e^{\rm c}_{\vecl{p}_2 s'})^{S_{1/2}} \\
	& = & \left( e_{\vecl{p}_1 s}^{S_{1/2}}\right)^\dagger \gamma_0 \left(\frac{1 \mp \gamma_5}{2}\right) i \gamma_2 
	\nba{e_{\vecl{p}_2 s'}^{S_{1/2}}}^*  \\
	& = & \left( g_{-1}^{(-)}(E_1, r) \chi_s^{\dagger} \mbox{ } \mbox{ }
	f_{1}^{(-)}(E_1, r)\chi_s^\dagger(\boldsymbol{\sigma}\cdot\vecl{\op{p}}_1) \right) \gamma_0 
	\left(\frac{1 \mp \gamma_5}{2}\right)	i \gamma_2  \\
	& \times & \begin{pmatrix}	g_{-1}^{(-)} (E_2, r) \chi_{s'} \\ f_{1}^{(-)} 
	(E_2, r)(\boldsymbol{\sigma}\cdot\vecl{\op{p}}_2) \chi_{s'} \end{pmatrix} \mbox{ }.
\end{array}
\end{equation}
After squaring and summing over spins, we obtain
\begin{equation}
	\begin{array}{rcl}
	\displaystyle \sum_{s,s'} \left| \bar{e}_1 \left(\frac{1 \mp \gamma_5}{2}\right) e_2^{\rm c} \right|^2 
	& = & \frac{1}{4} \left[ \tilde f_{11}^{(0)} + \tilde f_{11}^{(1)} (\hat{{\bf p}}_1 \cdot \hat{{\bf p}}_2) \right]
	\end{array}  \mbox{ },
\end{equation}
where $\vecl{\op p}_1\cdot\vecl{\op p}_2 = \cos\theta$ is the scalar product between the asymptotic momentum vectors of the two electrons.
\item[$\rhd$] $\omega $ terms $\left[ LR+RL\right] \left[
S_{1}^{1/2}S_{2}^{1/2}\right] $
\\Similarly we have
\begin{equation}
	\bar e_1 \gamma^0 e_2^{\rm c} \simeq (\bar{e}_{\vecl{p}_1 s})^{S_{1/2}} \gamma^0 
	(e^{\rm c}_{\vecl{p}_2 s'})^{S_{1/2}}
\end{equation}
and	
\begin{equation}
	\begin{array}{rcl}
	\displaystyle \sum_{s,s'} \left| \bar e_1 \gamma^0 e_2^{\rm c} \right|^2 
	& = & \tilde f_{33}^{(0)} + \tilde f_{33}^{(1)} (\hat{{\bf p}}_1 \cdot \hat{{\bf p}}_2)  
	\end{array}  \mbox{ },
\end{equation}
\item[$\rhd$] ${\bf k}$ terms $\left[ LR+RL\right] \left[
S_{1}^{1/2}P_{2}^{1/2}-P_{1}^{1/2}S_{2}^{1/2}\right] $
\begin{equation}
	\bar e_1 \gamma^j e_2^{\rm c} \simeq (\bar{e}_{\vecl{p}_1 s})^{S_{1/2}} \gamma^j 
	(e^{\rm c}_{\vecl{p}_2 s'})^{P_{1/2}} + (\bar{e}_{\vecl{p}_1 s})^{P_{1/2}} \gamma^j 
	(e^{\rm c}_{\vecl{p}_2 s'})^{S_{1/2}}
\end{equation}
and	
\begin{equation}
	\begin{array}{rcl}
	\displaystyle \sum_{s,s'} \left| \bar e_1 \gamma^j e_2^{\rm c} \right|^2 
	& = & \tilde f_{44}^{(0)} + \tilde f_{44}^{(1)} (\hat{{\bf p}}_1 \cdot \hat{{\bf p}}_2)  
	\end{array}  \mbox{ }.
\end{equation}
In a similar fashion, one can calculate the products
\begin{equation}
	\displaystyle \sum_{s,s'} \left( \bar e_1 \left(\frac{1 \pm \gamma_5}{2}\right) e_2^{\rm c} \right)^\dag
	\left( \bar e_1 \gamma^0 e_2^{\rm c} \right) = \frac{1}{2} \tilde f_{13}^{(0)}  \mbox{ },
\end{equation}
\begin{equation}
	\displaystyle \sum_{s,s'} \left( \bar e_1 \left(\frac{1 \pm \gamma_5}{2}\right) e_2^{\rm c} \right)^\dag
	\left( \bar e_1 \gamma^j e_2^{\rm c} \right) = \frac{1}{2} \left[ \tilde f_{14}^{(0)} + \tilde f_{14}^{(1)} 
	(\hat{{\bf p}}_1 \cdot \hat{{\bf p}}_2) \right] \mbox{ },
\end{equation}
and
\begin{equation}
	\displaystyle \sum_{s,s'} \left( \bar e_1 \gamma^0 e_2^{\rm c} \right)^\dag
	\left( \bar e_1 \gamma^j e_2^{\rm c} \right) = \tilde f_{34}^{(0)} + \tilde f_{34}^{(1)} (\hat{{\bf p}}_1 \cdot \hat{{\bf p}}_2) 
	\mbox{ }.
\end{equation}
\end{itemize}

%\color{red}
\item $\bar{\Pi}_{3}$ currents. Rank-1.

$j^{0}j^{k}-j^{k}j^{0}$:
\begin{itemize}
\item[$\rhd$] $\mathbf{k}$ terms $\left[ LR+RL\right] \left[
S_{1}^{1/2}P_{2}^{1/2}+P_{1}^{1/2}S_{2}^{1/2}\right] $%
\begin{equation}
\sum_{s,s^{\prime }}\left\vert \bar{e}_{1}^{S_{1/2}}\gamma ^{0}\gamma
^{j}\gamma ^{k}e_{2}^{cP_{1/2}}+\bar{e}_{1}^{P_{1/2}}\gamma ^{0}\gamma
^{k}\gamma ^{j}e_{2}^{cS_{1/2}}\right\vert ^{2}=\tilde{f}_{55}^{(0)}+\tilde{f%
}_{55}^{(1)}\left( \mathbf{\hat{p}}_{1}\mathbf{\cdot \hat{p}}_{2}\right)
\end{equation}
\end{itemize}
$j^{i}j^{k}-j^{k}j^{i},$ $i\neq k$:
\begin{itemize}
\item[$\rhd$] $\mathbf{k}$ terms $\left[ LR+RL\right] \left[
S_{1}^{1/2}S_{2}^{1/2}\right] $%
\begin{equation}
\sum_{s,s^{\prime }}\left\vert \bar{e}_{1}\gamma ^{k}e_{2}^{c}\right\vert
^{2}=\tilde{f}_{66}^{(0)}+\tilde{f}_{66}^{(1)}\left( \mathbf{\hat{p}}_{1}%
\mathbf{\cdot \hat{p}}_{2}\right)
\end{equation}%
and%
\begin{equation}
\sum_{s,s^{\prime }}\left( \bar{e}_{1}\gamma ^{k}e_{2}^{c}\right) ^{\dag
}\left( \bar{e}_{1}^{S_{1/2}}\gamma ^{0}\gamma ^{k}\gamma
^{j}e_{2}^{cP_{1/2}}+\bar{e}_{1}^{P_{1/2}}\gamma ^{0}\gamma ^{k}\gamma
^{j}e_{2}^{cS_{1/2}}\right) =\tilde{f}_{56}^{(0)}+\tilde{f}_{56}^{(1)}\left( 
\mathbf{\hat{p}}_{1}\mathbf{\cdot \hat{p}}_{2}\right) .
\end{equation}%
\end{itemize}
Also we have for the interference between rank-0 and rank-1%
\begin{equation}
\begin{array}{rcl}
&&\sum_{s,s^{\prime }}\left( \bar{e}_{1}^{S_{1/2}}\left( \frac{1\pm \gamma
_{5}}{2}\right) e_{2}^{cS_{1/2}}\right) ^{\dag }\left( \bar{e}%
_{1}^{S_{1/2}}\gamma ^{0}\gamma ^{k}\gamma ^{j}e_{2}^{cP_{1/2}}+\bar{e}%
_{1}^{P_{1/2}}\gamma ^{0}\gamma ^{k}\gamma ^{j}e_{2}^{cS_{1/2}}\right) 
 \\
&=&\frac{1}{2}\left[ \tilde{f}_{15}^{(0)}+\tilde{f}_{15}^{(1)}\left( \mathbf{%
\hat{p}}_{1}\mathbf{\cdot \hat{p}}_{2}\right) \right]
\end{array}
\end{equation}%
and%
\begin{equation}
\sum_{s,s^{\prime }}\left( \bar{e}_{1}^{S_{1/2}}\left( \frac{1\pm \gamma _{5}%
}{2}\right) e_{2}^{cS_{1/2}}\right) ^{\dag }\left( \bar{e}%
_{1}^{S_{1/2}}\gamma ^{k}e_{2}^{cS_{1/2}}\right) =\frac{1}{2}\left[ \tilde{f}%
_{16}^{(0)}+\tilde{f}_{16}^{(1)}\left( \mathbf{\hat{p}}_{1}\mathbf{\cdot 
\hat{p}}_{2}\right) \right] .
\end{equation}%
\end{itemize}
\color{black}
In these expressions, the quantities $\tilde f_{jk}^{(0)} = \tilde f_{jk}^{(0)}(E_1,E_2)$ and $\tilde f_{jk}^{(1)} = \tilde f_{jk}^{(1)}(E_1,E_2)$ and the associated normalized quantities $f_{jk}^{(0)}$ and $f_{jk}^{(1)}$ are given by
\begin{equation}
	\tilde f_{11}^{(0)} = \left|f^{-1-1}\right|^2+\left|f_{11}\right|^2+\left|f{^{-1}}_1\right|^2+\left|{f_1}^{-1}\right|^2
	\mbox{ , } \mbox{ } f_{11}^{(0)} = \tilde f_{11}^{(0)}
\end{equation}
\begin{equation}
	\label{eqn:f330}
	\tilde f_{33}^{(0)} = \left|{f^{-1}}_1\right|^2+\left|{f_1}^{-1}\right|^2
	\mbox{ , } \mbox{ } f_{33}^{(0)} = \left(\frac{E_1 - E_2}{m_e}\right)^2 \tilde f_{33}^{(0)}
\end{equation}
\begin{equation}
	\tilde f_{44}^{(0)} = \left| f^{-11} + f_{-11} \right|^2 + \left| f^{1-1} + f_{1-1} \right|^2
	\mbox{ , } \mbox{ } f_{44}^{(0)} = \left(\frac{1}{m_e R_A}\right)^2 \tilde f_{44}^{(0)}
\end{equation}
\begin{equation}
	\tilde f_{13}^{(0)} = \left|f{^{-1}}_1\right|^2 - \left|{f_1}^{-1}\right|^2
	\mbox{ , } \mbox{ } f_{13}^{(0)} = \left(\frac{E_1 - E_2}{m_e}\right) \tilde f_{13}^{(0)}
\end{equation}
\begin{equation}
	\tilde f_{14}^{(0)} = f{^{-1}}_1 \left( f^{-1,1} + f_{-1,1} \right) + {f_1}^{-1} \left( f_{1,-1} + f^{1,-1} \right)
	\mbox{ , } \mbox{ } f_{14}^{(0)} = \left(\frac{1}{m_e R_A}\right) \tilde f_{14}^{(0)}
\end{equation}
\begin{equation}
	\tilde f_{34}^{(0)} = f{^{-1}}_1 \left( f^{-1,1} + f_{-1,1} \right) - {f_1}^{-1} \left( f_{1,-1} + f^{1,-1} \right)
	\mbox{ , } \mbox{ } f_{34}^{(0)} = \left(\frac{E_1 - E_2}{m_e^2 R_A}\right) \tilde f_{34}^{(0)}
\end{equation}
and
\begin{equation}
	\tilde f_{11}^{(1)} = -2\left[{f^{-1}}_1 {f_1}^{-1} + f^{-1-1} {f_{11}}\right] 
	\mbox{ , } \mbox{ } f_{11}^{(1)} = \tilde f_{11}^{(1)}
\end{equation}
\begin{equation}
	\tilde f_{33}^{(1)} = 2\left[{f^{-1}}_1 {f_1}^{-1}\right]
	\mbox{ , } \mbox{ } f_{33}^{(1)} = \left(\frac{E_1 - E_2}{m_e}\right)^2 \tilde f_{33}^{(1)}
\end{equation}
\begin{equation}
	\tilde f_{44}^{(1)} = -2 \left[ \left( f^{-11} + f_{-11} \right) \left( f^{1-1} + f_{1-1} \right) \right]
	\mbox{ , } \mbox{ } f_{44}^{(1)} = \left(\frac{1}{m_e R_A}\right)^2 \tilde f_{44}^{(1)}
\end{equation}
\begin{equation}
	\tilde f_{13}^{(1)} = 0 \mbox{ , } \mbox{ } f_{13}^{(1)} = 0
\end{equation}
\begin{equation}
	\tilde f_{14}^{(1)} = - \left[ f{^{-1}}_1 \left( f^{1,-1} + f_{1,-1} \right) + {f_1}^{-1} \left( f_{-1,1} + f^{-1,1} \right) \right]
	\mbox{ , } \mbox{ } f_{14}^{(1)} = \left(\frac{1}{m_e R_A}\right) \tilde f_{14}^{(1)}
\end{equation}
\begin{equation}
	\tilde f_{34}^{(1)} = - \left[ f{^{-1}}_1 \left( f^{1,-1} + f_{1,-1} \right) - {f_1}^{-1} \left( f_{-1,1} + f^{-1,1} \right) \right]
	\mbox{ , } \mbox{ } f_{34}^{(1)} = \left(\frac{E_1 - E_2}{m_e^2 R_A}\right) \tilde f_{34}^{(1)}
\end{equation}
%\color{red}
 while for rank-1 and their inteference with rank-0 we have%
\begin{equation}
\tilde{f}_{55}^{(0)} =\left\vert f^{-1}\text{ }_{-1}+f_{-1}\text{ }%
^{-1}\right\vert ^{2}+\left\vert f_{1}\text{ }^{1}+f^{1}\text{ }%
_{1}\right\vert ^{2},\text{ \ \ }f_{55}^{(0)}=4\left( \frac{1}{m_{e}R_{A}}%
\right) ^{2}\tilde{f}_{55}^{(0)}  
\end{equation}
\begin{equation}
\tilde{f}_{66}^{(0)} =\left\vert f^{-1,-1}\right\vert ^{2}+\left\vert
f_{1,1}\right\vert ^{2},\text{ \ \ }f_{66}^{(0)}=16\left( \frac{1}{m_{e}R_{A}%
}\right) ^{2}\tilde{f}_{66}^{(0)}  
\end{equation}
\begin{equation}
\tilde{f}_{56}^{(0)} =-f_{1,1}\left( f_{1}\text{ }^{1}+f^{1}\text{ }%
_{1}\right) +f^{-1,-1}\left( f^{-1}\text{ }_{-1}+f_{-1}\text{ }^{-1}\right) ,%
\text{ \ \ }f_{56}^{(0)}=8\left( \frac{1}{m_{e}R_{A}}\right) ^{2}\tilde{f}%
_{56}^{(0)}   
\end{equation}
\begin{equation}
\tilde{f}_{15}^{(0)} =-f^{-1,-1}\left( f^{-1}\text{ }_{-1}+f_{-1}\text{ }%
^{-1}\right) -f_{1,1}\left( f_{1}\text{ }^{1}+f^{1}\text{ }_{1}\right) ,%
\text{ \ \ }f_{15}^{(0)}=2\left( \frac{1}{m_{e}R_{A}}\right) \tilde{f}%
_{15}^{(0)}   
\end{equation}
\begin{equation}
\tilde{f}_{16}^{(0)} =-\left\vert f^{-1,-1}\right\vert ^{2}+\left\vert
f_{1,1}\right\vert ^{2},\text{ \ \ }f_{16}^{(0)}=4\left( \frac{1}{m_{e}R_{A}}%
\right) \tilde{f}_{16}^{(0)}
\end{equation}%
and%
\begin{equation}
\tilde{f}_{55}^{(1)} =-2\left( f^{-1}\text{ }_{-1}+f_{-1}\text{ }%
^{-1}\right) \left( f_{1}\text{ }^{1}+f^{1}\text{ }_{1}\right) ,\text{ \ \ }%
f_{55}^{(1)}=4\left( \frac{1}{m_{e}R_{A}}\right) ^{2}\tilde{f}_{55}^{(1)} 
\end{equation}
\begin{equation}
\tilde{f}_{66}^{(1)} =2\left( f^{-1,-1}f_{1,1}\right) ,\text{ \ \ }%
f_{66}^{(1)}=16\left( \frac{1}{m_{e}R_{A}}\right) ^{2}\tilde{f}_{66}^{(1)} 
\end{equation}
\begin{equation}
\tilde{f}_{56}^{(1)} =f_{1,1}\left( f^{-1}\text{ }_{-1}+f_{-1}\text{ }%
^{-1}\right) -f^{-1,-1}\left( f_{1}\text{ }^{1}+f^{1}\text{ }_{1}\right) ,%
\text{ \ \ }f_{56}^{(1)}=8\left( \frac{1}{m_{e}R_{A}}\right) ^{2}\tilde{f}%
_{56}^{(1)}  
\end{equation}
\begin{equation}
\tilde{f}_{15}^{(1)} =f_{1,1}\left( f^{-1}\text{ }_{-1}+f_{-1}\text{ }%
^{-1}\right) +f^{-1,-1}\left( f_{1}\text{ }^{1}+f^{1}\text{ }_{1}\right) ,%
\text{ \ \ }f_{15}^{(1)}=2\left( \frac{1}{m_{e}R_{A}}\right) \tilde{f}%
_{15}^{(1)} 
\end{equation}
\begin{equation}
\tilde{f}_{16}^{(1)} =0,\text{ \ \ }f_{16}^{(1)}=0
\end{equation}%
In these equations, the phase space factors $\tilde{f}_{jk}$ have been
converted to the normalized phase space factors $f_{jk}$ of Tomoda \cite{Tomoda:1990rs} to
make them consistent with the neutrino potentials of Eqs. (\ref{eqn:vpot1})-(\ref{eqn:vpot3}) and with
the definition of the nuclear matrix elements by multiplying them by
dimensionless quantities. Note that $f_{66},f_{56},f_{16}$ contain extra
factors of $\left( 1/m_{e}R_{A}\right) $ compared with Tomoda \cite{Tomoda:1990rs} due to
our definition of the rank-1 recoil matrix elements.

%The phase space factors $\tilde f_{ij}$ have been converted to the normalized phase space factors $f_{ij}$ of Tomoda \cite{Tomoda:1990rs} to make them consistent with the neutrino potentials of Eqs. (\ref{eqn:vpot1}-\ref{eqn:vpot3}) by multiplying them by dimensionless quantities.
\color{black}
All the above listed phase-space factors are given in terms of the underlying energy-dependent wave functions of the two electrons \cite{Tomoda:1990rs,Kotila:2012zza,Graf:2018ozy} evaluated at the nuclear radius $R_A$:
\begin{equation}
	\begin{array}{c}
	f^{\kappa \kappa'}   = g_{\kappa}^{(-)}(E_1,R_A)g_{\kappa'}^{(-)}(E_2,R_A) \mbox{ }, \\
	f_{\kappa\kappa'}   = f_\kappa^{(-)}(E_1,R_A)f_{\kappa'}^{(-)}(E_2,R_A) \mbox{ },    \\
	{f^\kappa}_{\kappa'} = g_\kappa^{(-)}(E_1,R_A)f_{\kappa'}^{(-)}(E_2,R_A) \mbox{ },    \\
	{f_\kappa}^{\kappa'} = f_\kappa^{(-)}(E_1,R_A)g_{\kappa'}^{(-)}(E_2,R_A) \mbox{ }.
	\end{array}
\end{equation}
%\color{red}
\begin{itemize}
\item $\Pi_4,\Pi_5$ {\bf currents.}
For $0^+ \rightarrow 0^+$ transitions, we need only the matrix elements of the $\mu = 0$ component of the current.
The calculation proceeds as before, with results, for $\mu = 0$,
\begin{itemize}
\item[$\rhd$]  $m_{i}$ terms $\left[ LL\text{ and }RR\right] \left[
S_{1}^{1/2}S_{2}^{1/2}\right] $%
\begin{equation}
	\label{eqn:D40}
\sum_{s,s^{\prime }}\left\vert \bar{e}_{1}\gamma ^{0}\left( \frac{1\mp
\gamma _{5}}{2}\right) e_{2}^{c}\right\vert ^{2}=\frac{1}{4}\left[ \tilde{f}%
_{11}^{\prime (0)}+\tilde{f}_{11}^{\prime (1)}\left( \mathbf{\hat{p}}_{1}%
\mathbf{\cdot \hat{p}}_{2}\right) \right] ,
\end{equation}
\item[$\rhd$] $\omega $ terms $\left[ LR+RL\right] \left[
S_{1}^{1/2}S_{2}^{1/2}\right] $%
\begin{equation}
\sum_{s,s^{\prime }}\left\vert \bar{e}_{1}\left( \gamma ^{0}\right)
^{2}e_{2}^{c}\right\vert ^{2}=\tilde{f}_{33}^{\prime (0)}+\tilde{f}%
_{33}^{\prime (1)}\left( \mathbf{\hat{p}}_{1}\mathbf{\cdot \hat{p}}%
_{2}\right) ,
\end{equation}
\item[$\rhd$] $\mathbf{k}$ terms $\left[ LR+RL\right] \left[
S_{1}^{1/2}P_{2}^{1/2}-P_{1}^{1/2}S_{2}^{1/2}\right] $%
\begin{equation}
\sum_{s,s^{\prime }}\left\vert \bar{e}_{1}\gamma ^{0}\gamma
^{j}e_{2}^{c}\right\vert ^{2}=\tilde{f}_{44}^{\prime (0)}+\tilde{f}%
_{44}^{\prime (1)}\left( \mathbf{\hat{p}}_{1}\mathbf{\cdot \hat{p}}%
_{2}\right) ,
\end{equation}%
and products
\begin{equation}
\sum_{s,s^{\prime }}\left( \bar{e}_{1}\gamma ^{0}\left( \frac{1\mp \gamma
_{5}}{2}\right) e_{2}^{c}\right) ^{\dag }\left( \bar{e}_{1}\left( \gamma
^{0}\right) ^{2}e_{2}^{c}\right) =\frac{1}{2}\tilde{f}_{13}^{\prime (0)},
\end{equation}%
\begin{equation}
\sum_{s,s^{\prime }}\left( \bar{e}_{1}\gamma ^{0}\left( \frac{1\mp \gamma
_{5}}{2}\right) e_{2}^{c}\right) ^{\dag }\left( \bar{e}_{1}\gamma ^{0}\gamma
^{j}e_{2}^{c}\right) =\frac{1}{2}\left[ \tilde{f}_{14}^{\prime (0)}+\tilde{f}%
_{14}^{\prime (1)}\left( \mathbf{\hat{p}}_{1}\mathbf{\cdot \hat{p}}%
_{2}\right) \right]
\end{equation}%
and
\begin{equation}
	\label{eqn:D45}
\sum_{s,s^{\prime }}\left( \bar{e}_{1}\left( \gamma ^{0}\right)
^{2}e_{2}^{c}\right) ^{\dag }\left( \bar{e}_{1}\gamma ^{0}\gamma
^{j}e_{2}^{c}\right) =\tilde{f}_{34}^{\prime (0)}+\tilde{f}_{34}^{\prime
(1)}\left( \mathbf{\hat{p}}_{1}\mathbf{\cdot \hat{p}}_{2}\right) .
\end{equation}
\end{itemize}
\end{itemize}
\color{black}
In these expressions, the quantities $\tilde f_{jk}'^{(0)} = \tilde f_{jk}'^{(0)}(E_1,E_2)$ and $\tilde f_{jk}'^{(1)} = \tilde f_{jk}'^{(1)}(E_1,E_2)$ and the associated normalized quantities $f_{jk}'^{(0)}$ and $f_{jk}'^{(1)}$ are given by
\begin{equation}
	\tilde f_{11}'^{(0)} = \tilde f_{11}^{(0)} = \left|f^{-1-1}\right|^2+\left|f_{11}\right|^2
	+\left|f{^{-1}}_1\right|^2+\left|{f_1}^{-1}\right|^2
	\mbox{ , } \mbox{ } f_{11}'^{(0)} = \tilde f_{11}'^{(0)}
\end{equation}
\begin{equation}
	\tilde f_{33}'^{(0)} = \tilde f_{33}^{(0)} = \left|{f^{-1}}_1\right|^2+\left|{f_1}^{-1}\right|^2
	\mbox{ , } \mbox{ } f_{33}'^{(0)} = \left(\frac{E_1 - E_2}{m_e}\right)^2 \tilde f_{33}'^{(0)}
\end{equation}
\begin{equation}
	\tilde f_{44}'^{(0)} = \left| f^{-11} - f_{-11} \right|^2 + \left| f^{1-1} - f_{1-1} \right|^2
	\mbox{ , } \mbox{ } f_{44}'^{(0)} = \left(\frac{1}{m_e R_A}\right)^2 \tilde f_{44}'^{(0)}
\end{equation}
\begin{equation}
	\tilde f_{13}'^{(0)} = \tilde f_{13}^{(0)} = \left|f{^{-1}}_1\right|^2 - \left|{f_1}^{-1}\right|^2
	\mbox{ , } \mbox{ } f_{13}'^{(0)} = \left(\frac{E_1 - E_2}{m_e}\right) \tilde f_{13}'^{(0)}
\end{equation}
\begin{equation}
	\tilde f_{14}'^{(0)} = -\left[f{^{-1}}_1 \left( f_{-1,1} - f^{-1,1} \right) + {f_1}^{-1} \left( f^{1,-1} - f_{1,-1} \right)\right]
	\mbox{ , } \mbox{ } f_{14}'^{(0)} = \left(\frac{1}{m_e R_A}\right) \tilde f_{14}'^{(0)}
\end{equation}
\begin{equation}
	\tilde f_{34}'^{(0)} = f{^{-1}}_1 \left( f^{-1,1} - f_{-1,1} \right) - {f_1}^{-1} \left( f^{1,-1} - f_{1,-1} \right)
	\mbox{ , } \mbox{ } f_{34}'^{(0)} = \left(\frac{E_1 - E_2}{m_e^2 R_A}\right) \tilde f_{34}'^{(0)}
\end{equation}
and
\begin{equation}
	\tilde f_{11}'^{(1)} = -\tilde f_{11}^{(1)} = 2\left[{f^{-1}}_1 {f_1}^{-1} + f^{-1-1} {f_{11}}\right] 
	\mbox{ , } \mbox{ } f_{11}'^{(1)} = \tilde f_{11}'^{(1)}
\end{equation}
\begin{equation}
	\tilde f_{33}'^{(1)} = -\tilde f_{33}^{(1)} = -2\left[{f^{-1}}_1 {f_1}^{-1}\right]
	\mbox{ , } \mbox{ } f_{33}'^{(1)} = \left(\frac{E_1 - E_2}{m_e}\right)^2 \tilde f_{33}'^{(1)}
\end{equation}
\begin{equation}
	\tilde f_{44}'^{(1)} = - 2 \left[ \left( f^{-11} - f_{-11} \right) \left( f^{1-1} - f_{1-1} \right) \right]
	\mbox{ , } \mbox{ } f_{44}'^{(1)} = \left(\frac{1}{m_e R_A}\right)^2 \tilde f_{44}'^{(1)}
\end{equation}
\begin{equation}
	\tilde f_{13}'^{(1)} = \tilde f_{13}^{(1)} = 0
\end{equation}
\begin{equation}
	\tilde f_{14}'^{(1)} =  f{^{-1}}_1 \left( f_{1,-1} - f^{1,-1} \right) + {f_1}^{-1} \left( f^{-1,1} - f_{-1,1} \right) 
	\mbox{ , } \mbox{ } f_{14}'^{(1)} = \left(\frac{1}{m_e R_A}\right) \tilde f_{14}'^{(1)}
\end{equation}
\begin{equation}
	\tilde f_{34}'^{(1)} = - \left[ f{^{-1}}_1 \left( f^{1,-1} - f_{1,-1} \right) - {f_1}^{-1} \left( f^{-1,1} - f_{-1,1} \right) \right]
	\mbox{ , } \mbox{ } f_{34}'^{(1)} = \left(\frac{E_1 - E_2}{m_e^2 R_A}\right) \tilde f_{34}'^{(1)}
\end{equation}

\begin{itemize}
\item {\bf Interference} $\Pi_{1,2,3}-\Pi_{4,5}${\bf{.}}
The non-zero values are only for $\mu = 0$.
%\color{red}
\begin{itemize}
\item[$\rhd$] $m_{i}$ terms $\left[ LL\text{ and }RR\right] \left[
S_{1}^{1/2}S_{2}^{1/2}\right] $%
\begin{equation}
\sum_{s,s^{\prime }}\left( \bar{e}_{1}\gamma ^{0}\left( \frac{1\mp \gamma
_{5}}{2}\right) e_{2}^{c}\right) ^{\dag }\left( \bar{e}_{1}\left( \frac{1\mp
\gamma _{5}}{2}\right) e_{2}^{c}\right) =\frac{1}{4}\tilde{f}_{11}^{\prime
\prime (0)},
\end{equation}
\item[$\rhd$] $\omega $ terms $\left[ LR+RL\right] \left[
S_{1}^{1/2}S_{2}^{1/2}\right] $%
\begin{equation}
\sum_{s,s^{\prime }}\left( \bar{e}_{1}\left( \gamma ^{0}\right)
^{2}e_{2}^{c}\right) ^{\dag }\left( \bar{e}_{1}\gamma ^{0}e_{2}^{c}\right) =%
\tilde{f}_{33}^{\prime \prime },
\end{equation}
\item[$\rhd$] $\mathbf{k}$ terms $\left[ LR+RL\right] \left[
S_{1}^{1/2}P_{2}^{1/2}-P_{1}^{1/2}S_{2}^{1/2}\right] $%
\begin{equation}
\sum_{s,s^{\prime }}\left( \bar{e}_{1}\gamma ^{0}\gamma ^{j}e_{2}^{c}\right)
^{\dag }\left( \bar{e}_{1}\gamma ^{j}e_{2}^{c}\right) =\tilde{f}%
_{44}^{\prime \prime (0)}+\tilde{f}_{44}^{\prime \prime (1)}\left( \mathbf{%
\hat{p}}_{1}\mathbf{\cdot \hat{p}}_{2}\right) ,
\end{equation}%
\end{itemize}
\color{black}
and products
\begin{equation}
	\displaystyle \sum_{s,s'} \left( \bar e_1 \left(\frac{1 \mp \gamma_5}{2}\right) e_2^{\rm c} \right)^\dag
	\left( \bar e_1 (\gamma^0)^2 e_2^{\rm c} \right) = \frac{1}{2} \left[ \tilde f_{13}''^{(0)} 
	+ \tilde f_{13}''^{(1)} (\hat{{\bf p}}_1 \cdot \hat{{\bf p}}_2) \right]  \mbox{ },
\end{equation}
\begin{equation}
	\displaystyle \sum_{s,s'} \left( \bar e_1 \left(\frac{1 \mp \gamma_5}{2}\right) e_2^{\rm c} \right)^\dag
	\left( \bar e_1 \gamma^0 \gamma^j e_2^{\rm c} \right) = - \frac{1}{2} \left[ \tilde f_{14}''^{(0)} 
	+ \tilde f_{14}''^{(1)} (\hat{{\bf p}}_1 \cdot \hat{{\bf p}}_2) \right]  \mbox{ },
\end{equation}
\begin{equation}
\begin{array}{rcl}
	\displaystyle \sum_{s,s'} \left( \bar e_1 \gamma^0 e_2^{\rm c} \right)^\dag 
	\left( \bar e_1 \gamma^0 \gamma^j e_2^{\rm c} \right)	
	& = & \tilde f_{34}''^{(0)} + f_{34}''^{(1)} (\hat{{\bf p}}_1 \cdot \hat{{\bf p}}_2)
	\end{array}  \mbox{ },
\end{equation}
\begin{equation}
\begin{array}{rcl}
	\displaystyle \sum_{s,s'} \left( \bar e_1 \gamma^j e_2^{\rm c} \right)^\dag 
	\left( \bar e_1 (\gamma^0)^2 e_2^{\rm c} \right)	
	& = & \tilde f_{43}''^{(0)} + f_{43}''^{(1)} (\hat{{\bf p}}_1 \cdot \hat{{\bf p}}_2)
	\end{array}  \mbox{ },
\end{equation}
\end{itemize}
In these expressions
\begin{equation}
	\tilde f_{11}''^{(0)} = \left|f^{-1-1}\right|^2 + \left|f{^{-1}}_1\right|^2 - \left|f_{11}\right|^2
	- \left|{f_1}^{-1}\right|^2
	\mbox{ , } \mbox{ } f_{11}''^{(0)} = \tilde f_{11}''^{(0)}
\end{equation}
\begin{equation}
	\tilde f_{33}''^{(0)} = \left|{f^{-1}}_1\right|^2 - \left|{f_1}^{-1}\right|^2
	\mbox{ , } \mbox{ } f_{33}''^{(0)} = \left(\frac{E_1 - E_2}{m_e}\right)^2 \tilde f_{33}''^{(0)}
\end{equation}
\begin{equation}
	\tilde f_{44}''^{(0)} = \left| f^{-11}\right|^2 + \left| f^{1-1} \right|^2 - \left| f_{1-1} \right|^2
	 - \left| f_{-11} \right|^2
	\mbox{ , } \mbox{ } f_{44}''^{(0)} = \left(\frac{1}{m_e R_A}\right)^2 \tilde f_{44}''^{(0)}
\end{equation}
\begin{equation}
	\tilde f_{13}''^{(0)} = \left|f{^{-1}}_1\right|^2 + \left|{f_1}^{-1}\right|^2
	\mbox{ , } \mbox{ } f_{13}''^{(0)} = \left(\frac{E_1 - E_2}{m_e}\right) \tilde f_{13}''^{(0)}
\end{equation}
\begin{equation}
	\tilde f_{14}''^{(0)} =-\left[ f{^{-1}}_1 \left( - f^{-1,1} + f_{-1,1} \right) + {f_1}^{-1} \left( - f^{1,-1} + f_{1,-1} \right)\right]
	\mbox{ , } \mbox{ } f_{14}''^{(0)} = \left(\frac{1}{m_e R_A}\right) \tilde f_{14}''^{(0)}
\end{equation}
\begin{equation}
	\tilde f_{34}''^{(0)} = f{^{-1}}_1 \left( f^{-1,1} - f_{-1,1} \right) - {f_1}^{-1} \left( f^{1,-1} - f_{1,-1} \right)
	\mbox{ , } \mbox{ } f_{34}''^{(0)} = \left(\frac{E_1 - E_2}{m_e^2 R_A}\right) \tilde f_{34}''^{(0)}
\end{equation}
\begin{equation}
	\tilde f_{43}''^{(0)} = -f{^{-1}}_1 \left( f^{-1,1} + f_{-1,1} \right) + {f_1}^{-1} \left( f^{1,-1} + f_{1,-1} \right)
	\mbox{ , } \mbox{ } f_{43}''^{(0)} = \left(\frac{E_1 - E_2}{m_e^2 R_A}\right) \tilde f_{43}''^{(0)}
\end{equation}
and
\begin{equation}
	\tilde f_{11}''^{(1)} = 0 \mbox{ , } \mbox{ } f_{11}''^{(1)} = 0
\end{equation}
\begin{equation}
	\tilde f_{33}''^{(1)} = 0 \mbox{ , } \mbox{ } f_{33}''^{(1)} = 0
\end{equation}
\begin{equation}
	\tilde f_{44}''^{(1)} = -2 \left[ f^{-11} f^{1-1} - f_{-11} f_{1-1} \right]
	\mbox{ , } \mbox{ } f_{44}''^{(1)} = \left(\frac{1}{m_e R_A}\right)^2 \tilde f_{44}''^{(1)}
\end{equation}
\begin{equation}
	\tilde f_{13}''^{(1)} = 2 f{^{-1}}_1 {f_1}^{-1} 
	\mbox{ , } \mbox{ } f_{13}''^{(1)} = \left(\frac{E_1 - E_2}{m_e}\right) \tilde f_{13}''^{(1)}
\end{equation}
\begin{equation}
	\tilde f_{14}''^{(1)} = f{^{-1}}_1 \left( - f_{1,-1} + f^{1,-1} \right) + {f_1}^{-1} \left( - f_{-1,1} + f^{-1,1} \right)
	\mbox{ , } \mbox{ } f_{14}''^{(1)} = \left(\frac{1}{m_e R_A}\right) \tilde f_{14}''^{(1)}
\end{equation}
\begin{equation}
	\tilde f_{34}''^{(1)} = f{^{-1}}_1 \left( f_{1,-1} - f^{1,-1} \right) + {f_1}^{-1} \left( f^{-1,1} - f_{-1,1} \right)
	\mbox{ , } \mbox{ } f_{34}''^{(1)} = \left(\frac{E_1 - E_2}{m_e^2 R_A}\right) \tilde f_{34}''^{(1)}
\end{equation}
\begin{equation}
	\tilde f_{43}''^{(1)} = -  f{^{-1}}_1 \left( f_{1,-1} + f^{1,-1} \right) + {f_1}^{-1} \left( f^{-1,1} + f_{-1,1} \right) 
	\mbox{ , } \mbox{ } f_{43}''^{(1)} = \left(\frac{E_1 - E_2}{m_e^2 R_A}\right) \tilde f_{43}''^{(1)}
\end{equation}
%\color{red}
We do not consider here interference terms between rank-1 products $\bar{\Pi}%
_{3}$ and rank-0 products $\Pi _{4,5}$.

\end{appendix}

\end{document}